\documentclass[11pt]{article}
\usepackage{tikz}
\usetikzlibrary{patterns}
\usepackage[normalem]{ulem}
\usepackage{subfigure}
\usepackage{graphicx}  
\usepackage{hyperref}
\usepackage{epsfig,subfigure}  
\usepackage{color,soul}
\usepackage{enumerate}   
 
\usepackage{xcolor,cancel}
 
\usepackage{epsfig,subfigure}
\usepackage[cmex10,fleqn]{amsmath}
\usepackage{graphicx}
\usepackage{color}
\usepackage{amssymb}
\usepackage{amsfonts}
\usepackage{hyperref}
\usepackage{verbatim}
\usepackage{stfloats}
\usepackage[bookmarks=false]{}

\makeatletter
\newcommand{\crossout}[1]{%
  \begingroup
  \settowidth{\dimen@}{#1}%
  \setlength{\unitlength}{0.05\dimen@}%
  \settoheight{\dimen@}{#1}%
  \count@=\dimen@
  \divide\count@ by \unitlength
  \count0=20 \count4=\count@
  \loop
  \count2=\count0 
  \divide\count2\count4 \multiply\count2\count4
  \ifnum\count2<\count0
    \advance\count0 -\count2 
    \count2=\count0
    \count0=\count4
    \count4=\count2
  \repeat
  \count0=20 \divide\count0\count4
  \count2=\count@ \divide\count2\count4
  \begin{picture}(0,0)
  \put(0,0){\line(\count0,\count2){20}}
  \put(0,\count@){\line(\count0,-\count2){20}}
  \end{picture}%
  #1%
  \endgroup
}
\makeatother

\newtheorem{theorem}{Theorem}

\newtheorem{definition}{Definition}
\newtheorem{lemma}{Lemma}

\begin{document}
\date{}
\title{Sum-set Inequalities from Aligned Image Sets: \\ Instruments for Robust GDoF Bounds
}
\author{ \normalsize Arash Gholami Davoodi and Syed A. Jafar \\
{\small Center for Pervasive Communications and Computing (CPCC)}\\
{\small University of California Irvine, Irvine, CA 92697}\\
{\small \it Email: \{gholamid, syed\}@uci.edu}
}
\maketitle

\allowdisplaybreaks
\begin{abstract}
We present sum-set inequalities specialized to the generalized degrees of freedom (GDoF) framework. These are information theoretic lower bounds on the entropy of bounded density linear combinations of discrete, power-limited dependent random variables in terms of the joint entropies of arbitrary linear combinations of new random variables that are obtained by  power level partitioning of the original random variables. These bounds generalize the aligned image sets approach, and are useful instruments to obtain GDoF characterizations for wireless networks, especially with multiple antenna nodes, subject to arbitrary channel strength and channel uncertainty levels. To demonstrate the utility of these  bounds, we consider a non-trivial instance of wireless networks -- a two user interference channel with different number of antennas at each node, and different levels of partial channel knowledge available to the transmitters.  We obtain tight GDoF characterization for specific instance of this channel with the aid of sum-set inequalities.
\end{abstract}

\section{Introduction}
Originating in additive combinatorics, sum-set inequalities are bounds on the cardinalities of sum-sets (given $X_1,X_2$, the sumset $X_1+X_2\triangleq\{x_1+x_2: x_1\in X_1,x_2\in X_2\}$). Crossing over to network information theory, sum-set inequalities represent bounds on the entropies of sums of random variables, typically expressed in terms of the entropies of the constituent random variables. Prominent examples of such inequalities include Ruzsa's sum-triangle inequality in additive combinatorics \cite{Ruzsa} and the entropy power inequality in information theory \cite{Cover_Thomas}. Sum-set inequalities are essential to the study of the capacity of wireless interference networks. This is particularly true for the studies of capacity approximations known as generalized degrees of freedom (GDoF) \cite{Etkin_Tse_Wang} through deterministic models \cite{Avestimehr_Diggavi_Tse} which de-emphasize the additive noise to place the focus exclusively on the interactions between signals. Received signals in wireless networks are comprised of sums (more generally, linear combinations) of codewords from various codebooks, sent from various transmitters. GDoF optimal schemes seek  to maximize the entropy of received linear combinations of signals where they are desired, while simultaneously minimizing the entropy of received linear combinations of the same signals where they are undesired (e.g., by zero-forcing or interference alignment \cite{Cadambe_Jafar_int}). The fundamental constraints on the structure of sum-sets, as revealed by sum-set inequalities are therefore the critical determinants of the GDoF of wireless interference networks. However, in spite of much recent progress in translating sum-set inequalities from additive combinatorics to network information theory \cite{Madiman}, the structure of sum sets  remains scarcely understood, and continues to be an impediment for GDoF characterizations. In fact, the intricacies of the sum-set structure are such that even a coarse metric like the degrees of freedom (DoF) for  constant channel realizations turns out to be sensitive to fragile details of no conceivable practical relevance -- e.g., whether the channel coefficients take rational or irrational values \cite{Etkin_Ordentlich}.  

Useful insights need robust models and metrics which respond predominantly to those parameters that are known to be of the greatest practical significance. For  wireless interference networks, the most significant aspects include the interplay of spatial dimensions (especially if multiple antennas are involved) with channel strengths and channel uncertainty levels \cite{Jafar_FnT}. Fortunately, the GDoF framework incorporates all three -- spatial dimensions, channel uncertainties and channel strength levels. Furthermore, the fragile aspects of the GDoF metric may be avoided by restricting channel state information at the transmitters (CSIT) to finite precision. 

The study of DoF under finite precision channel knowledge was initiated by Lapidoth et al. in  \cite{Lapidoth_Shamai_Wigger_BC},  leading to a conjecture on the collapse of DoF. In spite of various attempts at proving or disproving the conjecture the conjecture remained open for a decade. It was ultimately settled using an approach based on a combinatorial accounting of the size of the aligned image sets (AIS) under finite precision channel knowledge, in short the AIS approach in \cite{Arash_Jafar_PN}. The AIS approach modeled finite precision channel knowledge as the assumption that from the transmitters' perspective, all joint and conditional probability density functions of channel coefficients exist and are bounded. The bounded density assumption was found to be compatible with various levels of channel strengths and channel knowledge. The AIS approach was further developed to fully characterize the GDoF of the $2$ user MISO BC (broadcast channel with two antennas at the transmitter and one antenna at each of the two receivers) for arbitrary channel strength levels and arbitrary channel uncertainty levels for each channel coefficient, establishing the GDoF optimality of robust schemes in all cases \cite{Arash_Jafar_TC}. It has also led to GDoF characterizations for the $K$ user symmetric IC under finite precision CSIT \cite{Arash_Jafar_IC}, symmetric instances of $K$ user MISO BC  \cite{Arash_Bofeng_Jafar_BC}, symmetric DoF of interference networks with finite precision CSIT and perfect CSIR \cite{Arash_Jafar_Coherence}, and GDoF of $2$ user symmetric MIMO IC with partial CSIT \cite{Arash_Jafar_MIMOsym_ArXiv}. Indeed, there exists the distant but exciting possibility that the AIS approach may ultimately lead us to the GDoF characterizations of broad classes of wireless networks. If so, then the resulting comprehensive and fundamental understanding of  these complex networks -- the interplay between spatial dimensions, channel strengths, and channel uncertainty levels -- would be invaluable. However, in order to get there, it is evident that a robust understanding of sum-sets will be needed. Specifically, there is the need to identify the key sum-set inequalities for signals subject to arbitrary power levels under the robust bounded density assumption. This is the goal that we pursue in this work. 

 The paper is organized as follows. Section $2$ provides the necessary definitions. The main results, i.e., the sum-set inequalities  are presented in Section 3 through progressively generalized theorems for ease of exposition, starting from a single-letter single-antenna form to the general multi-letter multi-antenna form that is needed to derive GDoF outer bounds for  MIMO networks. In Section 4, we present an example to show how these sum-set inequalities allow us to obtain new GDoF characterizations for non-trivial networks under partial CSIT\footnote{Channel uncertainty and channel strengths are interchangeable to a certain extent in MIMO interference networks, because the channel uncertainty level governs the strength of residual interference when signals are zero-forced. This is previously noted in \cite{Gou_Jafar}.} that were previously open. The  example is comprised of a $2$ user MIMO interference channel (IC) where the two transmitters are equipped with $M_1=5$ and $M_2=5$ antennas, their corresponding receivers  with $N_1=2$ and $N_2=3$ antennas, the channel strength parameters are chosen to be $(\alpha_{11},\alpha_{12},\alpha_{21},\alpha_{22})=(1,\frac{3}{4},\frac{2}{3},1)$ and partial CSIT parameters are chosen to be $\beta_{12}=1/4$ and $\beta_{21}=1/3$. Remarkably, building upon these insights,  in \cite{Bofeng_Arash_Jafar_ArXiv} we have found that the sum-set inequalities allow us to fully characterize the GDoF region of the MIMO IC with arbitrary antenna configurations $(M_1,M_2,N_1,N_2)$ under arbitrary levels of partial CSIT.  Moreover, sum-set inequalities allowed the authors to characterize the full GDoF region of the two user MIMO BC with  arbitrary antenna configurations $(M,N_1,N_2)$ under arbitrary levels of partial CSIT in \cite{Arash_Jafar_MIMOBC_Region}.

{\it Notation:} For $n\in\mathbb{N}$, define the notation $[n]=\{1,2,\cdots,n\}$. The cardinality of a set $A$ is denoted as $|A|$. The notation~ $X^{[n]}$ stands~ for $\{X(1), X(2), \cdots, X(n)\}$. Moreover,  $X_{i}^{[n]}$ also stands for $\{X_i(t): \forall t\in[n]\}$. The
support of a random variable $X$ is denoted as supp$(X)$. The sets $\mathbb{R}$ and $\mathbb{R}^n$ stand for the set of real numbers and the set of all $n$-tuples of real numbers respectively. Moreover, the set $\mathbb{R}^{2+}$ is defined as the set of all pairs of non-negative numbers. If $A$ is a set of random variables, then $H(A)$ refers to the joint entropy of the random variables in $A$. Conditional entropies, mutual information and joint and conditional probability densities of sets of random variables are similarly interpreted. Moreover, we use the Landau $O(\cdot)$ and $o(\cdot)$ notations as follows. For  functions $f(x), g(x)$ from $\mathbb{R}$ to $\mathbb{R}$, $f(x)=O(g(x))$ denotes that $\limsup_{x\rightarrow\infty}\frac{|f(x)|}{|g(x)|}<\infty$.  $f(x)=o(g(x))$ denotes that $\limsup_{x\rightarrow\infty}\frac{|f(x)|}{|g(x)|}=0$. We use $\mathbb{P}(\cdot)$ to denote the probability function $\mbox{Prob}(\cdot)$. For any real number $x$ we define $\lfloor x\rfloor$ as the largest integer that is smaller than or equal to $x$ when $x>0$,  the smallest integer that is larger than or equal to $x$ when $x<0$, and $x$ itself when $x$ is an integer. We also define $(x)^+$ as maximum of the number $x$ and $0$, i.e., $\max(x,0)$. The number $X_{r,s}$ may be represented as $X_{rs}$ if there is no cause for ambiguity.

\section{Definitions}
The information theoretic sum-set inequalities that we seek are motivated by the GDoF framework. Since in the next section we present general statements of sum-set inequalities, here we only present definitions needed for Section \ref{Results}. The definitions needed for the MIMO IC  setting that we use as an example, are presented in Section \ref{MIMOIC}.

\begin{definition}[Power Levels] Consider integer valued  variables $X_i$ over alphabet $\mathcal{X}_{\lambda_i}$,
\begin{eqnarray}
\mathcal{X}_{\lambda_i}&\triangleq&\{0,1,2,\cdots,\bar{P}^{\lambda_i}-1\}
\end{eqnarray}
where $\bar{P}^{\lambda_i}$ is a compact notation for $\left\lfloor\sqrt{P^{\lambda_i}}\right\rfloor$. We refer to $P\in\mathbb{R}_+$ as \emph{power}, and are primarily interested in limits as  $P\rightarrow\infty$. Quantities that do not depend on $P$ will be referred to as constants. The constant $\lambda_i\in\mathbb{R}_+$ denotes the \emph{power level} of $X_i$. 
\end{definition}
{We are interested in sum-set inequalities in terms of entropies of random variables such as $X_i$, normalized by $\log{\bar{P}}$ as ${P}\rightarrow\infty$, while the power levels $\lambda_i$ are held fixed. All the sumset inequalities in this work hold in this asymptotic sense, i.e., while disregarding terms that are negligible relative to $\log({P})$. Such terms are denoted as $o(\log({P}))$ terms.}

\begin {definition}\label{powerlevel} For any nonnegative real numbers $X$,  $\lambda_1$ and $\lambda_2$, define  $(X)_{\lambda_1}$ and $(X)^{\lambda_2}_{\lambda_1}$  as,
 \begin{eqnarray}
(X)_{\lambda_1}&\triangleq& X-\bar{P}^{\lambda_1} \left \lfloor \frac{X}{\bar{P}^{\lambda_1}} \right \rfloor\\
(X)^{\lambda_2}_{\lambda_1}&\triangleq&\left \lfloor \frac{X-\bar{P}^{\lambda_2}\left \lfloor\frac{X}{ {\bar{P}}^{\lambda_2}}\right \rfloor }{{\bar{P}}^{\lambda_1}} \right \rfloor\label{mid}
\end{eqnarray}
\end {definition}

\begin{figure}[h]
\begin{center}
\begin{tikzpicture}
\draw [black, thick] (1,0) rectangle (2,6) node[midway] {$X$};
\draw[thick, <->] (2.18,0)--(2.18,2.5) node[midway, right]{$\lambda_1$};
\draw[thick, <->] (2.18,2.5)--(2.18,4) node[midway, right]{$\lambda_2$};
\draw[thick, <->] (2.18,4)--(2.18,6) node[midway, right]{$\lambda_3$};
\draw [help lines] (2,0)--(3,0);
\draw [help lines] (2,2.5)--(5,2.5);
\draw [help lines] (2,4)--(7,4);
\draw [help lines] (2,6)--(7,6);
\draw [thick](3,0)  -| (4,2.5) 
    node[pos=0.75,right] {$(X)_{\lambda_1}$}
    -| (3,0);

\draw [thick](5,2.5)  -| (6,4) 
    node[pos=0.75,right] {$(X)_{\lambda_1}^{\lambda_1+\lambda_2}$}
    -| (5,2.5);

\draw [thick](7,4)  -| (8,6) 
    node[pos=0.75,right] {$(X)_{\lambda_1+\lambda_2}^{\lambda_1+\lambda_2+\lambda_3}$}
    -| (7,4);
\end{tikzpicture}
\caption[]{Conceptual depiction of an arbitrary variable $X\in\mathcal{X}_{\lambda_1+\lambda_2+\lambda_3}$, and its power-level partitions $(X)_{\lambda_1}$, $(X)_{\lambda_1}^{\lambda_1+\lambda_2}$ and $(X)_{\lambda_1+\lambda_2}^{\lambda_1+\lambda_2+\lambda_3}$.}\label{tg}
\end{center}
\end{figure}
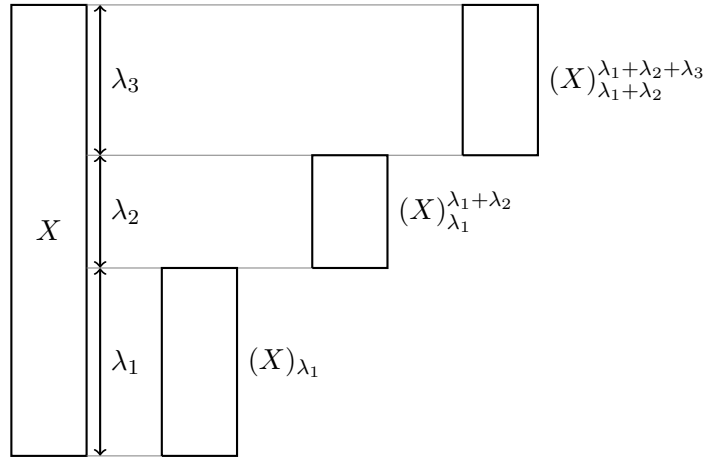

In words, for any $X\in\mathcal{X}_{\lambda_1+\lambda_2}$, $(X)^{\lambda_1+\lambda_2}_{\lambda_1}$ retrieves the top $\lambda_2$ power levels of $X$, while $(X)_{\lambda_1}$ retrieves the bottom $\lambda_1$ levels of $X$.  $(X)^{\lambda_3}_{\lambda_1}$ retrieves only the part of $X$ that lies between power levels $\lambda_1$ and $\lambda_3$.  Note that $X\in \mathcal{X}_\lambda$ can be expressed as $X={\bar{P}^{\lambda_1}}{(X)}_{\lambda_1}^{\lambda}+{(X)}_{\lambda_1}$ for $0\leq\lambda_1<\lambda$.  Equivalently, suppose $X_1\in\mathcal{X}_{\lambda_1}$, $X_2\in\mathcal{X}_{\lambda_2}$, $0<\lambda_2$ and $X=X_1+X_2\bar{P}^{\lambda_1}$. Then $X_1={(X)}_{\lambda_1}$, $X_2={(X)}^{\lambda_1+\lambda_2}_{\lambda_1}$.  A conceptual illustration of power level partitions is shown in 
Figure \ref{tg}.

\begin {definition}
 For  the vector ${\bf V}=\begin{bmatrix}v_1&v_2&\cdots&v_k\end{bmatrix}^T$, we define  $({\bf V})_{\lambda_1}$ and $({\bf V})^{\lambda_2}_{\lambda_1}$ as,
 \begin{eqnarray}
({\bf V})_{\lambda_1}&\triangleq& \begin{bmatrix}(v_1)_{\lambda_1}&(v_2)_{\lambda_1}&\cdots&(v_k)_{\lambda_1}\end{bmatrix}^T\\
({\bf V})^{\lambda_2}_{\lambda_1}&\triangleq& \begin{bmatrix}(v_1)^{\lambda_2}_{\lambda_1}&(v_2)^{\lambda_2}_{\lambda_1}&\cdots&(v_k)^{\lambda_2}_{\lambda_1}\end{bmatrix}^T
\end{eqnarray}
\end{definition}

\begin{definition}[Bounded Density Channel Coefficients]\label{defbounded} Bounded density channels are represented by a set of real valued random variables, $\mathcal{G}$ such that the magnitude of each random variable $g\in\mathcal{G}$ is bounded away from zero and infinity, $0<\Delta_1\leq |g|\leq\Delta_2<\infty$, for some constants $\Delta_1,\Delta_2$, and there exists a finite positive constant $f_{\max}$, such that for all finite cardinality disjoint subsets $\mathcal{G}_1, \mathcal{G}_2$ of $\mathcal{G}$, the joint probability density function of all random variables in $\mathcal{G}_1$, conditioned on all random variables in $\mathcal{G}_2$, exists and is bounded above by $f_{\max}^{|\mathcal{G}_1|}$. 
\end{definition}

\begin{definition}[Arbitrary Channel Coefficients] Let $\mathcal{H}$ be a set of arbitrary constant values that are bounded above by $\Delta_2$, i.e., if $h\in\mathcal{H}$ then $|h|\leq\Delta_2<\infty$.
\end{definition}

\begin {definition}\label{deflc} For {real}  numbers {\color{black}$x_1\in\mathcal{X}_{\eta_1},x_2\in\mathcal{X}_{\eta_2},\cdots,x_k\in\mathcal{X}_{\eta_k}$} and the vectors $\vec{\gamma}=(\gamma_1,\gamma_2,\cdots, \gamma_k)$ and $\vec{\delta}=(\delta_1,\delta_2,\cdots, \delta_k)$ define the notations $L_j^b(x_i,1\le i\le k)$, $L_j(x_i,1\le i\le k)$, {\color{black}$L_j^{b\vec{\gamma}\vec{\delta}}(x_i,1\le i\le k)$ and $L_j^{\vec{\gamma}\vec{\delta}}(x_i,1\le i\le k)$ } to represent,
\begin {eqnarray}
L^b_j(x_1,x_2,\cdots,x_k )&=&\sum_{1\le i\le k} \lfloor g_{j_i}x_i\rfloor\\
L_j(x_1,x_2,\cdots,x_k)&=&\sum_{1\le i\le k} \lfloor h_{j_i}x_i\rfloor\\
L_j^{b\vec{\gamma}\vec{\delta}}(x_1,x_2,\cdots,x_k)&=&\sum_{1\le i\le k} \lfloor g_{j_i}(x_i)^{\gamma_i}_{\delta_i}\rfloor\\
L_j^{\vec{\gamma}\vec{\delta}}(x_1,x_2,\cdots,x_k)&=&\sum_{1\le i\le k} \lfloor h_{j_i}(x_i)^{\gamma_i}_{\delta_i}\rfloor
\end{eqnarray}
for  distinct random variables $g_{j_i}\in\mathcal{G}$, some arbitrary  real valued and finite constants {$h_{j_i}\in\mathcal{H}$} and  some arbitrary non-negative real valued constants $\delta_i,\gamma_i$. For  the vector $V=\begin{bmatrix}v_1&v_2&\cdots&v_k\end{bmatrix}^T$ we also define the notations $L^b_j(V)$ and $L_j(V)$ to represent,
\begin {eqnarray}
L^b_j(V)=\sum_{1\le i\le k} \lfloor g_{j_i}v_i\rfloor\\
L_j(V)=\sum_{1\le i\le k} \lfloor h_{j_i}v_i\rfloor
\end{eqnarray}
for  distinct random variables $g_{j_i}\in\mathcal{G}$ and $h_{j_i}\in\mathcal{H}$.
\end {definition}

\begin{figure}[!h]
\begin{tikzpicture}[scale=0.9, baseline=(current bounding box.center)]

\begin{scope}[shift={(-0.5,0)}]
\draw[ thick, pattern = crosshatch dots, pattern color =blue] (0,0) rectangle (1,2) node[fill=white, midway]{$x_2$};
\draw[ thick, pattern = crosshatch, pattern color =blue] (0,2) rectangle (1,5)  node[fill=white, midway]{$x_1$};
\draw (0.5, 5) node[above]{$X_1$};
\draw[thick, <->](-0.2,0)--(-0.2,2) node[midway, left]{$\eta_2$};
\draw[thick, <->](1.2,0)--(1.2,0.5) -- node[ right]{$\gamma_2$} (1.2,1.5);
\draw[thick, <->](1.4,0)--(1.4,0.5) node[midway, right]{$\delta_2$};
\draw[thick, pattern = crosshatch dots, pattern color=blue](2,0.5) rectangle (3,1.5);
\draw (2.5,1.5) node[above]{$(x_2)^{\gamma_2}_{\delta_2}$};
\draw[help lines] (1,0.5)--(2,0.5);
\draw[help lines] (1,1.5)--(2,1.5);
\draw[thick, <->](-0.2,2)--(-0.2,5) node[midway, left]{$\eta_1$};
\draw[thick, <->](1.2,2)--(1.2,2.5) -- node[ right]{$\gamma_1$} (1.2,4.5);
\draw[thick, <->](1.4,2)--(1.4,2.5) node[midway, right]{$\delta_1$};
\draw[thick, pattern = crosshatch, pattern color=blue](2,2.5) rectangle (3,4.5);
\draw[help lines] (1,2.5)--(2,2.5);
\draw[help lines] (1,4.5)--(2,4.5);
\draw (2.5,4.5) node[above]{$(x_1)^{\gamma_1}_{\delta_1}$};
\end{scope}

\begin{scope}[shift={(4.3,1.5)}]
\draw[very thick, rounded corners] (-0.5, -1.5) rectangle (5.5, 3.5);

\draw[thick, pattern = crosshatch, pattern color=blue](0,0) rectangle (1,2);
\draw[thick, pattern = crosshatch dots, pattern color=blue](1,0) rectangle (2,1);
\draw[thick, pattern = grid, pattern color=red](2,0) rectangle (3,2.4);
\draw[thick, pattern = vertical lines, pattern color=red](3,0) rectangle (4,1.2);
\draw[thick, |-|] (0,-0.2)--(4,-0.2) node[midway, below]{$A=L^{\vec{\gamma}\vec{\delta}}$};
\draw[thick,<->] (4.2,0)--(4.2,2.4) node[midway,right]{\footnotesize $\mathcal{T}(A)$};
\draw[help lines] (0,2.4)--(4.2,2.4);
\end{scope}

\begin{scope}[shift={(11,0)}]
\draw (2.5, 6) node[above]{$X_2$};

\draw[ thick, pattern = vertical lines, pattern color =red] (2,3) rectangle (3,6) node[fill=white, midway]{$x_3$};
\draw[thick, <->](3.2,3)--(3.2,6) node[midway, right]{$\eta_3$};

\draw[ thick, pattern = vertical lines, pattern color =red] (0,4.5) rectangle (1,5.7);
\draw[thick, <->](1.8,3)--(1.8,4.5) -- node[left]{$\gamma_3$} (1.8,5.7);
\draw[thick, <->](1.6,3)-- node[left]{$\delta_3$} (1.6,4.5);
\draw[help lines] (1,4.5)--(2,4.5);
\draw[help lines] (1,5.7)--(2,5.7);
\draw (0.5, 5.7) node[above]{$(x_3)^{\gamma_3}_{\delta_3}$};

\draw[ thick, pattern = grid, pattern color =red] (2,0) rectangle (3,3) node[fill=white, midway]{$x_4$};
\draw[thick, <->](3.2,0)--(3.2,3) node[midway, right]{$\eta_4$};

\draw[ thick, pattern = grid, pattern color =red] (0,0.4) rectangle (1,2.8);
\draw (0.5, 2.8) node[above]{$(x_4)^{\gamma_4}_{\delta_4}$};
\draw[thick, <->](1.6,0)--(1.6,0.4) node[midway, left]{$\delta_4$};
\draw[thick, <->](1.8,0)--(1.8,2.8) node[midway, left]{$\gamma_4$};
\draw[help lines] (1,0.4)--(2,0.4);
\draw[help lines] (1,2.8)--(2,2.8);
\end{scope}
\end{tikzpicture}
\caption{Visual illustration of $L^{\vec{\gamma}\vec{\delta}}$ and $\mathcal{T}(A)$. In this example, $x_1\in \mathcal{X}_{\eta_1}$ and $x_2\in\mathcal{X}_{\eta_2}$ are obtained as partitions of $X_1\in\mathcal{X}_{\eta_1+\eta_2}$. Similarly, $x_3\in \mathcal{X}_{\eta_3}$ and $x_4\in\mathcal{X}_{\eta_4}$ are  obtained as partitions of $X_2\in\mathcal{X}_{\eta_3+\eta_4}$. Note that $(\gamma_i, \delta_i)$ are only used to further trim the size of $x_i$, yielding $(x_i)_{\delta_i}^{\gamma_i}$ as the trimmed versions. These trimmed variables are then combined with arbitrary coefficients to produce $A=L^{\vec{\gamma}\vec{\delta}}$. Finally, note that  $\mathcal{T}(A)$ represents the size (power level) of the largest trimmed variable involved in $L^{\vec{\gamma}\vec{\delta}}$.}\label{fig:TA}
\end{figure}
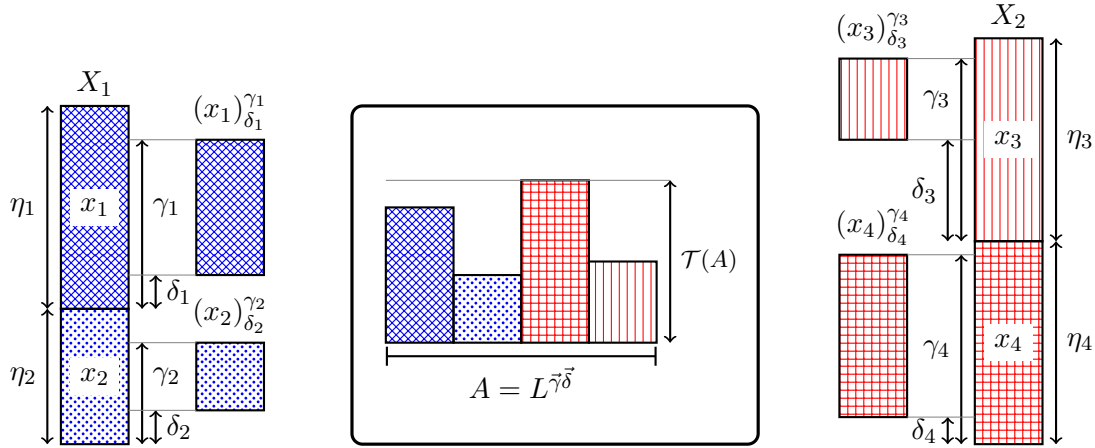

Note that, the subscript $j$ is used to distinguish among multiple linear combinations, and may be dropped if there is no potential for ambiguity. We refer to the $L^b$ functions as  bounded density linear combinations. 
{\color{black}
\begin {definition}\label{def:length} For the linear combinations $A=L^{b\vec{\gamma}\vec{\delta}}(x_i,1\le i\le k)$ and $B=L^{\vec{\gamma}\vec{\delta}}(x_i,1\le i\le k)$ where $x_1\in\mathcal{X}_{\eta_1},x_2\in\mathcal{X}_{\eta_2},\cdots,x_k\in\mathcal{X}_{\eta_k}$ we define $\mathcal{T}(A)$ and $\mathcal{T}(B)$ as,
\begin {eqnarray}
\mathcal{T}(A)=\mathcal{T}(B)=\max_{j\in[k]}\min(\eta_j,(\gamma_{j}-\delta_{j})^+).\label{total}
\end{eqnarray}
\end {definition}}
Note that the terminology from Definition \eqref{deflc} is invoked in Definition \eqref{def:length}. Figure \ref{fig:TA} provides a visual illustration of $L^{\vec{\gamma}\vec{\delta}}$ and $\mathcal{T}(A)$. From the definition of $\mathcal{T}(A)$ and $\mathcal{T}(B)$ in \eqref{total}, it follows that,
\begin {eqnarray}
A&\in&\{a:a\in\mathbb{Z}, |a| \le k\Delta_2\bar{P}^{\mathcal{T}(A)}\}\label{re1}\\
B&\in&\{b:b\in\mathbb{Z}, |b| \le k\Delta_2\bar{P}^{\mathcal{T}(B)}\}\label{re2}
\end{eqnarray}
This is because all elements of $\mathcal{G},\mathcal{H}$ are bounded from above by $\Delta_2$.

\begin{definition}\label{defvec} For any vector $V=\begin{bmatrix}v_1&\cdots&v_k\end{bmatrix}^T$ and non-negative integer numbers $m$ and $n$ less than $k$, define 
\begin{eqnarray}
V_{m,n}&\triangleq&\left\{
\begin{array}{ll}
\begin{bmatrix}v_{m+1}&\cdots&v_{m+n}\end{bmatrix}^T,& m+n\le k\\
\begin{bmatrix}v_{m+1}&\cdots&v_k&v_1&\cdots&v_{m+n-k}\end{bmatrix}^T,& k<m+n
\end{array}
\right.
\end{eqnarray}
Moreover, for the two vectors $V=\begin{bmatrix}v_1&\cdots&v_{k_1}\end{bmatrix}^T$ and $W=\begin{bmatrix}w_1&\cdots&w_{k_2}\end{bmatrix}^T$ define $V\bigtriangledown W$ as $\begin{bmatrix}v_1&\cdots&v_{k_1}&w_1&\cdots&w_{k_2}\end{bmatrix}^T$.
\end{definition} 

\section{Results}\label{Results}
\begin{theorem} \label{Theorem AIS01} 
For $\lambda_1\geq \lambda_2\geq 0$, consider random variables $X_{1},X_{2} \in \mathcal{X}_{\lambda_1+\lambda_2}$, all independent of $\mathcal{G}$,  and define 
\begin {eqnarray}
Z&=&L^b(X_1,X_2)\\
Z_{1}&=& L_{1}((X_{1})_{\lambda_1}^{\lambda_1+\lambda_2},(X_{2})_{\lambda_1}^{\lambda_1+\lambda_2})\\
Z_{2}&=& L_{2}((X_{1})_{\lambda_1},(X_{2})_{\lambda_1},(X_{1})_{\lambda_1}^{\lambda_1+\lambda_2},(X_{2})_{\lambda_1}^{\lambda_1+\lambda_2})
\end{eqnarray}
then
\begin {eqnarray}
H(Z\mid \mathcal{G})&\geq&H(Z_1,Z_2)+o(\log{\bar{P}})\label{dsd1old}
\end{eqnarray} 
\end{theorem}
The  following remarks place Theorem \ref{Theorem AIS01} in perspective and discuss some of its generalizations.
\begin{enumerate}
\item Let $\mathcal{G}(Z)\subset\mathcal{G}$ denote the set of all bounded density channel coefficients that appear in $Z=L^b(X_2,X_2)$, and let $W$ be a random variable such that conditioned on any $\mathcal{G}_o\subset (\mathcal{G}/\mathcal{G}(Z))\cup \{W\}$, the channel coefficients $\mathcal{G}(Z)$ satisfy the bounded density assumption. Then (\ref{dsd1}) generalizes to the following conditional form.
\begin{eqnarray}
H(Z\mid \mathcal{G},W)&\geq&H(Z_1,Z_2|W)+o(\log{\bar{P}})\label{dsd1}
\end{eqnarray}
 The proof presented in Appendix \ref{proof:AIS01} covers this generalization. In various applications of these sum-set inequalities, the conditioning variable $W$ could represent terms such as  $L_3(X_1,X_2)$, $(X_1)_{\delta}^{\gamma}$ or $L_4^b((X_1)_{\frac{1}{2}},X_2)$.
 

\item A typical restriction in information theoretic sum-set inequalities is the independence of random variables. In contrast, note that the statement of Theorem \ref{Theorem AIS01} also holds for dependent random variables.

\item Since the linear combining coefficients $h_i$ involved in $L_1$ and $L_2$ can take arbitrary (including zero) values, several specializations of Theorem \ref{Theorem AIS01} follow immediately, e.g.,
\begin{eqnarray}
H(Z|\mathcal{G})&\geq&H(Z_1, L_{2}((X_{1})_{\lambda_1},(X_{2})_{\lambda_1}))+o(\log{\bar{P}})\\
H(Z|\mathcal{G})&\geq&H((X_{1})_{\lambda_1}^{\lambda_1+\lambda_2},(X_{2})_{\lambda_1}^{\lambda_1+\lambda_2})+o(\log{\bar{P}})
\end{eqnarray}
Figure \ref{eexample1} visually illustrates these inequalities in terms of the power levels.

\begin{figure}[!h] 
\begin{eqnarray*}
H\left(\begin{tikzpicture}[scale=0.9, baseline=(current bounding box.center)]
\draw[ thick, pattern=crosshatch dots, pattern color=blue] (0,0) rectangle (1,1.3);
\draw[ thick, pattern=crosshatch, pattern color=blue] (0,1.3) rectangle (1,2.3);
\draw[ thick, pattern=crosshatch dots, pattern color=red] (1,0) rectangle (2,1.3);
\draw[ thick, pattern=crosshatch, pattern color=red] (1,1.3) rectangle (2,2.3);
\draw[thick, <->] (2.2,0)--(2.2,1.3) node[midway, right]{$\lambda_1$};
\draw[thick, <->] (2.2,1.3)--(2.2,2.3) node[midway, right]{$\lambda_2$};
\draw[thick, |-|] (0,-0.2)--(2,-0.2) node[midway, below]{$Z=L^b$};
\draw (0.5, 2.3) node[above]{$X_1$};
\draw (1.5, 2.3) node[above]{$X_2$};
\end{tikzpicture}
\right)
&\geq&
H\left(
\begin{tikzpicture}[scale=0.9, baseline=(current bounding box.center)]
\draw[ thick, pattern=crosshatch dots, pattern color=blue] (0,0) rectangle (1,1.3);
\draw[ thick, pattern=crosshatch dots, pattern color=red] (1,0) rectangle (2,1.3);
\draw[ thick, pattern=crosshatch, pattern color=blue] (2,0) rectangle (3,1);
\draw[ thick, pattern=crosshatch, pattern color=red] (3,0) rectangle (4,1);
\draw[thick, |-|] (0,-0.2)--(4,-0.2) node[midway, below]{$Z_1=L_1$};
\path (4.2,0 ) node[right]{\Huge ,};
\draw[ thick, pattern=crosshatch, pattern color=blue] (5,0) rectangle (6,1);
\draw[ thick, pattern=crosshatch, pattern color=red] (6,0) rectangle (7,1);
\draw (0.5,1.3) node[above]{\tiny $(X_1)_{\lambda_1}$};
\draw (1.5,1.3) node[above]{\tiny $(X_2)_{\lambda_1}$};
\draw (2.5,1) node[above]{\tiny $(X_1)_{\lambda_1}^{\lambda_2}$};
\draw (3.5,1) node[above]{\tiny $(X_2)_{\lambda_1}^{\lambda_2}$};
\draw (5.5,1) node[above]{\tiny $(X_1)_{\lambda_1}^{\lambda_2}$};
\draw (6.5,1) node[above]{\tiny $(X_2)_{\lambda_1}^{\lambda_2}$};
\draw[thick, <->] (4.2,0)--(4.2,1.3) node[midway, right]{$\lambda_1$};
\draw[thick, <->] (7.2,0)--(7.2,1) node[midway, right]{$\lambda_2$};
\draw[thick, |-|] (5,-0.2)--(7,-0.2) node[midway, below]{$Z_2=L_2$};
\end{tikzpicture}
\right)\\
H\left(\begin{tikzpicture}[scale=0.9, baseline=(current bounding box.center)]
\draw[ thick, pattern=crosshatch dots, pattern color=blue] (0,0) rectangle (1,1.3);
\draw[ thick, pattern=crosshatch, pattern color=blue] (0,1.3) rectangle (1,2.3);
\draw[ thick, pattern=crosshatch dots, pattern color=red] (1,0) rectangle (2,1.3);
\draw[ thick, pattern=crosshatch, pattern color=red] (1,1.3) rectangle (2,2.3);
\draw[thick, <->] (2.2,0)--(2.2,1.3) node[midway, right]{$\lambda_1$};
\draw[thick, <->] (2.2,1.3)--(2.2,2.3) node[midway, right]{$\lambda_2$};
\draw[thick, |-|] (0,-0.2)--(2,-0.2) node[midway, below]{$Z=L^b$};
\draw (0.5,2.3) node[above]{$X_1$};
\draw (1.5,2.3) node[above]{$X_2$};
\end{tikzpicture}
\right)
&\geq&
H\left(
\begin{tikzpicture}[scale=0.9, baseline=(current bounding box.center)]
\draw[ thick, pattern=crosshatch dots, pattern color=blue] (0,0) rectangle (1,1.3);
\draw[ thick, pattern=crosshatch dots, pattern color=red] (1,0) rectangle (2,1.3);
\path (2.2,0 ) node[right]{\Huge ,};
\draw[ thick, pattern=crosshatch, pattern color=blue] (3,0) rectangle (4,1);
\draw[ thick, pattern=crosshatch, pattern color=red] (4,0) rectangle (5,1);
\draw (0.5,1.3) node[above]{\tiny $(X_1)_{\lambda_1}$};
\draw (1.5,1.3) node[above]{\tiny $(X_2)_{\lambda_1}$};
\draw (3.5,1) node[above]{\tiny $(X_1)_{\lambda_1}^{\lambda_2}$};
\draw (4.5,1) node[above]{\tiny $(X_2)_{\lambda_1}^{\lambda_2}$};
\draw[thick, <->] (2.2,0)--(2.2,1.3) node[midway, right]{$\lambda_1$};
\draw[thick, <->] (5.2,0)--(5.2,1) node[midway, right]{$\lambda_2$};
\draw[thick, |-|] (0,-0.2)--(2,-0.2) node[midway, below]{$Z_1=L_1$};
\draw[thick, |-|] (3,-0.2)--(5,-0.2) node[midway, below]{$Z_2=L_2$};
\end{tikzpicture}
\right)\\
H\left(\begin{tikzpicture}[scale=0.9, baseline=(current bounding box.center)]
\draw[ thick, pattern=crosshatch dots, pattern color=blue] (0,0) rectangle (1,1.3);
\draw[ thick, pattern=crosshatch, pattern color=blue] (0,1.3) rectangle (1,2.3);
\draw[ thick, pattern=crosshatch dots, pattern color=red] (1,0) rectangle (2,1.3);
\draw[ thick, pattern=crosshatch, pattern color=red] (1,1.3) rectangle (2,2.3);
\draw[thick, <->] (2.2,0)--(2.2,1.3) node[midway, right]{$\lambda_1$};
\draw[thick, <->] (2.2,1.3)--(2.2,2.3) node[midway, right]{$\lambda_2$};
\draw[thick, |-|] (0,-0.2)--(2,-0.2) node[midway, below]{$Z=L^b$};
\draw (0.5,2.3) node[above]{$X_1$};
\draw (1.5,2.3) node[above]{$X_2$};
\end{tikzpicture}
\right)
&\geq&
H\left(
\begin{tikzpicture}[scale=0.9, baseline=(current bounding box.center)]
\path (1.2,0 ) node[right]{\Huge ,};
\draw[ thick, pattern=crosshatch, pattern color=blue] (0,0) rectangle (1,1);
\draw[ thick, pattern=crosshatch, pattern color=red] (2,0) rectangle (3,1);
\draw (0.5,1) node[above]{\tiny $(X_1)_{\lambda_1}^{\lambda_2}$};
\draw (2.5,1) node[above]{\tiny $(X_2)_{\lambda_1}^{\lambda_2}$};
\draw[thick, <->] (1.2,0)--(1.2,1) node[midway, right]{$\lambda_2$};
\draw[thick, <->] (3.2,0)--(3.2,1) node[midway, right]{$\lambda_2$};
\end{tikzpicture}
\right)
\end{eqnarray*}
\caption[]{Illustration of various specializations of Theorem \ref{Theorem AIS01}. On the left is the entropy of  a sum (bounded density linear combination) of two dependent random variables, which is bounded below by joint entropy of two arbitrary linear combinations of constituent random variables. The bounded density assumption for the left hand side is critical. Without it, for example, interference alignment or zero-forcing could be used to immediately violate the last inequality.}\label{eexample1}
\end{figure}
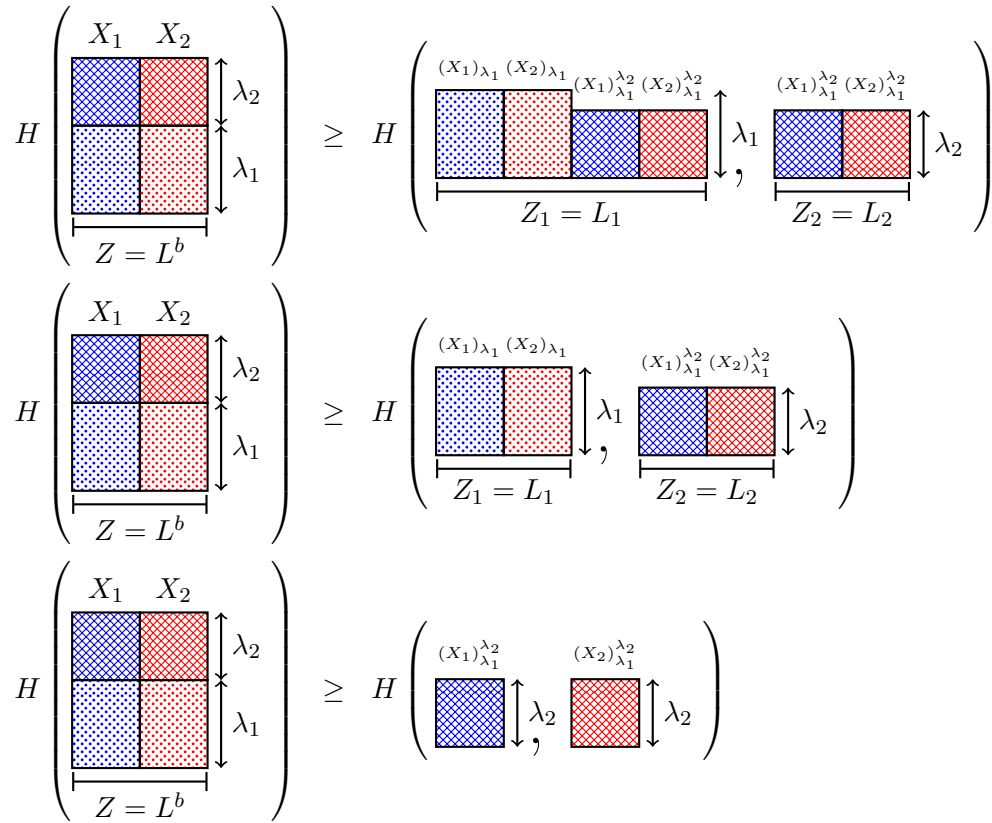

\item Theorem \ref{Theorem AIS01} also holds if $L_1, L_2$ are replaced with bounded density linear combinations, i.e., $L_1^{b}, L_2^{b}$.
\item While in the GDoF framework, Theorem \ref{Theorem AIS01} is typically used when $\lambda_1\geq\lambda_2$ as assumed, it is possible to generalize the result of Theorem \ref{Theorem AIS01} to  allow $\lambda_2\geq\lambda_1$. In that case, the inequality (\ref{dsd1}) becomes $H(Z\mid W,\mathcal{G})\geq H(Z_1, Z_2\mid W)-(\lambda_2-\lambda_1)^+\log(\bar{P})+o(\log{\bar{P}})$. The proof presented in Appendix \ref{proof:AIS01} covers this generalization. 

\item The result of Theorem \ref{Theorem AIS01} lends itself to extensive generalizations in terms of the number of random variables, and the number of power level partitions. Such a generalization is presented in the following theorem. 

\begin{theorem} \label{Theorem AIS02}
Consider {\color{black}$M$ non-negative numbers  $\lambda_1,\cdots,\lambda_M$ and} random variables $X_j \in \mathcal{X}_{\lambda_1+\lambda_2+\cdots+\lambda_M}$, $j\in[N]$ independent of $\mathcal{G}$,  and define 
\begin {eqnarray}
Z&=&L^b(X_1,X_2, \cdots, X_N)\\
Z_{1}&=& L_{1}^{\vec{\gamma}_1\vec{\delta}_1}((X_{j})_{\sum_{r=1}^{i-1}\lambda_r}^{\sum_{r=1}^i\lambda_r}, i\in I_1, j\in[N])\\
Z_{2}&=& L_{2}^{\vec{\gamma}_2\vec{\delta}_2}((X_{j})_{\sum_{r=1}^{i-1}\lambda_r}^{\sum_{r=1}^i\lambda_r}, i\in I_2, j\in[N])\\
&\vdots&\notag\\
Z_{l}&=&L_{l}^{\vec{\gamma}_l\vec{\delta}_l}((X_{j})_{\sum_{r=1}^{i-1}\lambda_r}^{\sum_{r=1}^i\lambda_r}, i\in I_l, j\in[N])
\end{eqnarray}
$I_1, I_2, \cdots, I_l\subset [M]$ such that  $\forall a,b\in[M]$, 
$a<b\Rightarrow m(a)\geq m(b)$, where we define
\begin{eqnarray}
m(a)&=& \min\{i: i\in I_a\}
\end{eqnarray}
If for each $s\in\{1,2,\cdots, l-1\}$,
\begin {eqnarray}
\lambda_1+\lambda_2+\cdots+\lambda_{(m(s)-1)}&\geq&\mathcal{T}(Z_{s+1})+\mathcal{T}(Z_{s+2})+\cdots+\mathcal{T}(Z_l)\label{con1}
\end{eqnarray}
then, 
\begin {eqnarray}
H(Z\mid \mathcal{G}, W)&\geq& H(Z_1,Z_2, \cdots, Z_l|W)+o(\log{\bar{P}})\label{dsd2}
\end{eqnarray}

\end{theorem}
Recall that for any real number $x$, we define $(x)^+=\max(x,0)$.
\begin{figure}[!h] 
\begin{eqnarray*}
H\left(\begin{tikzpicture}[scale=0.7, baseline=(current bounding box.center)]
\draw[ thick, pattern=crosshatch dots, pattern color=blue] (0,0) rectangle (1,3);
\draw[ thick, pattern=crosshatch, pattern color=blue] (0,3) rectangle (1,5);
\draw[ thick, pattern=grid, pattern color=blue] (0,5) rectangle (1,7.5);
\draw[ thick, pattern=vertical lines, pattern color=blue] (0,7.5) rectangle (1,9);
\draw[ thick, pattern=crosshatch dots, pattern color=red] (1,0) rectangle (2,3);
\draw[ thick, pattern=crosshatch, pattern color=red] (1,3) rectangle (2,5);
\draw[ thick, pattern=grid, pattern color=red] (1,5) rectangle (2,7.5);
\draw[ thick, pattern=vertical lines, pattern color=red] (1,7.5) rectangle (2,9);
\draw[thick, <->] (2.2,0)--(2.2,3) node[midway, right]{$\lambda_1$};
\draw[thick, <->] (2.2,3)--(2.2,5) node[midway, right]{$\lambda_2$};
\draw[thick, <->] (2.2,5)--(2.2,7.5) node[midway, right]{$\lambda_3$};
\draw[thick, <->] (2.2,7.5)--(2.2,9) node[midway, right]{$\lambda_4$};
\draw[thick, |-|] (0,-0.2)--(2,-0.2) node[midway, below]{$Z=L^b$};
\draw (0.5, 9) node[above]{$X_1$};
\draw (1.5, 9) node[above]{$X_2$};
\end{tikzpicture}
\right)
&\geq&
H\left(
\begin{tikzpicture}[scale=0.7, baseline=(current bounding box.center)]
\draw[ thick, pattern=vertical lines, pattern color=blue] (0,9) rectangle (1,9.5);
\draw[ thick, pattern=vertical lines, pattern color=red] (1,9) rectangle (2,9.7);
\draw[thick,|-|](0,8.8)--(2,8.8) node[midway, below]{ $Z_1=L_1^{\vec{\gamma_1}\vec{\delta_1}}$};
\draw[thick,<->](-0.2,9)--(-0.2,10.5) node[midway, left]{$\lambda_4$};
\draw[help lines](0,9.7)--(2,9.7);
\draw[thick,<->](2.2,9)--(2.2,9.7) node[midway, right]{ $\mathcal{T}(Z_1)$};
\path (4, 9) node[right]{\Huge ,};

\draw[ thick, pattern=vertical lines, pattern color=blue] (0,5) rectangle (1,6);
\draw[ thick, pattern=vertical lines, pattern color=red] (1,5) rectangle (2,5.5);
\draw[ thick, pattern=grid, pattern color=blue] (2,5) rectangle (3,7.2);
\draw[ thick, pattern=crosshatch, pattern color=blue] (3,5) rectangle (4,6.5);
\draw[ thick, pattern=crosshatch, pattern color=red] (4,5) rectangle (5,6);
\draw[thick,<->](-0.2,5)--(-0.2,6.5) node[right]{$\lambda_4$};
\draw[thick,<->](-0.4,5)--(-0.4,7.5) node[left]{$\lambda_3$};
\draw[thick,<->](-0.6,5)--(-0.6,7) node[left]{$\lambda_2$};
\draw[thick,<->](5.2,5)--(5.2,7.2) node[midway, right]{ $\mathcal{T}(Z_2)$};
\draw[help lines](0,7.2)--(5,7.2);
\path (6, 5) node[right]{\Huge ,};
\draw[thick,|-|](0,4.8)--(5,4.8) node[midway, below]{ $Z_2=L_2^{\vec{\gamma_2}\vec{\delta_2}}$};

\draw[ thick, pattern=vertical lines, pattern color=red] (0,0) rectangle (1,1);
\draw[ thick, pattern=grid, pattern color=blue] (1,0) rectangle (2,0.8);
\draw[ thick, pattern=grid, pattern color=red] (2,0) rectangle (3,2.1);
\draw[ thick, pattern=crosshatch, pattern color=blue] (3,0) rectangle (4,1.3);
\draw[ thick, pattern=crosshatch, pattern color=red] (4,0) rectangle (5,1.6);
\draw[ thick, pattern=crosshatch dots, pattern color=blue] (5,0) rectangle (6,2.8);
\draw[ thick, pattern=crosshatch dots, pattern color=red] (6,0) rectangle (7,2.6);
\draw[thick,<->](-0.2,0)--(-0.2,1.5) node[right]{$\lambda_4$};
\draw[thick,<->](-0.4,0)--(-0.4,3) node[left]{$\lambda_1$};
\draw[thick,<->](-0.6,0)--(-0.6,2.5) node[left]{$\lambda_3$};
\draw[thick,<->](-0.8,0)--(-0.8,2) node[left]{$\lambda_2$};
\draw[thick,<->](7.2,0)--(7.2,2.8) node[midway, right]{ $\mathcal{T}(Z_3)$};
\draw[help lines](0,2.8)--(7,2.8);
\draw[thick,|-|](0,-0.2)--(7,-0.2) node[midway, below]{ $Z_3=L_3^{\vec{\gamma_3}\vec{\delta_3}}$};

\end{tikzpicture}
\right)
\end{eqnarray*}
\caption[]{Illustration of an application of Theorem \ref{Theorem AIS02}. On the left is the entropy of  the sum (bounded density linear combination) of $N=2$ dependent random variables, $X_1, X_2\in\mathcal{X}_{\lambda_1+\lambda_2+\lambda_3+\lambda_4}$, $(M=4)$, which is bounded below by joint entropy of $l=3$ arbitrary linear combinations, $Z_1, Z_2, Z_3$, of power level partitions of the two random variables. In this example, $I_1=\{4\}, I_2=\{2,3,4\}, I_3=\{1,2,3,4\}$. Therefore, ${m(1)}=4, {m(2)}=2$, and  ${m(3)}=1$. Condition (\ref{con1}) is verified as $\lambda_1+\lambda_2+\lambda_3\geq \mathcal{T}(Z_2)+\mathcal{T}(Z_3)$ and  $\lambda_1\geq \mathcal{T}(Z_3)$.} \label{grant}
\end{figure}
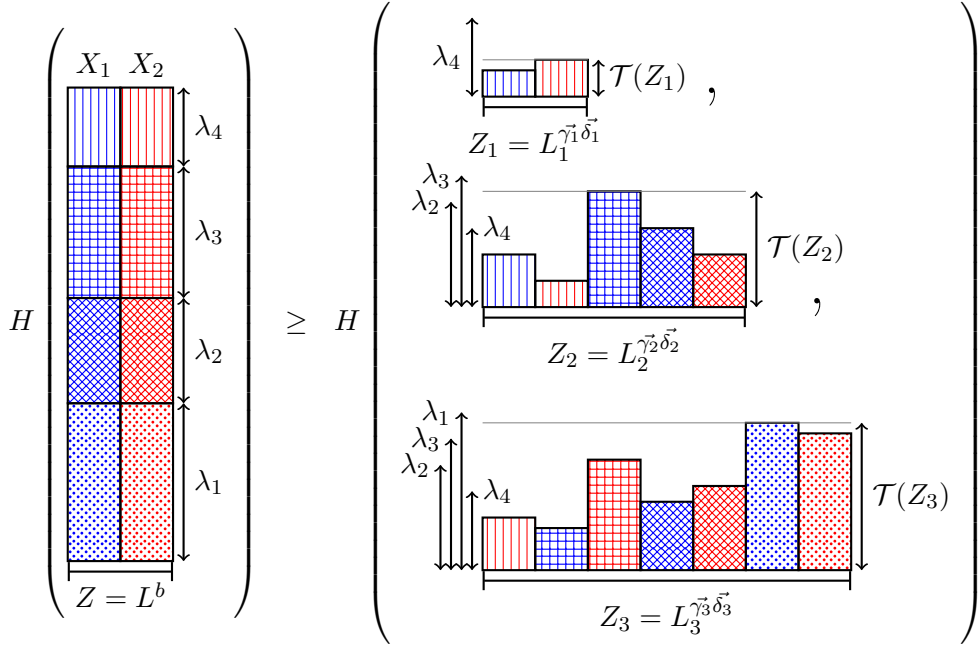

\item Theorem \ref{Theorem AIS01} is recovered as a special case of Theorem \ref{Theorem AIS02} if $M=N=2$, $I_1=\{2\}$, $I_2=\{1,2\}$, $\delta_{kij}=0$ and $\gamma_{kij}=\max_{q\in[M]}\lambda_q$ for any $k,i,j\in\{1,2\}$.

\item While applying Theorem \ref{Theorem AIS02} in the GDoF framework, a multi-letter extension is required. Such a generalization is presented in the following theorem. The same applies for extensions to complex valued random variables which can be obtained along the same lines as  previous bounds based on the AIS approach, e.g., Section VII in \cite{Arash_Jafar_PN}. 

\begin{theorem} \label{Theorem AIS03}
Consider $M$ non-negative numbers  $\lambda_1,\cdots,\lambda_M$ and random variables $X_j (t) \in \mathcal{X}_{\lambda_1+\lambda_2+\cdots+\lambda_M}$, $j\in[N]$, $t\in\mathbb{N}$ independent of $\mathcal{G}$,  and define 
\begin {eqnarray}
Z(t)&=&L^b(t)(X_1(t),X_2(t), \cdots, X_N(t))\label{mn1}\\
Z_{1}(t)&=& L_{1}^{\vec{\gamma}_1\vec{\delta}_1}(t)((X_{j}(t))_{\sum_{r=1}^{i-1}\lambda_r}^{\sum_{r=1}^i\lambda_r}, i\in I_1, j\in[N])\label{mn2}\\
Z_{2}(t)&=& L_{2}^{\vec{\gamma}_2\vec{\delta}_2}(t)((X_{j}(t))_{\sum_{r=1}^{i-1}\lambda_r}^{\sum_{r=1}^i\lambda_r}, i\in I_2, j\in[N])\label{mn3}\\
&\vdots&\notag\\
Z_{l}(t)&=&L_{l}^{\vec{\gamma}_l\vec{\delta}_l}(t)((X_{j}(t))_{\sum_{r=1}^{i-1}\lambda_r}^{\sum_{r=1}^i\lambda_r}, i\in I_l, j\in[N])\label{mn4}
\end{eqnarray}
The channel uses are indexed by {$t$}$\in\mathbb{N}$. $I_1, I_2, \cdots, I_l$ are  subsets of $\{1,2,\cdots, M\}$ such that 
$m(a)\geq m(b)$ whenever $a,b\in\{1,2,\cdots, M\}$ and $a<b$, then
\begin {eqnarray}
H(Z^{[n]}\mid W,\mathcal{G})&\geq& H(Z_1^{[n]},Z_2^{[n]}, \cdots, Z_l^{[n]}\mid W)+n~o(\log{\bar{P}})\label{dssd2}
\end{eqnarray}
if for each $s\in\{1,2,\cdots, l-1\}$,
\begin {eqnarray}
\mathcal{T}(Z_{s+1}(t))+\mathcal{T}(Z_{s+2}(t))+\cdots+\mathcal{T}(Z_l(t))&\leq&\lambda_1+\lambda_2+\cdots+\lambda_{(m(s)-1)}\label{t3con3}
\end{eqnarray}
\end{theorem}
{\color{black}Note that, for any $i\in[l]$ the set $I_i$ indicates what power levels are used by each $Z_i(t)$. For instance $I_3=\{1\}$ enforces $Z_3(t)$ to be a linear combination of bottom $\lambda_1$ part of $X_j(t)$ for all $j\in[N]$, i.e., $Z_{3}(t)= L_{3}^{\vec{\gamma}_3\vec{\delta}_3}(t)((X_{j}(t))_{\lambda_1}, j\in[N])$.} 
\item While applying Theorem \ref{Theorem AIS03} in the GDoF framework, a multi-antenna extension is required. The results of Theorem \ref{Theorem AIS03} can be generalized as follows,
\begin{theorem} \label{Theorem AIS04}
Consider $KM$ non-negative numbers {$\{\lambda_{km}: k\in[K],m\in[M]\}$}  and random variables $X_j (t) \in \mathcal{X}_{\max_{k\in[K]}\{\lambda_{k,1}+\lambda_{k,2}+\cdots+\lambda_{k,M}\}}$, $j\in[N]$, $t\in\mathbb{N}$,  independent of $\mathcal{G}$,  and  $\forall k\in[K], K\le N$, define 
\begin {eqnarray}
Z_k(t)&=&L_k^b(t)(X_1(t),X_2(t), \cdots, X_N(t))\label{mn1}\\
Z_{k,1}(t)&=& L_{k1}^{\vec{\gamma}_{k1}\vec{\delta}_{k1}}(t)((X_{j}(t))_{\sum_{r=1}^{i-1}\lambda_{kr}}^{\sum_{r=1}^i\lambda_{kr}}, i\in I_{k,1}, j\in[N])\label{mn2}\\
Z_{k,2}(t)&=& L_{k2}^{\vec{\gamma}_{k2}\vec{\delta}_{k2}}(t)((X_{j}(t))_{\sum_{r=1}^{i-1}\lambda_{kr}}^{\sum_{r=1}^i\lambda_{kr}}, i\in I_{k,2}, j\in[N])\label{mn3}\\
&\vdots&\notag\\
Z_{k,l_k}(t)&=&L_{kl_k}^{\vec{\gamma}_{kl_k}\vec{\delta}_{kl_k}}(t)((X_{j}(t))_{\sum_{r=1}^{i-1}\lambda_{kr}}^{\sum_{r=1}^i\lambda_{kr}}, i\in I_{k,l_k}, j\in[N])\label{mn4}
\end{eqnarray}
The channel uses are indexed by $t\in\mathbb{N}$. $I_{kk'}\subset [M], k\in[K], k'\in[l_k],$  such that $i<j\Rightarrow m(k,i)\geq m(k,j)$, where $$m(a,b)\triangleq \min\{m: m\in I_{a,b}\}.$$ If for all $k\in[K]$ and for each $s\in\{1,2,\cdots, l_k-1\}$,
\begin {eqnarray}
\mathcal{T}(Z_{k,s+1})+\mathcal{T}(Z_{k,s+2})+\cdots+\mathcal{T}(Z_{k,l_k})&\leq& \lambda_{k,1}+\lambda_{k,2}+\cdots+\lambda_{k,(m(k,s)-1)}\label{condt4}
\end{eqnarray}
then
\begin {eqnarray}
H(Z_1^{[n]},\cdots,Z_K^{[n]}\mid W,\mathcal{G})&\geq& H(Z_{1,1}^{[n]},\cdots,Z_{K,l_K}^{[n]}\mid W)+Kn~o(\log{\bar{P}})\label{dssd4}
\end{eqnarray}
\end{theorem}

\begin{figure}[!h] 
\begin{eqnarray*}
H\left(\begin{tikzpicture}[scale=0.7, baseline=(current bounding box.center)]
\draw[ thick, pattern=crosshatch dots, pattern color=blue] (0,0+11) rectangle (1,3+11);
\draw[ thick, pattern=crosshatch, pattern color=blue] (0,3+11) rectangle (1,5+11);
\draw[ thick, pattern=grid, pattern color=blue] (0,5+11) rectangle (1,7.5+11);
\draw[ thick, pattern=vertical lines, pattern color=blue] (0,7.5+11) rectangle (1,9+11);
\draw[ thick, pattern=crosshatch dots, pattern color=red] (1,0+11) rectangle (2,3+11);
\draw[ thick, pattern=crosshatch, pattern color=red] (1,3+11) rectangle (2,5+11);
\draw[ thick, pattern=grid, pattern color=red] (1,5+11) rectangle (2,7.5+11);
\draw[ thick, pattern=vertical lines, pattern color=red] (1,7.5+11) rectangle (2,9+11);
\draw[ thick, pattern=crosshatch dots, pattern color=green] (2,0+11) rectangle (3,3+11);
\draw[ thick, pattern=crosshatch, pattern color=green] (2,3+11) rectangle (3,5+11);
\draw[ thick, pattern=grid, pattern color=green] (2,5+11) rectangle (3,7.5+11);
\draw[ thick, pattern=vertical lines, pattern color=green] (2,7.5+11) rectangle (3,9+11);
\draw[thick, <->] (3.2,0+11)--(3.2,3+11) node[midway, right]{$\lambda_{11}$};
\draw[thick, <->] (3.2,3+11)--(3.2,5+11) node[midway, right]{$\lambda_{12}$};
\draw[thick, <->] (3.2,5+11)--(3.2,7.5+11) node[midway, right]{$\lambda_{13}$};
\draw[thick, <->] (3.2,7.5+11)--(3.2,9+11) node[midway, right]{$\lambda_{14}$};
\draw[thick, |-|] (0,-0.2+11)--(3,-0.2+11) node[midway, below]{$Z_1=L_1^b$};
\draw (0.5, 9+11) node[above]{$X_1$};
\draw (1.5, 9+11) node[above]{$X_2$};
\draw (2.5, 9+11) node[above]{$X_3$};
\path (3, 10) node[right]{\Huge ,};
\draw[ thick, pattern= dots, pattern color=blue] (0,0) rectangle (1,3.8);
\draw[ thick, pattern=north east lines, pattern color=blue] (0,3.8) rectangle (1,5.2);
\draw[ thick, pattern=north west lines, pattern color=blue] (0,5.2) rectangle (1,8.1);
\draw[ thick, pattern=horizontal lines, pattern color=blue] (0,8.1) rectangle (1,9);
\draw[ thick, pattern= dots, pattern color=red] (1,0) rectangle (2,3.8);
\draw[ thick, pattern=north east lines, pattern color=red] (1,3.8) rectangle (2,5.2);
\draw[ thick, pattern=north west lines, pattern color=red] (1,5.2) rectangle (2,8.1);
\draw[ thick, pattern=horizontal lines, pattern color=red] (1,8.1) rectangle (2,9);
\draw[ thick, pattern= dots, pattern color=green] (2,0) rectangle (3,3.8);
\draw[ thick, pattern=north east lines, pattern color=green] (2,3.8) rectangle (3,5.2);
\draw[ thick, pattern=north west lines, pattern color=green] (2,5.2) rectangle (3,8.1);
\draw[ thick, pattern=horizontal lines, pattern color=green] (2,8.1) rectangle (3,9);
\draw[thick, <->] (3.2,0)--(3.2,3.8) node[midway, right]{$\lambda_{21}$};
\draw[thick, <->] (3.2,3.8)--(3.2,5.2) node[midway, right]{$\lambda_{22}$};
\draw[thick, <->] (3.2,5.2)--(3.2,8.1) node[midway, right]{$\lambda_{23}$};
\draw[thick, <->] (3.2,8.1)--(3.2,9) node[midway, right]{$\lambda_{24}$};
\draw[thick, |-|] (0,-0.2)--(3,-0.2) node[midway, below]{$Z_2=L_2^b$};
\draw (0.5, 9) node[above]{$X_1$};
\draw (1.5, 9) node[above]{$X_2$};
\draw (2.5, 9) node[above]{$X_3$};
\end{tikzpicture}
\right)
&\geq&
H\left(
\begin{tikzpicture}[scale=0.8, baseline=(current bounding box.center)]
\draw[ thick, pattern=vertical lines, pattern color=blue] (-1,0.1+9) rectangle (0,0.1+9.6);
\draw[ thick, pattern=vertical lines, pattern color=red] (0,0.1+9) rectangle (1,0.1+9.4);
\draw[ thick, pattern=vertical lines, pattern color=green] (1,0.1+9) rectangle (2,0.1+9.7);
\draw[thick,|-|](-1,0.1+8.8)--(2,0.1+8.8) node[midway, below]{ $Z_{11}=L_{11}^{\vec{\gamma_{11}}\vec{\delta_{11}}}$};
\draw[thick,<->](-1.2,0.1+9)--(-1.2,0.1+10.5) node[midway, left]{$\lambda_{14}$};
\draw[help lines](-1,0.1+9.7)--(2,0.1+9.7);
\draw[thick,<->](2.2,0.1+9)--(2.2,0.1+9.7) node[midway, right]{ $\mathcal{T}(Z_{11})$};
\path (3.6, 0.1+9) node[right]{\Huge ,};
\draw[ thick, pattern=horizontal lines, pattern color=blue] (4.9+0,0.1+9) rectangle (4.9+1,0.1+9.4);
\draw[ thick, pattern=horizontal lines, pattern color=red] (4.9+1,0.1+9) rectangle (4.9+2,0.1+9.2);
\draw[thick,|-|](4.9+0,0.1+8.8)--(4.9+2,0.1+8.8) node[midway, below]{ $Z_{21}=L_{21}^{\vec{\gamma_{21}}\vec{\delta_{21}}}$};
\draw[thick,<->](4.9+-0.1,0.1+9)--(4.9+-0.1,0.1+9.9) node[midway, left]{$\lambda_{24}$};
\draw[help lines](4.9+0,0.1+9.4)--(4.9+2,0.1+9.4);
\draw[thick,<->](4.9+2.2,0.1+9)--(4.9+2.2,0.1+9.4) node[midway, right]{ $\mathcal{T}(Z_{21})$};
\path (4.9+3.8, 0.1+9) node[right]{\Huge ,};

\draw[ thick, pattern=vertical lines, pattern color=blue] (0-0.5,5) rectangle (1-0.5,6);
\draw[ thick, pattern=vertical lines, pattern color=red] (1-0.5,5) rectangle (2-0.5,5.5);
\draw[ thick, pattern=crosshatch, pattern color=blue] (2-0.5,5) rectangle (3-0.5,6.9);
\draw[thick,<->](-0.2-0.5,5)--(-0.2-0.5,6.5) node[right]{$\lambda_{14}$};
\draw[thick,<->](-0.4-0.5,5)--(-0.4-0.5,7) node[left]{$\lambda_{12}$};
\draw[thick,<->](3.2-0.5,5)--(3.2-0.5,6.9) node[midway, right]{ $\mathcal{T}(Z_{12})$};
\draw[help lines](0,6.9)--(3-0.5,6.9);
\path (3.6-0.5, 5) node[right]{\Huge ,};
\draw[thick,|-|](0-0.5,4.8)--(3-0.5,4.8) node[midway, below]{ $Z_{12}=L_{12}^{\vec{\gamma_{12}}\vec{\delta_{12}}}$};
\draw[ thick, pattern=horizontal lines, pattern color=green] (4.8+0,5) rectangle (4.8+1,5.3);\draw[ thick, pattern=north west lines, pattern color=blue] (4.8+1,5) rectangle (4.8+2,6.2);
\draw[ thick, pattern=north west lines, pattern color=red] (4.8+2,5) rectangle (4.8+3,7.2);
\draw[thick,<->](4.8+-0.2,5)--(4.8+-0.2,5.9) node[right]{$\lambda_{24}$};
\draw[thick,<->](4.8+-0.4,5)--(4.8+-0.4,7.5) node[left]{$\lambda_{23}$};
\draw[thick,<->](4.8+3.2,5)--(4.8+3.2,7.2) node[midway, right]{ $\mathcal{T}(Z_{22})$};
\draw[help lines](4.8+0,7.2)--(4.8+3,7.2);
\path (4.8+3.5, 5) node[right]{\Huge ,};
\draw[thick,|-|](4.8+0,4.8)--(4.8+3,4.8) node[midway, below]{ $Z_{22}=L_{22}^{\vec{\gamma_{22}}\vec{\delta_{22}}}$};

\draw[ thick, pattern=vertical lines, pattern color=red] (0,0) rectangle (1,1);
\draw[ thick, pattern=vertical lines, pattern color=blue] (1,0) rectangle (2,0.8);
\draw[ thick, pattern=grid, pattern color=green] (2,0) rectangle (3,2.1);
\draw[ thick, pattern=crosshatch, pattern color=blue] (3,0) rectangle (4,1.3);
\draw[ thick, pattern=crosshatch, pattern color=red] (4,0) rectangle (5,1.6);
\draw[ thick, pattern=crosshatch dots, pattern color=green] (5,0) rectangle (6,2.8);
\draw[ thick, pattern=crosshatch dots, pattern color=red] (6,0) rectangle (7,2.6);
\draw[thick,<->](-0.2,0)--(-0.2,1.5) node[right]{$\lambda_{14}$};
\draw[thick,<->](-0.4,0)--(-0.4,3) node[left]{$\lambda_{11}$};
\draw[thick,<->](-0.6,0)--(-0.6,2.5) node[left]{$\lambda_{13}$};
\draw[thick,<->](-0.8,0)--(-0.8,2) node[left]{$\lambda_{12}$};
\draw[thick,<->](7.2,0)--(7.2,2.8) node[midway, right]{ $\mathcal{T}(Z_{13})$};
\draw[help lines](0,2.8)--(7,2.8);
\path (6, -1) node[right]{\Huge ,};
\draw[thick,|-|](0,-0.2)--(7,-0.2) node[midway, below]{ $Z_{13}=L_{13}^{\vec{\gamma_{13}}\vec{\delta_{13}}}$};

\draw[ thick, pattern=horizontal lines, pattern color=red] (0,-5) rectangle (1,-4.4);
\draw[ thick, pattern=horizontal lines, pattern color=green] (1,-5) rectangle (2,-4.6);
\draw[ thick, pattern=north west lines, pattern color=green] (2,-5) rectangle (3,-3.9);
\draw[ thick, pattern=north east lines, pattern color=blue] (3,-5) rectangle (4,-4.7);
\draw[ thick, pattern=north east lines, pattern color=green] (4,-5) rectangle (5,-3.8);
\draw[ thick, pattern= dots, pattern color=blue] (5,-5) rectangle (6,-2.6);
\draw[ thick, pattern= dots, pattern color=green] (6,-5) rectangle (7,-2.9);
\draw[thick,<->](-0.2,-5)--(-0.2,-4.1) node[right]{$\lambda_{24}$};
\draw[thick,<->](-0.4,-5)--(-0.4,-1.2) node[left]{$\lambda_{21}$};
\draw[thick,<->](-0.6,-5)--(-0.6,-2.1) node[left]{$\lambda_{23}$};
\draw[thick,<->](-0.8,-5)--(-0.8,-3.6) node[left]{$\lambda_{22}$};
\draw[thick,<->](7.2,-5)--(7.2,-2.6) node[midway, right]{ $\mathcal{T}(Z_{23})$};
\draw[help lines](0,-2.6)--(7,-2.6);
\draw[thick,|-|](0,-5.2)--(7,-5.2) node[midway, below]{ $Z_{23}=L_{23}^{\vec{\gamma_{23}}\vec{\delta_{23}}}$};

\end{tikzpicture}
\right)
\end{eqnarray*}
\caption[]{Illustration of an application of Theorem \ref{Theorem AIS04}. {\color{black}Note that in this figure we dropped the time index $(t)$ for convenience}. On the left is the joint entropy of  the sum (bounded density linear combination) of $N=3$ dependent random variables, $X_1(t), X_2(t),X_3(t)\in\mathcal{X}_{\max_{k\in[2]}\{\lambda_{k1}+\lambda_{k2}+\lambda_{k3}+\lambda_{k4}\}}$, $(M=4)$, which is bounded below by joint entropy of $l_1+l_2=6$ arbitrary linear combinations, $Z_{11}, Z_{12}, Z_{13},Z_{21}, Z_{22}, Z_{23}$, of power level partitions of the two random variables. In this example, $I_{11}=I_{21}=\{4\}, I_{12}=\{2,4\}, I_{22}=\{3,4\}, I_{13}=I_{23}=\{1,2,3,4\}$. Condition (\ref{condt4}) is verified as $\lambda_{11}+\lambda_{12}+\lambda_{13}\geq \mathcal{T}(Z_{12})+\mathcal{T}(Z_{13})$, $\lambda_{21}+\lambda_{22}+\lambda_{23}\geq \mathcal{T}(Z_{22})+\mathcal{T}(Z_{23})$, $\lambda_{21}\geq \mathcal{T}(Z_{23})$ and  $\lambda_{11}\geq \mathcal{T}(Z_{13})$.} \label{grant}
\end{figure}
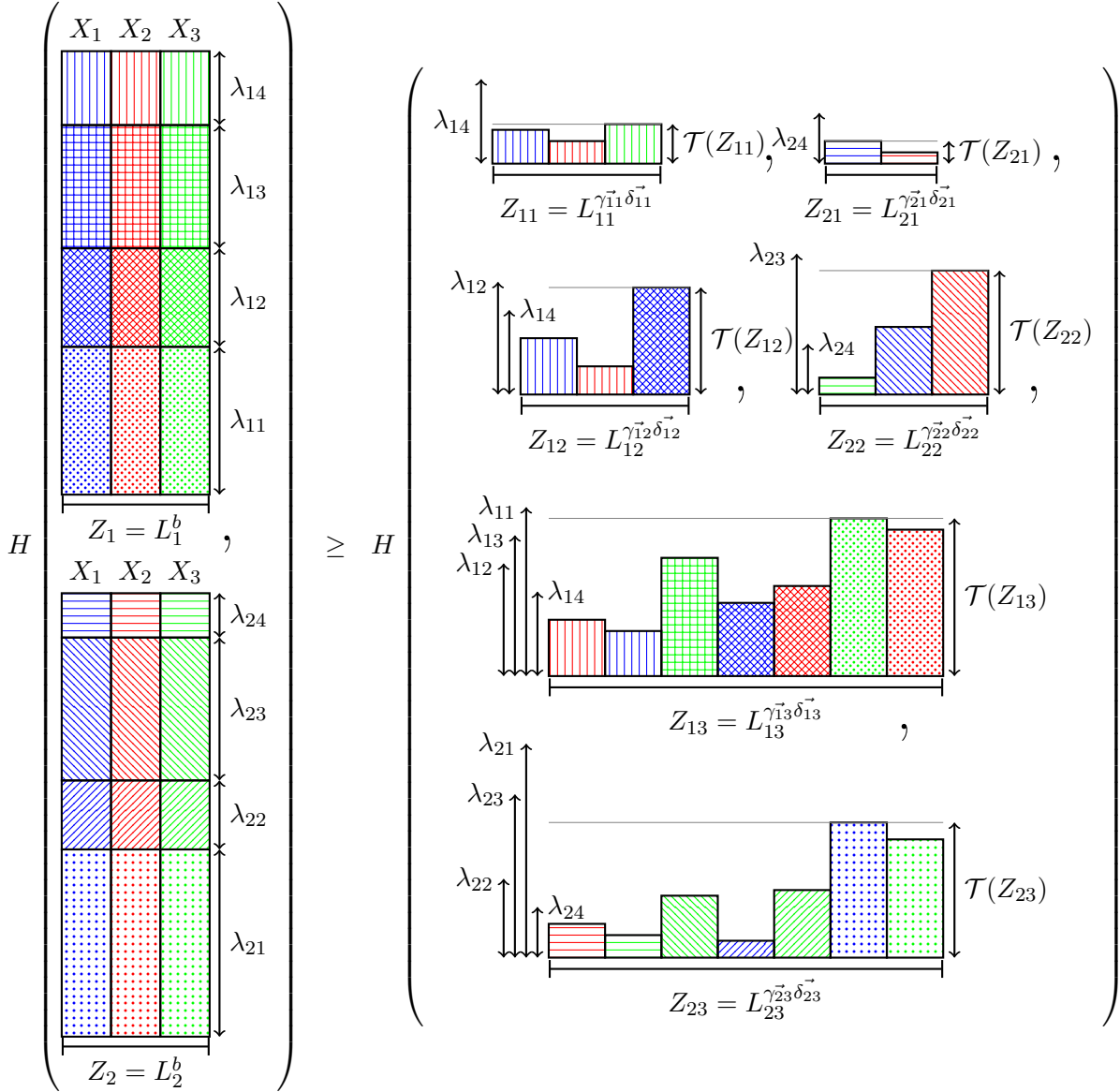

See Appendix \ref{prooft4} for the proof of Theorem \ref{Theorem AIS04}. Note that the proof of Theorem \ref{Theorem AIS04} also proves Theorem \ref{Theorem AIS02} and Theorem \ref{Theorem AIS03} which may be recovered as specializations of Theorem \ref{Theorem AIS04}.

A visual illustration of an application of Theorem \ref{Theorem AIS04} is provided in Figure \ref{grant}.

\item The results of Theorem \ref{Theorem AIS01} and its generalization in Theorem \ref{Theorem AIS04} can be further combined with sub-modularity properties of the entropy function\footnote{If  $\Omega$  is a finite set, a submodular function is a set function $ f:2^{\Omega}\rightarrow \mathbb{R}$, where $ 2^\Omega$ denotes the power set of $ \Omega$, which satisfies the following property;\\

For every  $S, T \subseteq \Omega$ we have that  $f(S)+f(T)\geq f(S\cup T)+f(S\cap T)$ \cite{subm}.} to obtain a variety of sum-set inequalities specialized for different GDoF settings. 

\item To show how the new sum-set inequalities presented in Theorem \ref{Theorem AIS04} are useful to obtain tight GDoF bounds in conjunction with submodularity properties of entropy, an example that arise in the context of the $2$ user MIMO IC is presented in Section \ref{MIMOIC}.
\end{enumerate}

\section{GDoF Outer Bound for a $2$ User MIMO IC under Partial CSIT}\label{MIMOIC}

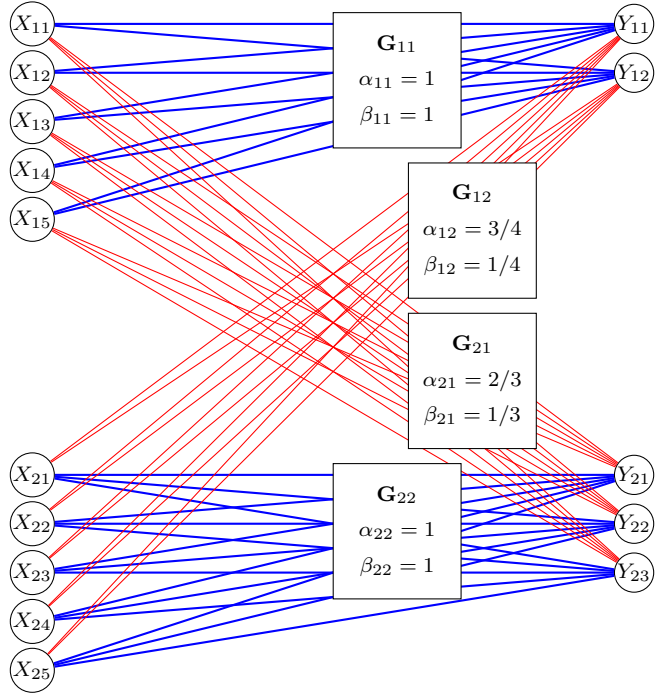
\begin{figure}[h] 
\centering
\begin{tikzpicture}[scale=1]
\foreach \m in {1,..., 5}
{
\node[circle, draw=black, inner sep = 0.2] (M1\m) at (1, -0.65*\m) {\scriptsize $X_{1\m}$};
};

\foreach \m in {1,..., 5}
{
\node[circle, draw=black,  inner sep = 0.2] (M2\m) at (1, -6-0.65*\m) {\scriptsize $X_{2\m}$};
};

\begin{scope}[shift={(1,0)}]
\foreach \n in {1,..., 2}
{
\node[circle, draw=black,  inner sep = 0.2] (N1\n) at (8, -0.65*\n) {\scriptsize $Y_{1\n}$};
};

\foreach \n in {1,...,3}
{
\node[circle, draw=black,  inner sep = 0.2] (N2\n) at (8, -6-0.65*\n) {\scriptsize $Y_{2\n}$};
};
\end{scope}

\foreach \m in {1, ..., 5}
{ 
	\foreach \n in {1, ..., 2}
	{
	\draw[thick, color=blue](M1\m)--(N1\n);
	}
}

\foreach \m in {1, ..., 5}
{ 
	\foreach \n in {1, ..., 3}
	{
	\draw[thick, color=blue](M2\m)--(N2\n);
	}
}

\foreach \m in {1, ..., 5}
{ 
	\foreach \n in {1, ..., 3}
	{
	\draw[thin, color=red](M1\m)--(N2\n);
	}
}

\foreach \m in {1, ..., 5}
{ 
	\foreach \n in {1, ..., 2}
	{
	\draw[thin, color=red](M2\m)--(N1\n);
	}
}

\draw[thin, fill=white] (5,-0.5) rectangle (6.7,-2.3) node[midway, align=center] {\scriptsize ${\bf G}_{11}$ \\ \scriptsize $\alpha_{11}=1$\\ \scriptsize $\beta_{11}=1$};

\draw[thin, fill=white] (6,-2.5) rectangle (7.7,-4.3) node[midway, align=center] {\scriptsize ${\bf G}_{12}$ \\ \scriptsize $\alpha_{12}=3/4$\\ \scriptsize $\beta_{12}=1/4$};

\draw[thin, fill=white] (6,-4.5) rectangle (7.7,-6.3) node[midway, align=center] {\scriptsize ${\bf G}_{21}$ \\ \scriptsize $\alpha_{21}=2/3$\\ \scriptsize $\beta_{21}=1/3$};

\draw[thin, fill=white] (5,-6.5) rectangle (6.7,-8.3) node[midway, align=center] {\scriptsize ${\bf G}_{22}$ \\ \scriptsize $\alpha_{22}=1$\\ \scriptsize $\beta_{22}=1$};

\end{tikzpicture}

\caption[]{MIMO IC setting under consideration.}
\label{regionxc}
\end{figure}

In this section, as an example of the use of the sum-set inequalities, we obtain a tight GDoF  outer bound for a non-trivial two user MIMO IC setting with  asymmetric antenna configuration and asymmetric partial CSIT.  Specifically, we consider the two user MIMO IC with $(M_1,M_2,N_1,N_2)=(5,5,2,3)$ as shown in Figure \ref{regionxc}. We assume, $(\alpha_{11},\alpha_{12},\alpha_{21},\alpha_{22})=(1,\frac{3}{4},\frac{2}{3},1)$ and $\beta_{12}=\frac{1}{4},\beta_{21}=\frac{1}{3}$ and derive a tight GDoF bound for this channel using our sum-set inequalities. Achievability for this case is already known from \cite{Bofeng_Arash_Jafar_ArXiv} and \cite{Hao_Rassouli_Clerckx}.
\subsection{The Channel}
The channel model for the two user $(M_1,M_2,N_1,N_2)=(5,5,2,3)$ MIMO IC  with ~{$(\alpha_{11},\alpha_{12},\alpha_{21},\alpha_{22})=(1,\frac{3}{4},\frac{2}{3},1)$}  is defined by the following input-output equations.
\begin{eqnarray}
\mathbf{Y}_{1}(t)&=&\sqrt{P}\mathbf{G}_{11}(t)\mathbf{X}_{1}(t)+\sqrt{P^{\frac{3}{4}}}\mathbf{G}_{12}(t)\mathbf{X}_{2}(t)+\mathbf{\Gamma}_{1}(t), \label{eq::received1}\\
\mathbf{Y}_{2}(t)&=&\sqrt{P^{\frac{2}{3}}}\mathbf{G}_{21}(t)\mathbf{X}_{1}(t)+\sqrt{P}\mathbf{G}_{22}(t)\mathbf{X}_{2}(t)+\mathbf{\Gamma}_{2}(t),\label{eq::received2}
\end{eqnarray}
Here, $\mathbf{X}_{1}(t)=[{X}_{11}(t)\ {X}_{12}(t)\ {X}_{13}(t)\ {X}_{14}(t)\ {X}_{15}(t)]^T$  and $\mathbf{X}_{2}(t)=[{X}_{21}(t)\ {X}_{22}(t)\ {X}_{23}(t)\ {X}_{24}(t)\  {X}_{25}(t) ]^T$ are the  ${5\times1}$ signal vectors sent from the first and second transmitters respectively, normalized so that each is subject to unit power
constraint. $\mathbf{Y}_{1}(t)=[{Y}_{11}(t)\ {Y}_{12}(t)]^T$ and $\mathbf{Y}_{2}(t)=[{Y}_{21}(t)\ {Y}_{22}(t)\ {Y}_{23}(t)]^T$ are the ${2\times1}$ and ${3\times1}$ received signal vectors at the first and second receivers, respectively. $\mathbf{\Gamma}_{1}(t)$ and $\mathbf{\Gamma}_{2}(t)$ are the ${2\times1}$ and ${3\times1}$ vectors whose components are zero-mean unit-variance additive white Gaussian noise (AWGN).  The $N_r\times M_s$ matrix ${\bf G}_{rs}(t)$ is the channel fading coefficient matrix between the $r$-th receiver and the  $s$-th transmitter for any $r,s\in\{1,2\}$. The entry in the $n$-th row and $m$-th column of the matrix ${\bf G}_{rs}(t)$ is ${G}_{rsnm}(t)$.

\subsubsection{Partial CSIT} \label{defp}
Under partial CSIT, the channel coefficients are represented as
\begin{eqnarray*}
G_{rsnm}(t)&=&\hat{G}_{rsnm}(t)+\sqrt{P^{-\beta_{rs}}}\tilde{G}_{rsnm}(t)
\end{eqnarray*}
Recall that $G_{rsnm}(t)$ is the channel fading coefficient between the $n$-th antenna of the $r$-th receiver and the $m$-th antenna of the $s$-th transmitter. $\hat{G}_{rsnm}(t)$ is the  channel estimate and $\tilde{G}_{rsnm}(t)$ is the  estimation error term. To avoid degenerate conditions, for each $N_r\times M_s$ channel matrix ${\bf G}_{rs}(t)$, we require that all its $N_r\times N_r$ submatrices are non-singular, i.e., their determinants are bound away from zero. To this end, for all $t\in[n],~r,s\in\{1,2\}$, and for all choices of $N_r$ transmit antenna indices $\{m_1,m_2,\cdots,m_{N_r}:m_i\in [M_s]\}$ define the determinant $D(t)$ as
\begin{eqnarray}
D(t)\triangleq \begin{vmatrix}
 G_{rs1m_1}(t)& G_{rs1m_2}(t)& \cdots & G_{rs1m_{N_r}}(t)\\ 
 \vdots&\vdots  & \ddots  &\vdots \\ 
 G_{rsN_rm_1}(t)& G_{rsN_rm_2}(t)&  \cdots& G_{rsN_rm_{N_r}}(t)
\end{vmatrix}.\label{deter}
\end{eqnarray}
Then we require that there exists a positive constant $\Delta_1>0$, such that $|D(t)|\geq \Delta_1$, for all $t\in[n],~r,s\in\{1,2\},\{m_1,m_2,\cdots,m_{N_r}:m_i\in [M_s]\}.$
The channel variables $\hat{G}_{rsnm}(t), \tilde{G}_{rsnm}(t)$ are distinct random variables drawn from the set $\mathcal{G}$. The  realizations of $\hat{G}_{rsnm}(t)$ are known to the transmitter, but the realizations of $\tilde{G}_{rsnm}(t)$ are not available to the transmitter. We also assume that the channel coefficients $|{G}_{rsnm}(t)|$ are bounded away from zero, i.e., 
\begin{eqnarray}
\Delta_1&\le&|{G}_{rsnm}(t)|, \forall t\in[n],~r,s\in\{1,2\},m\in [M_s],n\in [N_r]\label{Deter2}
\end{eqnarray}
 Note that under the partial CSIT model, the variance of the channel coefficients $G_{rsnm}(t)$  behaves as $\sim P^{-\beta_{rs}}$ and the peak of the probability density function  behaves as $\sim\sqrt{P^{\beta_{rs}}}$. 

For any $r,s\in\{1,2\}$, in order to span the full range of partial channel knowledge at the transmitters, the corresponding range of $\beta_{rs}$ parameters, assumed throughout this work, is $0\leq\beta_{rs}\leq1$. $\beta_{rs}=0$ and $\beta_{rs}=1$ correspond to the two extremes where the CSIT is essentially absent, or perfect, respectively. Note that the value of $\beta_{11}$ and $\beta_{22}$ will not affect the GDoF.

\subsubsection{GDoF}
The definitions of achievable rates $R_i(P)$ and capacity region $\mathcal{C}(P)$ are standard. The DoF region is defined as
\begin{eqnarray}
\mathcal{D}&=&\{(d_1,d_2): \exists (R_1(P),R_2(P))\in\mathcal{C}(P), \mbox{ s.t. } d_k=\lim_{P\rightarrow\infty}\frac{R_k(P)}{\frac{1}{2}\log{(P)}}, \forall k\in\{1,2\}\} \label {region}
\end{eqnarray}

\subsection{Channel Model}

The channel model is derived similar to the channel model for the general two user IC with arbitrary number of antennas in \cite{Arash_Jafar_IC}. We will avoid repetition of explanations
for those steps that are essentially identical to \cite{Arash_Jafar_IC}, and focus
instead on the deviations from the original proof. As in \cite{Arash_Jafar_IC},
the starting point is to bound the problem with  deterministic
model, such that a GDoF outer bound on the deterministic
model is also a GDoF outer bound for the original problem. Since the derivation of the deterministic model is essentially
identical to \cite{Arash_Jafar_IC}, here we simply state the resulting equivalent deterministic
model.

\subsubsection{Equivalent Deterministic Model}\label{DM_1}
As in \cite{Arash_Jafar_IC}, without loss of generality for DoF characterizations, we will use the deterministic model for the equivalent channel.
\begin{eqnarray}
\bar{\bf Y}_{1}(t)&=&L_{1}^b(t)\left(\bar{\bf X}_{1c}(t)\bigtriangledown (\bar{\mathbf{X}}_{2a}(t))^1_{\frac{1}{4}}\bigtriangledown (\bar{\mathbf{X}}_{2c}(t))^1_{\frac{1}{2}}\right)\label{rrre2}\\
\bar{\mathbf{Y}}_{2}(t)&=&L_{2}^b(t)\left(\bar{\mathbf{X}}_{2c}(t)\bigtriangledown (\bar{\mathbf{X}}_{1a}(t))^1_{\frac{1}{3}}\bigtriangledown (\bar{\bf X}_{1c}(t))^1_{\frac{2}{3}}\right)\label{rre3}
\end{eqnarray}
for all $t\in[n]$. $\bar{\mathbf{X}}_{1a}(t)$, $\bar{{X}}_{1b}(t)$, $\bar{{X}}_{1c}(t)$,  $\bar{\mathbf{X}}_{2a}(t)$, $\bar{\mathbf{X}}_{2c}(t)$ and $\bar{\mathbf{Y}}_1(t)$ are defined as,
\begin{eqnarray}
\bar{\mathbf{X}}_{1}(t)&=&\begin{bmatrix}\bar{X}_{11}(t)&\bar{X}_{12}(t)&\bar{X}_{13}(t)&\bar{X}_{14}(t)&\bar{X}_{15}(t)\end{bmatrix}^T\\
\bar{\mathbf{X}}_{1a}(t)&=&\begin{bmatrix}\bar{X}_{11}(t)&\bar{X}_{12}(t)&\bar{X}_{13}(t)\end{bmatrix}^T\\
\bar{\mathbf{X}}_{1c}(t)&=&\begin{bmatrix}\bar{X}_{14}(t)&\bar{X}_{15}(t)\end{bmatrix}^T\\
\bar{\mathbf{X}}_{2}(t)&=&\begin{bmatrix}\bar{X}_{21}(t)&\bar{X}_{22}(t)&\bar{X}_{23}(t)&\bar{X}_{24}(t)&\bar{X}_{25}(t)\end{bmatrix}^T\\
\bar{\mathbf{X}}_{2a}(t)&=&\begin{bmatrix}\bar{X}_{21}(t)&\bar{X}_{22}(t)\end{bmatrix}^T\\
\bar{\mathbf{X}}_{2c}(t)&=&\begin{bmatrix}\bar{X}_{23}(t)&\bar{X}_{24}(t)&\bar{X}_{25}(t)\end{bmatrix}^T
\end{eqnarray}
and $\bar{X}_{1m}(t),\bar{X}_{2m}(t)\in\{0, 1, \cdots, {\bar{P}}\}$, $\forall m\in[5]$.
\subsection{GDoF region of the two user MIMO IC} \label{essboundIC}
\begin{theorem}\label{theoremIC} The GDoF region of the two user~ $(M_1,M_2,N_1,N_2)=(5,5,2,3)$ MIMO IC with $(\alpha_{11},\alpha_{12},\alpha_{21},\alpha_{22})=(1,\frac{3}{4},\frac{2}{3},1)$ and $\beta_{12}=\frac{1}{4},\beta_{21}=\frac{1}{3}$, is as follows
\begin{eqnarray}
\left\{\right.(d_1,d_2)\in {\color{olive}\mathbb{R}^{2+}},~d_1\le2,~d_2\le3,~\frac{d_1}{2}+\frac{d_2}{3}\le{\frac{3}{2}},~d_1+d_2\le3+\frac{7}{9}\left.\right\}\label{L_new1}
\end{eqnarray}
\end{theorem}
Note that, these bounds turns out to be tight, i.e., the achievability and the outer bounds coincide with each others,  see \cite{Bofeng_Arash_Jafar_ArXiv}.\\

\begin{figure}[h] 
\centering
\includegraphics[width=0.8\textwidth]{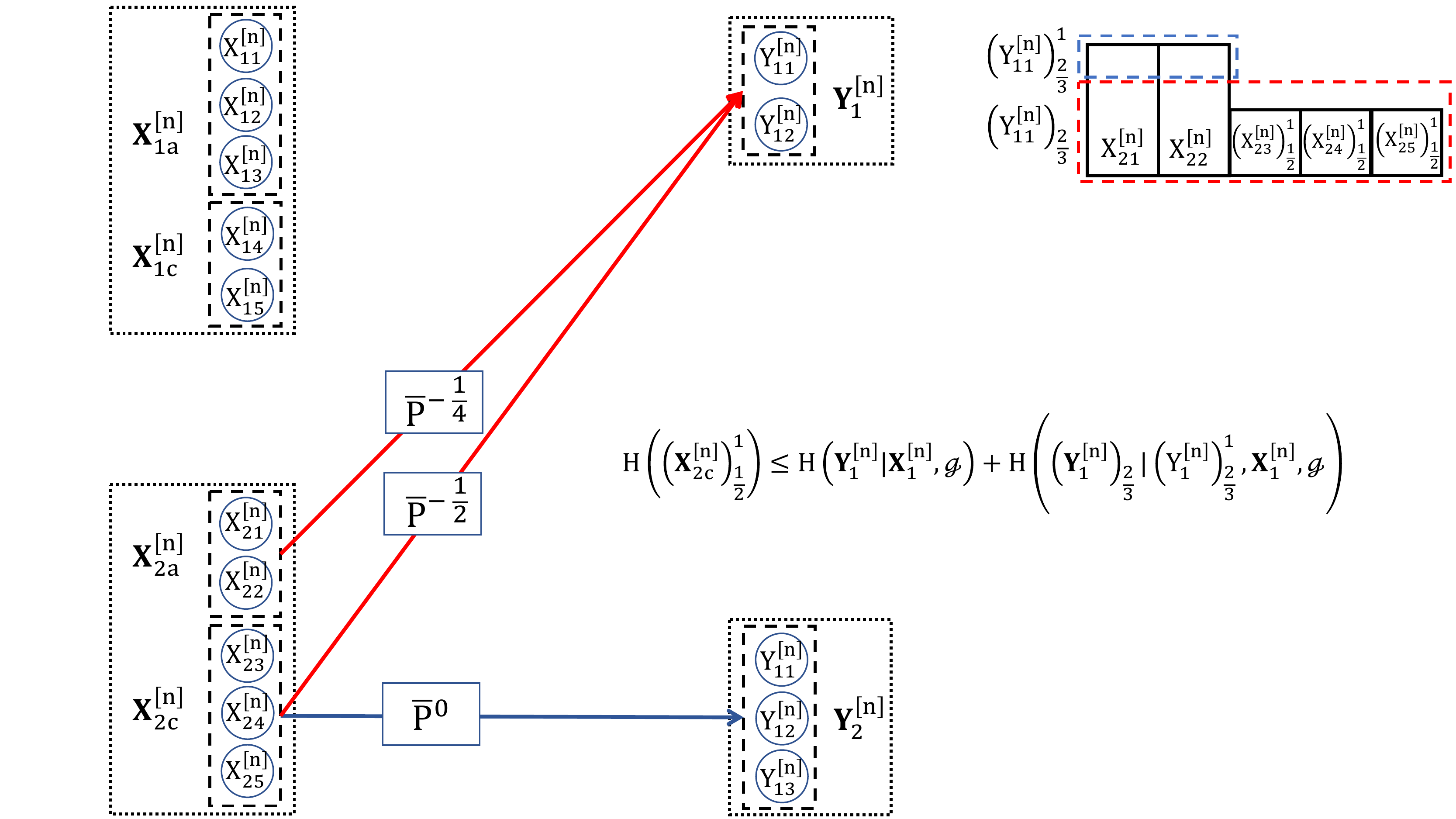}
\caption[]{The two sides of (\ref{firstlem}) are compared.}
\label{lemmaxxv}
\end{figure}

Proof of Theorem \ref{theoremIC}  is relegated to Appendix \ref {PIC} and is straightforward except for the following lemma which is the main novelty of the outer bound proof. It is in deriving this key lemma that we require both the sum-set inequalities of Theorem \ref{Theorem AIS04} and the sub-modularity of entropy functions. 
\begin{lemma}\label{lemma1} For the two user $(M_1,M_2,N_1,N_2)=(5,5,2,3)$ MIMO IC with $(\alpha_{11},\alpha_{12},\alpha_{21},\alpha_{22})=(1,\frac{3}{4},\frac{2}{3},1)$ and $\beta_{12}=\frac{1}{4},\beta_{21}=\frac{1}{3}$ levels of partial CSIT, we have, 
\begin{eqnarray}
2H((\bar{\mathbf{X}}^{[n]}_{2c})^1_{\frac{1}{2}})&{\le}&2H(\bar{\bf Y}^{n}_{1}\mid \bar{\mathbf{X}}^{[n]}_{1},\mathcal{G}){+}H((\bar{\bf Y}^{[n]}_{1})_{\frac{2}{3}}\mid (\bar{\bf Y}^{[n]}_{1})_{\frac{2}{3}}^{1},\bar{\mathbf{X}}^{[n]}_{1},\mathcal{G})+n~o~(\log{\bar{P}})\label{firstlem}
\end{eqnarray}
See Figure \ref{lemmaxxv} for the comparison of the two sides of (\ref{firstlem}).
\end{lemma}
See Appendix \ref{appendix1} for proof of Lemma \ref{lemma1}. 

\pagebreak

\section{Conclusion} 
We present a class of sum-set inequalities for bounded set linear combinations of random variables typically encountered in the GDoF framework. The bounds are obtained by building upon the aligned image sets (AIS) approach.  Through an example, we showed that these inequalities are useful for obtaining tight GDoF bounds for MIMO interference networks with arbitrary antenna configurations and arbitrary levels of channel uncertainty for each channel. Indeed, we expect these inequalities to be broadly useful for obtaining tight GDoF bounds for MIMO wireless interference and broadcast networks under varying levels of channel strengths and channel uncertainty.

\appendix

\section{Proof of  Theorems \ref{Theorem AIS01}, \ref{Theorem AIS02}, \ref{Theorem AIS03}, \ref{Theorem AIS04} and  \ref{theoremIC}}
\subsection{Proof of Theorem \ref{Theorem AIS01}}\label{proof:AIS01}
\subsubsection{Sketch of the proof}\label{sketch1}
Let us start with a summary of the Aligned Image Sets approach that we use in this proof. 
We are only interested in maximum of difference of entropies of $Z'=(Z_1,Z_2)$ and $Z$ conditioned on $W,\mathcal{G}$, i.e., $H(Z'\mid W)-H(Z\mid W,\mathcal{G})$. Following directly along the AIS approach \cite{Arash_Jafar_PN}, from the functional dependence argument it follows that without loss of generality $Z$ can be assumed to be a function of $Z',W,\mathcal{G}$. So, it follows that, 
\begin {eqnarray}
&&H(Z\mid W,\mathcal{G})+H(Z'\mid Z,W,\mathcal{G})\nonumber\\
&=&H(Z',Z\mid W,\mathcal{G})\label{ew1}\\
&=&H(Z'\mid W)\label{ew2}
\end{eqnarray}
where $(\ref{ew1})$ follows from chain rule and $(\ref{ew2})$ is true as $Z$ is a function of $Z',W,\mathcal{G}$. Thus, the difference of entropies is equal to $H(Z'|Z,W,\mathcal{G})$. Now, for a given $W$ and channel realization $\mathcal{G}$, define aligned image set $S_{\nu}(W,\mathcal{G})$ as the set of all $Z'$ which result in the same $Z$. In the other words, since $Z$ is a function of $Z',W,\mathcal{G}$, we define the set $S_{\nu}(W,\mathcal{G})$ as the set of all values of $Z'$ which produce  the same value for $Z$, as is produced by $Z'=\nu$. Since uniform distribution maximizes the entropy,
\begin{eqnarray}
\mathcal{D}_{\Delta}&\triangleq&H(Z'\mid W)-H(Z\mid W,\mathcal{G})\nonumber\\
&\le& H(Z'\mid Z,W,\mathcal{G})\nonumber\\
&\le&\mbox{E}_{\mathcal{G}}\{\log{\left|S_{\nu}(W,\mathcal{G})\right| \}}\label{jen0}\\
&=&\mbox{E}_{W}\left\{\mbox{E}_{\mathcal{G}}\{\log{|S_{\nu}(W,\mathcal{G})}|\mid W\}\right\}\label{expw}\\
&\le&\max_{w\in\mathcal{W}}\mbox{E}_{\mathcal{G}}\{\log{|S_{\nu}(W,\mathcal{G})}|\mid W=w\}\label{expw2}\\
&\le&\max_{w\in\mathcal{W}}\log{\left\{\mbox{E}_{\mathcal{G}}\{|S_{\nu}(W,\mathcal{G})|\mid W=w\}\right\}}\label{jen}
\end{eqnarray}  
where $\mathcal{W}$ is support of the random variable $W$. (\ref{jen}) comes from the Jensen's Inequality. Thus, the difference of the entropies is bounded by the log of expected value of cardinality of the aligned image set. Now, the most crucial step is to bound the cardinality of $S_{\nu}(W,\mathcal{G})$ where we need to use Bounded Density Assumption of $\mathcal{G}$ to bound the cardinality of $S_{\nu}(W,\mathcal{G})$.  So, from the equation  (\ref{jen}),  $\mbox{E}_{\mathcal{G}}\{|S_{\nu}(W,\mathcal{G})|\mid {\color{blue}W=w}\}$ is what needed to be calculated.
Expected value of size of the cardinality of aligned image set is equal to the summation of probability of alignment over all $Z'$, or in the other words,
\begin{eqnarray}
\mbox{E}_{\mathcal{G}}\{\left|S_{\nu}(W,\mathcal{G})\right| \mid {\color{blue}W=w}\}=\sum_{Z'}P_a(Z') \label{vbcc}
\end{eqnarray}
where $P_a$ is defined as the probability that $Z'$ and $\nu$ correspond to the same $Z$. In the proof, we prove that $\mbox{E}_{\mathcal{G}}\{|S_{\nu}(W,\mathcal{G})|\mid {\color{blue}W=w}\}$ is bounded by $(c_1+c_2\bar{P}^{{(\lambda_2-\lambda_1)}^+})(c_3+c_4\log {\bar{P}})$ from above for constants $c_1,c_2,c_3$ and $c_4$. So, from the inequality (\ref{jen}), we have,
\begin{eqnarray}
\mathcal{D}_{\Delta}&\triangleq&H(Z'\mid W)-H(Z\mid W,\mathcal{G})\nonumber\\
&\le&\max_{w\in\mathcal{W}}\log{\{\mbox{E}_{\mathcal{G}}\{\left|S_{\nu}(W,\mathcal{G})\right| \mid {\color{blue}W=w}\}\}}\\
&\le&c_3+\log{(\log{\bar{P}})} \label {xs1}
\end{eqnarray}
for some constant $c_5$ if $\lambda_2\le\lambda_1$. As $\log{(\log{\bar{P}})}=o(\log{\bar{P}})$, (\ref{dsd1}) is concluded. The detailed arguments are presented next.

\subsubsection{Functional Dependence $Z(Z',W,\mathcal{G})$} \label{funcdep}
We start by showing that there is no loss of generality in the assumption that $(X_1,X_2)$ is a function of $Z'$ and $W$, and therefore $Z$ is a function of $Z',W,\mathcal{G}$. Recall that $(X_1,X_2)$ is independent of $\mathcal{G}$. However, there may be multiple values of $(X_1,X_2)$ that cast the same image in $Z',W$. So the mapping from $Z',W$ to $(X_1,X_2)$ is in general random. Let us denote it by $\mathcal{L}$ i.e.
\begin{eqnarray}
(X_1,X_2)&=&\mathcal{L}(Z',W)
\end{eqnarray} 
In general, because the mapping may be random, $\mathcal{L}$ is a random variable. Conditioning cannot increase entropy, therefore,
\begin{eqnarray}
H(Z\mid W,\mathcal{G})&\geq&H(Z\mid W,\mathcal{G}, \mathcal{L})\nonumber\\
&\geq&\min_{L\in\{\mathcal{L}\}}H(Z\mid W,\mathcal{G},  \mathcal{L}=L)\nonumber\\
\end{eqnarray}
Let $L_o\in\mathcal{L}$ be the mapping that minimizes the entropy term. Fix this as the deterministic mapping,
\begin{eqnarray}
(X_1,X_2)&=&{L_o}(Z',W)
\end{eqnarray}
so that now $Z$ is a function of $Z',W,\mathcal{G}$, and since $Z'$ is a function of $(X_1, X_2)$, $Z$ is equivalently a function of $(X_1,X_2,W,\mathcal{G})$. Based on convenience, we may indicate the functional dependence in any of these forms as
\begin{eqnarray}
Z=Z(Z',W,\mathcal{G})=Z(X_1,X_2,W,\mathcal{G})
\end{eqnarray}
We note that the choice of mapping does not affect the positive entropy
term $H(Z'\mid W)$ but it minimizes $H(Z\mid W,\mathcal{G})$.

\subsubsection { Definition of Aligned Image Sets} \label {defrt}
The aligned image set containing the codeword $\nu\in\mbox{supp}(Z')$ for a given $W=w$ and realization $\mathcal{G}=G$ is defined as the set of all values of $Z'$ that produces the same $Z$ value as is produced by $Z'=\nu$. Mathematically,
\begin{eqnarray}
S_{\nu}(w,{G})&\triangleq&\{z'\in\mbox{supp}(Z'): Z(\nu,w,{G})=Z(z',w,{G})\}
\end{eqnarray}
Since we will need the average (over $\mathcal{G}$) of the cardinality of an aligned image set, E$|S_{\nu}(W,\mathcal{G})|$, it is worthwhile to point out that the cardinality $|S_{\nu}(W,\mathcal{G})|$ as a function of $\mathcal{G}$, is a bounded simple function, and therefore measurable.\footnote{A simple function is a finite sum of indicator functions of measurable sets \cite{stein}.} It is bounded because its values are restricted to  the set of natural numbers not greater than $c\bar{P}^{\lambda_2+\max(\lambda_1,\lambda_2)}$, where $c$ depends on coefficients of linear combinations $L_{1}$ and $L_{2}$. Following the same steps as \cite{Arash_Jafar_PN} it is a simple function too. 
\subsubsection{Bounding the Probability of Image Alignment}
From (\ref{vbcc}), we have $\mbox{E}_{\mathcal{G}}\{\left|S_{\nu}(W,\mathcal{G})\right| \mid W=w\}=\sum_{Z'}P_a(Z')$. Given $W=w,\mathcal{G}$, consider two distinct realizations of $(Z_1,Z_2)$, say $(z_1,z_2)$, and $(z'_1,z'_2)$, which are produced by two distinct  realizations of $(X_1,X_2)$, denoted as  $(\mu_{1},\mu_{2})$ and $(\nu_{1},\nu_{2})$. For  $i\in\{1,2\}$, define $\mu_{1i}$, $\mu_{2i}$, $\nu_{1i}$ and $\nu_{2i}$ as  $(\mu_{i})_{\lambda_1}$, $(\mu_{i})_{\lambda_1}^{\lambda_1+\lambda_2}$, $(\nu_{i})_{\lambda_1}$, and $(\nu_{i})_{\lambda_1}^{\lambda_1+\lambda_2}$ respectively.
\begin{eqnarray}
(z_1,z_2)&=&(\sum_{1\le j\le 2}\lfloor h_{2j} \mu_{2j} \rfloor,\sum_{1\le i,j\le 2}\lfloor h'_{ij} \mu_{ij} \rfloor)\label{eqrs1}\\
(z'_1,z'_2)&=&(\sum_{1\le j\le 2}\lfloor h_{2j} \nu_{2j} \rfloor, \sum_{1\le i,j\le 2}\lfloor h'_{ij} \nu_{ij} \rfloor)\label{eqrs2}
\end{eqnarray}
We wish to bound the probability that the images of these two codewords align, or in other words $Z(z_1,z_2,W,\mathcal{G})=Z(z'_1,z'_2,W,\mathcal{G})$,
\begin{eqnarray}
\lfloor g_{1} \mu_{21} \bar{P}^{\lambda_1}+ g_{1}\mu_{11}\rfloor+\lfloor g_{2} \mu_{22} \bar{P}^{\lambda_1}+g_{2}\mu_{12}\rfloor&=&\lfloor g_{1} \nu_{21} \bar{P}^{\lambda_1}+ g_{1}\nu_{11}\rfloor+\lfloor g_{2} \nu_{22} \bar{P}^{\lambda_1}+g_{2}\nu_{12}\rfloor\\
\Rightarrow |g_{1} (\mu_{21}-\nu_{21}) \bar{P}^{\lambda_1}+ g_{1}(\mu_{11}-\nu_{11})&+& g_{2} (\mu_{22}-\nu_{22}) \bar{P}^{\lambda_1}+g_{2}(\mu_{12}-\nu_{12})|\le 2 \label{dfg1}
~~~~~~~
\end{eqnarray}
defining $C$ as $g_{2} (\mu_{22}-\nu_{22}) \bar{P}^{\lambda_1}-g_{2}(\mu_{12}-\nu_{12})$ we have
\begin{eqnarray}
-2-C\le g_{1} (\mu_{21}-\nu_{21}) \bar{P}^{\lambda_1}&+& g_{1}(\mu_{11}-\nu_{11})\le 2-C
\end{eqnarray}
So for fixed values of $g_{2}$ the random variable $g_{1} (\mu_{21}-\nu_{21}) \bar{P}^{\lambda_1}+ g_{1}(\mu_{11}-\nu_{11})$ must take values within an interval of length no more than $4$. If $|\mu_{21}-\nu_{21}|+|\mu_{11}-\nu_{11}| \neq 0$, then $g_1$ must take values in an interval of length no more than $\frac{4}{|(\mu_{21}-\nu_{21}) \bar{P}^{\lambda_1}+ \mu_{11}-\nu_{11}|}$, the probability of which is no more than $\frac{4f_{\max}}{|(\mu_{21}-\nu_{21}) \bar{P}^{\lambda_1}+ \mu_{11}-\nu_{11}|}$. Note that the integral of any real-valued measurable function $h(x)$ over any measurable set $S$ can be bounded above by $\max_{x\in \mathcal {R}}h(x)$ times the measure of the set $S$, which for the interval $I$ reduces to the length of the interval $I$ \cite{stein}. Similarly, for fixed values of $g_1$ probability of alignment will be bounded by $\frac{4f_{\max}}{|(\mu_{22}-\nu_{22}) \bar{P}^{\lambda_1}+ \mu_{12}-\nu_{12}|}$. As $(z_1,z_2)$, and $(z'_1,z'_2)$ are two distinct realizations of $(Z_1,Z_2)$, at least one of $\mu_{ij}-\nu_{ij}$ for $i,j\in\{1,2\}$ is nonzero. So, it can be concluded that the probability is no more than $P_a(z'_1,z'_2)$ where $P_a(z'_1,z'_2)$ is defined as follows.
\begin{eqnarray}
P_a(z'_1,z'_2)&=&\min(1,\frac{4f_{\max}}{\max({|(\mu_{21}-\nu_{21}) \bar{P}^{\lambda_1}+ \mu_{11}-\nu_{11}|},{|(\mu_{22}-\nu_{22}) \bar{P}^{\lambda_1}+ \mu_{12}-\nu_{12}|})})
\end{eqnarray}
\subsubsection{Bounding the Average Size of Aligned Image Sets}
From (\ref{vbcc}) we have to compute the following summation,
\begin{eqnarray}
&&\mbox{E}_{\mathcal{G}}\{\left|S_{z_1,z_2}(w,\mathcal{G})\right| \}\le\sum_{(z'_1,z'_2),(z'_1,z'_2)\ne(z_1,z_2)}P_a(z'_1,z'_2)
\end{eqnarray}
Note that, from (\ref{eqrs1}), and (\ref{eqrs2}) the terms $|z_1-z'_1|$, and $|z_2-z'_2|$ can be bounded from above by $3+\lfloor 2\Delta_2\bar{P}^{\lambda_2}\rfloor$, and $5+\lfloor 4\Delta_2\bar{P}^{\max(\lambda_1,\lambda_2)}\rfloor$, respectively as $|h_{2j}|$ and $|h'_{ij}|$ are less than $\Delta_2$ for all $i,j\in\{1,2\}$ and, $|\mu_{1j}-\nu_{1j}|$ and $|\mu_{2j}-\nu_{2j}|$ are also less than $\bar{P}^{\lambda_1}$ , $\bar{P}^{\lambda_2}$ respectively. Using the definition of Aligned Image Sets from \ref{defrt}, we have,
\begin{eqnarray}
&&\mbox{E}_{\mathcal{G}}\{\left|S_{z_1,z_2}(w,\mathcal{G})\right| \}\nonumber\\
&\le&\sum_{|z_1-z'_1|\in S_1,|z_2-z'_2|\in S_2}P_a(z'_1,z'_2)\nonumber\\
&\le&\sum_{|\mu_{21}-\nu_{21}|\notin\{0,1\},|z_1-z'_1|\in S_1,|z_2-z'_2|\in S_2}P_a(z'_1,z'_2)+\sum_{|\mu_{22}-\nu_{22}|\notin\{0,1\},|z_1-z'_1|\in S_1,|z_2-z'_2|\in S_2}P_a(z'_1,z'_2)\nonumber\\
&&+\sum_{|\mu_{21}-\nu_{21}|\in\{0,1\},|\mu_{22}-\nu_{22}|\in\{0,1\},|z_2-z'_2|\in S_2}P_a(z'_1,z'_2)\label{hhq}
\end{eqnarray}
where the sets $S_1$, $S'_1$ and $S_2$ are defined as $\{0,1,\cdots,3+\lfloor 2\Delta_2\bar{P}^{\lambda_2}\rfloor\}$, $\{h_o,\cdots,3+\lfloor 2\Delta_2\bar{P}^{\lambda_2}\rfloor\}$ and $\{0,1,\cdots,5+\lfloor 4\Delta_2\bar{P}^{\max(\lambda_1,\lambda_2)}\rfloor\}$, respectively. For simplicity we defined $h_{o}$ as the constant $4+\lfloor |h_{21}|+|h_{22}|\rfloor$. Now, let us bound each term in  (\ref{hhq}) separately. Note that our ultimate goal is to prove $\mbox{E}_{\mathcal{G}}\{\left|S_{z_1,z_2}(\mathcal{G})\right|\}\leq\bar{P}^{(\lambda_2-\lambda_1)^+}(c_3+c_4{(\log{(\bar{P})})})$ for some constants $c_3$ and $c_4$, which along with (\ref{jen}) leads to the conclusion  (\ref{dsd1}) when $\lambda_2<\lambda_1$.
\begin{enumerate}
\item{} Let us compute the first summation in (\ref{hhq}).

\begin{eqnarray}
&&\sum_{|\mu_{21}-\nu_{21}|\notin\{0,1\},|z_1-z'_1|\in S_1,|z_2-z'_2|\in S_2}P_a(z'_1,z'_2)\nonumber\\
&\le&\sum_{|\mu_{21}-\nu_{21}|\notin\{0,1\},|z_1-z'_1|\in S_1,|z_2-z'_2|\in S_2}\frac{4f_{\max}}{\bar{P}^{\lambda_1}\times\max((|\mu_{21}-\nu_{21}|-1) ,(|\mu_{22}-\nu_{22}|-1) )}\label{poi1}\\
&\le&\sum_{|z_2-z'_2|\in S_2}\frac{4f_{\max}}{\bar{P}^{\lambda_1} }\times\sum_{|\mu_{21}-\nu_{21}|\notin\{0,1\},|z_1-z'_1|\in S_1}\frac{1}{\max((|\mu_{21}-\nu_{21}|-1) ,(|\mu_{22}-\nu_{22}|-1) )}\label{poi101}\\
&\le&\sum_{|z_2-z'_2|\in S_2}\frac{4f_{\max}}{\bar{P}^{\lambda_1} }\times\left\{h_o\right.\nonumber\\
&&\left.+\sum_{|\mu_{21}-\nu_{21}|\notin\{0,1\},|z_1-z'_1|\in S'_1}\frac{1}{\max((|\mu_{21}-\nu_{21}|-1) ,(|\mu_{22}-\nu_{22}|-1) )}\right\}\nonumber\\
&&\label{poi11}\\
&\le&\frac{4(5+\lfloor4\Delta_2\bar{P}^{\max(\lambda_1,\lambda_2)}\rfloor)f_{\max}}{\bar{P}^{\lambda_1} }\times\left\{h_o+\sum_{|\mu_{21}-\nu_{21}|\notin\{0,1\},
|z_1-z'_1|\in S'_1}\frac{|h_{21}|+|h_{22}|}{|z_1-z'_1|-h_o+1}\right\}\label{poi12}\nonumber\\
&&\\
&\le&\frac{4(5+\lfloor4\Delta_2\bar{P}^{\max(\lambda_1,\lambda_2)}\rfloor)f_{\max}}{\bar{P}^{\lambda_1} }\times h_o(3+\ln(3+\lfloor 2\Delta_2\bar{P}^{\lambda_2}\rfloor))\label{poi2}
\end{eqnarray}
(\ref{poi1}) follows by bounding the terms $|(\mu_{2i}-\nu_{2i}) \bar{P}^{\lambda_1}+ \mu_{1i}-\nu_{1i}|$ from below by $(|\mu_{2i}-\nu_{2i}|-1) \bar{P}^{\lambda_1}$ if $|\mu_{2i}-\nu_{2i}|>1$. (\ref{poi101}) is breaking the summation into multiplication of two summations, and (\ref{poi11}) is true because $P_a$ is bounded by one, so the summation of $h_o$ such terms can be at most $h_o$. Finally, (\ref{poi12}) follows by bounding $|z_1-z'_1|$ from (\ref{eqrs1}) and (\ref{eqrs2}) as,
\begin{eqnarray}
&&|z_1-z'_1|\nonumber\\
&=&|h_{21}(\mu_{21}-\nu_{21})+h_{22}(\mu_{22}-\nu_{22})+I|\label{I1}\\
&\le &h_o-1+(|h_{21}|+|h_{22}|)\times({\max((|\mu_{21}-\nu_{21}|-1) ,(|\mu_{22}-\nu_{22}|-1) )}) \label{qwe}
\end{eqnarray}
where $I$ is a random variable which takes values in the interval  $(-2,2)$. (\ref{poi2}) is true as the partial sum of harmonic series can be bounded above by logarithmic function, i.e., $\sum_{i=1}^n\frac{1}{i}\le 1+\ln{(n)}$. 
\item{} The second term in (\ref{hhq}), i.e., $\sum_{|\mu_{22}-\nu_{22}|\notin\{0,1\},|z_1-z'_1|\in S_1,|z_2-z'_2|\in S_2}P_a$ is bounded similarly by the exact term in equation (\ref{poi2}) as the inequalities (\ref{poi1})-(\ref{poi2}) remain true whether the summation is over $|\mu_{21}-\nu_{21}|\notin\{0,1\}$ or $|\mu_{22}-\nu_{22}|\notin\{0,1\}$. 
\item{} Finally, the third term in (\ref{hhq}) is bounded from above by splitting the summation into four summations where in each summation the terms $|\mu_{21}-\nu_{21}|$ and $|\mu_{22}-\nu_{22}|$ are fixed to either zero or one. First of all let us write $z_2-z'_2$ from $(\ref{eqrs1})$, and $(\ref{eqrs2})$ as,
\begin{eqnarray}
&&z_2-z'_2\nonumber\\
&=&h'_{11}(\mu_{11}-\nu_{11})+h'_{12}(\mu_{12}-\nu_{12})+h'_{21}(\mu_{21}-\nu_{21})+h'_{22}(\mu_{22}-\nu_{22})+I\label{I1} \label{trg}
\end{eqnarray}
where $I$ is a random variable depending on $\mu_{ij},\nu_{ij},\forall i,j\in\{1,2\}$ which takes values in the interval  $(-4,4)$.
\begin{eqnarray}
&&\sum_{|\mu_{21}-\nu_{21}|\in\{0,1\},|\mu_{22}-\nu_{22}|\in\{0,1\},|z_2-z'_2|\in S_2}P_a(z'_1,z'_2)\nonumber\\
&\le&\sum_{r,s\in\{0,1\}}\sum_{|\mu_{21}-\nu_{21}|=r,|\mu_{22}-\nu_{22}|=s,|z_2-z'_2|\in S_2}\nonumber\\
&&\min{(1,\frac{4f_{\max}}{\max(r\bar{P}^{\lambda_1}-\hat{r}|\mu_{11}-\nu_{11}|, s\bar{P}^{\lambda_1}-\hat{s}|\mu_{12}-\nu_{12}|)})}\nonumber\\
&&\label{ppjj}\\
&\le&\sum_{r,s\in\{0,1\}}\sum_{|\mu_{21}-\nu_{21}|=r,|\mu_{22}-\nu_{22}|=s,|z_2-z'_2|\in S_2}\nonumber\\
&&\min{(1,\frac{4f_{\max}(|h'_{11}|+|h'_{12}|)}{|r\hat{r}u_{11}h'_{11}\bar{P}^{\lambda_1}+s\hat{s}u_{12}h'_{12}\bar{P}^{\lambda_1}-h'_{11}(\mu_{11}-\nu_{11})-h'_{12}(\mu_{12}-\nu_{12})|})}
\label{ppjjcn}\\
&=&\sum_{r,s\in\{0,1\}}\sum_{|\mu_{21}-\nu_{21}|=r,|\mu_{22}-\nu_{22}|=s,|z_2-z'_2|\in S_2}\min(1,\frac{4f_{\max}(|h'_{11}|+|h'_{12}|)}{|P_t+I-(z_2-z'_2)|})\label{ppjjcb}\label{jhg}\\
&=&\sum_{r,s\in\{0,1\}}\sum_{|\mu_{21}-\nu_{21}|=r,|\mu_{22}-\nu_{22}|=s,|z_2-z'_2|\in S_2,|z_2-z'_2-I-P_t|<1}\min(1,\frac{4f_{\max}(|h'_{11}|+|h'_{12}|)}{|P_t+I-(z_2-z'_2)|})\nonumber\\
&+&\sum_{r,s\in\{0,1\}}\sum_{|\mu_{21}-\nu_{21}|=r,|\mu_{22}-\nu_{22}|=s,|z_2-z'_2|\in S_2,|z_2-z'_2-I- P_t|\ge1 }\min(1,\frac{4f_{\max}(|h'_{11}|+|h'_{12}|)}{|P_t+I-(z_2-z'_2)|})\nonumber\\
&&\label{ppjjcb}\\
&\le&\sum_{r,s\in\{0,1\}}\sum_{|\mu_{21}-\nu_{21}|=r,|\mu_{22}-\nu_{22}|=s,|z_2-z'_2|\in S_2,|z_2-z'_2-I-P_t|<1}1\nonumber\\
&+&{4f_{\max}(|h'_{11}|+|h'_{12}|)}\nonumber\\
&&\times\sum_{r,s\in\{0,1\}}\sum_{|\mu_{21}-\nu_{21}|=r,|\mu_{22}-\nu_{22}|=s,|z_2-z'_2|\in S_2,|z_2-z'_2-I- P_t|\ge1 }\frac{1}{\lfloor|P_t+I-(z_2-z'_2)|\rfloor}\label{klb1}\\
&\le&11+{72f_{\max}(|h'_{11}|+|h'_{12}|)}\sum_{r,s\in\{0,1\}}\sum_{\hat{z}\in S_3}\frac{1}{\hat{z}}\label{klb2}\\
&\le&11+{576f_{\max}\Delta_2}(1+\ln{(1+P_r)})\label{poii4}
\end{eqnarray}
where the function $sgn(x)$ is defined as $1$ if $x\ge0$ and $-1$ if $x<0$. The numbers $\hat{r}$, $\hat{s}$, $u_{11}$, $u_{12}$, $P_r$ and $P_t$ are also defined as $2r-1$, $2s-1$, $sgn(\mu_{11}-\nu_{11})$, $sgn(\mu_{12}-\nu_{12})$, $4+\lfloor 4\Delta_2\rfloor+\lfloor 4\Delta_2\bar{P}^{\lambda_1}\rfloor$, and $r\hat{r}u_{11}h'_{11}\bar{P}^{\lambda_1}+s\hat{s}u_{12}h'_{12}\bar{P}^{\lambda_1}+rh'_{21}+sh'_{22}$. The set $S_3$ is defined as the set of integer numbers $\{1,2,\cdots,P_r\}$. (\ref{ppjj}) is derived by replacing $|\mu_{21}-\nu_{21}|,|\mu_{22}-\nu_{22}|\in\{0,1\}$ in $P_a$.  (\ref{ppjjcn}) follows as,
\begin{eqnarray}
&&{|r\hat{r}u_{11}h'_{11}\bar{P}^{\lambda_1}+s\hat{s}u_{12}h'_{12}s\bar{P}^{\lambda_1}-h'_{11}(\mu_{11}-\nu_{11})-h'_{12}(\mu_{12}-\nu_{12})|}\nonumber\\
&\le&{(|\hat{r}u_{11}h'_{11}|+|\hat{s}u_{12}h'_{12}|)}\times{\max(r\bar{P}^{\lambda_1}-\hat{r}|\mu_{11}-\nu_{11}|, s\bar{P}^{\lambda_1}-\hat{s}|\mu_{12}-\nu_{12}|)}\nonumber\\
&=&{(|h'_{11}|+|h'_{12}|)}\times{\max(r\bar{P}^{\lambda_1}-\hat{r}|\mu_{11}-\nu_{11}|, s\bar{P}^{\lambda_1}-\hat{s}|\mu_{12}-\nu_{12}|)}\label{gfds}
\end{eqnarray}
where $(\ref{gfds})$ is true as the $\hat{r},\hat{s},u_{11},u_{12}\in\{-1,1\}$. (\ref{jhg}) is derived by replacing $z_2-z'_2$ from (\ref{trg}), (\ref{ppjjcb}) follows by breaking summation into two summations, and, (\ref{klb1}) is true as minimum of any two numbers can be bounded above by one of them. Note that the first summation in (\ref{klb1}) can be at most $11$ as $|z_2-z'_2-I- P_t|<1$ is true only if $-5+P_t<z_2-z'_2<P_t+5$. Moreover, $-5+P_t<z_2-z'_2<P_t+5$ can be true for at most $11$ integer numbers of $|z_2-z'_2|$. So, the first summation is at most 11. Second summation in (\ref{klb1}) is bounded as,
 \begin {eqnarray}
&&\sum_{r,s\in\{0,1\}}\sum_{|\mu_{21}-\nu_{21}|=r,|\mu_{22}-\nu_{22}|=s,|z_2-z'_2|\in S_2,|z_2-z'_2-I- P_t|\ge1 }\frac{1}{\lfloor|P_t+I-(z_2-z'_2)|\rfloor}\nonumber\\
 &\le&18\sum_{r,s\in\{0,1\}}\sum_{\hat{z}\in S_3}\frac{1}{\hat{z}} \label{bvc}
 \end{eqnarray}
where (\ref{bvc}) is concluded from the following two points,
 \begin {enumerate}
 \item{} Each summand in the left summation (\ref{bvc}) is reciprocal of a positive integer number, and reciprocal of any positive integer number, i.e., $n$ can be repeated in the left summation at most $18$ times as $\lfloor|P_t+I-(z_2-z'_2)|\rfloor=n$ can have at most $18$ solutions in the set of $z_2-z'_2\in\mathcal{Z}$ for any fixed integer number $n$. 
\begin{eqnarray}
&&\lfloor|P_t+I-(z_2-z'_2)|\rfloor=n\\
\Rightarrow&& n-4+P_t<z_2-z'_2<n+5+P_t \label{mvn}\\
\text{or}&&-n-5+P_t<z_2-z'_2<-n+4+P_t \label{mvn1}
 \end{eqnarray}
(\ref{mvn}), (\ref{mvn1}) can be true only for $18$ integers of $z_2-z'_2$ in the set of $z_2-z'_2\in\mathcal{Z}$ for any fixed integer number of $n$. So, any reciprocal of any positive integer number, i.e., $n$ can be repeated at most 18 times in the left summation.
 \item{} $\hat{z}$ can only get integer numbers from the set $S_3$, as ${\lfloor|P_t+I-(z_2-z'_2)|\rfloor}$ is bounded from above by $4+\lfloor 4\Delta_2\rfloor+\lfloor 4\Delta_2\bar{P}^{\lambda_1}\rfloor$.
 \end{enumerate}
Finally, (\ref{poii4}) is true as the partial sum of harmonic series can be bounded above by logarithmic function.  

\end{enumerate}
\subsubsection{Combining the Bounds to Complete the Proof}
Now, from (\ref{hhq}), (\ref{poi2}), and (\ref{poii4}) since constant terms and $\log{\log{{P}}}$ are $o(\log{{P}})$, we have,
\begin{eqnarray}
&&\log\{\mbox{E}_{\mathcal{G}}\{\left|S_{z_1,z_2}(w,\mathcal{G})\right| \}\}\le{(\lambda_2-\lambda_1)}^+\log{\bar{P}}+o(\log{{P}}) \label{sa}
\end{eqnarray}
as (\ref{sa}) is true for all $W=w$, from (\ref{vbcc}) we have,
\begin{eqnarray}
&&H(Z'\mid W)-H(Z\mid W,\mathcal{G})\le {(\lambda_2-\lambda_1)}^+\log{\bar{P}}+o(\log{P})
\end{eqnarray}
Note that, (\ref{dsd1}) is concluded when $\lambda_2\le\lambda_1$.

\subsection{Proof of  Theorem \ref{Theorem AIS02},  \ref{Theorem AIS03} and  \ref{Theorem AIS04}}\label{prooft4} 
In this section we only present  proof of Theorem \ref{Theorem AIS04} as Theorem \ref{Theorem AIS02} and  \ref{Theorem AIS03} are obtained as special cases of Theorem \ref{Theorem AIS04}.\\

Recall that,
 \begin {eqnarray}
Z_k(t)&=&L_k^b(t)(X_1(t),X_2(t), \cdots, X_N(t))\label{mn1+}\\
Z_{k,1}(t)&=& L_{k1}^{\vec{\gamma}_{k1}\vec{\delta}_{k1}}(t)((X_{j}(t))_{\sum_{r=1}^{i-1}\lambda_{kr}}^{\sum_{r=1}^i\lambda_{kr}}, i\in I_{k,1}, j\in[N])\label{mn2+}\\
Z_{k,2}(t)&=& L_{k2}^{\vec{\gamma}_{k2}\vec{\delta}_{k2}}(t)((X_{j}(t))_{\sum_{r=1}^{i-1}\lambda_{kr}}^{\sum_{r=1}^i\lambda_{kr}}, i\in I_{k,2}, j\in[N])\label{mn3+}\\
&\vdots&\notag\\
Z_{k,l_k}(t)&=&L_{kl_k}^{\vec{\gamma}_{kl_k}\vec{\delta}_{kl_k}}(t)((X_{j}(t))_{\sum_{r=1}^{i-1}\lambda_{kr}}^{\sum_{r=1}^i\lambda_{kr}}, i\in I_{k,l_k}, j\in[N])\label{mn4+}
\end{eqnarray}
The channel uses are indexed by $t\in\mathbb{N}$. $I_{kk'}\subset [M], k\in[K], k'\in[l_k],$  such that $i<j\Rightarrow m(k,i)\geq m(k,j)$, where $$m(a,b)\triangleq \min\{m: m\in I_{a,b}\}.$$ If for all $k\in[K]$ and for each $s\in\{1,2,\cdots, l_k-1\}$,
\begin {eqnarray}
\mathcal{T}(Z_{k,s+1})+\mathcal{T}(Z_{k,s+2})+\cdots+\mathcal{T}(Z_{k,l_k})&\leq& \lambda_{k,1}+\lambda_{k,2}+\cdots+\lambda_{k,(m(k,s)-1)}\label{condt4++}
\end{eqnarray}
then
\begin {eqnarray}
H(Z_1^{[n]},\cdots,Z_K^{[n]}\mid W,\mathcal{G})&\geq& H(Z_{1,1}^{[n]},\cdots,Z_{K,l_K}^{[n]}\mid W)+Kn~o(\log{\bar{P}})\label{dssd4+}
\end{eqnarray}
Note that, \eqref{condt4++} is equivalent to,
\begin {eqnarray}
\sum_{s<r\le l_k}\max_{j\in[N],k'\in I_{kr}}\min(\lambda_{kk'},(\gamma_{krk'j}-\delta_{krk'j})^+)\le {\sum_{1\le k'<\min_{m\in I_{ks}}m} \lambda_{kk'}}\label{conth4+}
\end{eqnarray}
where $\gamma_{krk'j}$ and $\delta_{krk'j}$ are elements of the vectors $\vec{\gamma}_{kr}$, $\vec{\delta}_{kr}$. Without loss of generality we assume $K< N$ since for $K\ge N$ the left hand side of \eqref{dssd4+} is equal to $H(X_1^{[n]},\cdots,X_N^{[n]}\mid W,\mathcal{G})$. Therefore, \eqref{dssd4+} is immediate. Moreover, we assume $\delta_{krk'j}\le\gamma_{krk'j}$  for any $k\in[K],r\in[l_k],j\in[N],k'\in I_{kr}$ \footnote{Note that $(x)_{\delta}^{\gamma}=0$ if $\gamma\le\delta$, see Definition \ref{deflc}.}.
\subsubsection{Sketch of the proof}\label{sketch2}
The first steps to prove (\ref{dssd4+}) follows from from the same lines of proof of Theorem \ref{Theorem AIS01} and it is straightforward based on it, see \ref{sketch1}. To avoid repetition we only go over the parts that are different from proof of Theorem \ref{Theorem AIS01}. Similar to proof of Theorem \ref{Theorem AIS01}, we are only interested in maximum of difference of entropies of ${Z'}^n=(Z_{11}^{[n]},\cdots,Z_{Kl_K}^{[n]})$ and $Z^n=(Z_1^{[n]},\cdots,Z_K^{[n]})$ conditioned on $W$ and $\mathcal{G}$, i.e., $H({Z'}^n\mid W)-H(Z^n\mid W,\mathcal{G})$. Similar to proof of Theorem \ref{Theorem AIS01}, from the functional dependence argument it follows that without loss of generality $Z^n$ can be made a function of ${Z'}^n,W,\mathcal{G}$. For given $W$ and channel realization $\mathcal{G}$, define aligned image set $S_{\nu^n}(W,\mathcal{G})$ as the set of all ${Z'}^n$ which result in the same $Z^n$. Thus, we have,
\begin{eqnarray}
\mathcal{D}_{\Delta}&\triangleq&H(Z'^n\mid W)-H(Z^n\mid W,\mathcal{G})\nonumber\\
&\le& H(Z'^n\mid Z^n,W,\mathcal{G})\nonumber\\
&\le&\mbox{E}_{\mathcal{G}}\{\log{\left|S_{\nu^n}(W,\mathcal{G})\right| \}}\label{jen0+}\\
&=&\mbox{E}_{W}\left\{\mbox{E}_{\mathcal{G}}\{\log{|S_{\nu^n}(W,\mathcal{G})}|\mid W\}\right\}\label{expwq}\\
&\le&\max_{w\in\mathcal{W}}\mbox{E}_{\mathcal{G}}\{\log{|S_{\nu^n}(W,\mathcal{G})}|\mid W=w\}\label{expwq2}\\
&\le&\max_{w\in\mathcal{W}}\log{\left\{\mbox{E}_{\mathcal{G}}\{|S_{\nu^n}(W,\mathcal{G})|\mid W=w\}\right\}}\label{jen+}
\end{eqnarray}  
where (\ref{jen+}) comes from the Jensen's Inequality. Thus, the difference of the entropies is bounded by the log of expected value of cardinality of the aligned image set. Similar to proof of Theorem \ref{Theorem AIS01}, the key step is to bound the cardinality of $S_{\nu^n}(W,\mathcal{G})$ where we need to use Bounded Density Assumption of $\mathcal{G}$ to bound the cardinality of $S_{\nu^n}(W,\mathcal{G})$. So, from (\ref{jen+}), $\mbox{E}_{\mathcal{G}}\{|S_{\nu^n}(W,\mathcal{G})|\mid W=w\}$ is what needed to be calculated. Expected value of size of the cardinality of aligned image set is equal to the summation of probability of alignment over all ${Z'}^n$, or in the other words,
\begin{eqnarray}
\mbox{E}_{\mathcal{G}}\{|S_{\nu^n}(W,\mathcal{G})|{\color{blue}\mid W=w}\}=\sum_{{Z'}^n}\mathbb{P}({Z'}^n) \label{vbcc+}
\end{eqnarray}
where $\mathbb{P}({Z'}^n) $ is defined as the probability that ${Z'}^n$ and $\nu^n$ correspond to the same $Z^n$. In the proof, we prove that for any $w\in\mathcal{W}$, $\mbox{E}_{\mathcal{G}}\{|S_{\nu^n}(W,\mathcal{G})|\mid W=w\}$ is bounded by ${(c_1+c_2\log {\bar{P}})}^{Kn}$ from above for some positive constants $c_1,c_2$. Note that, $\mathcal{W}$ was defined as the support of $W$. So, from the inequality (\ref{jen}), we have,
\begin{eqnarray}
\mathcal{D}_{\Delta}&\triangleq&H({Z'}^n\mid W)-H(Z^n\mid W,\mathcal{G})\nonumber\\
&\le&\max_{w\in\mathcal{W}}\log{\{\mbox{E}_{\mathcal{G}}\{|S_{\nu^n}(W,\mathcal{G})|{\color{blue}\mid W=w}\}\}}\\
&\le&Kn~c_3+Kn~\log{(\log{\bar{P}})} \label {xs1}
\end{eqnarray}
for some positive constant $c_3$. As $\log{(\log{\bar{P}})}=o(\log{\bar{P}})$, (\ref{dssd4}) is concluded. The detailed arguments are presented next.

\subsubsection{Bounding the Probability of Image Alignment}

Given ~$\mathcal{G}$ and $W=w$, ~consider~ two ~distinct ~instances ~of ~${Z' }^n$ ~denoted as $\mu^{[n]}=(\mu_{11}^{[n]},\cdots,\mu_{Kl_K}^{[n]})$ and $\nu^{[n]}=(\nu_{11}^{[n]},\cdots,\nu_{Kl_K}^{[n]})$ produced by corresponding  realizations of codewords $(X_1^n,X_2^n,\cdots,X_N^n)$ denoted by $(E_1^n,E_2^n,\cdots,E_N^n)$ and $(F_1^n,F_2^n,\cdots,F_N^n)$, respectively. For any $k\in[K]$, $l\in [l_k]$, $t\in[n]$, the random variables $\mu_{kl}(t)$ and $\nu_{kl}(t)$ are derived as, 
\begin{eqnarray}
\mu_{kl}(t)&=&L_{kl}^{\vec{\gamma}_{kl}\vec{\delta}_{kl}}((E_{j}(t))_{\sum_{r=1}^{i-1}\lambda_{k,r}}^{\sum_{r=1}^i\lambda_{kr}}, i\in I_{kl}, j\in[N])\label{+xff2f}\\
\nu_{kl}(t)&=&L_{kl}^{\vec{\gamma}_{kl}\vec{\delta}_{kl}}((F_{j}(t))_{\sum_{r=1}^{i-1}\lambda_{kr}}^{\sum_{r=1}^i\lambda_{kr}}, i\in I_{kl}, j\in[N])\label{+xff2f2}
\end{eqnarray}
In the next step we bound $\mathbb{P}(\mu^{[n]}\in \mathcal{S}_{\nu^{[n]}})$ from above. We wish to bound the probability that the images of these two codewords align, or in other words $Z^n({\mu}^n,W,\mathcal{G})=Z^n({\nu}^n,W,\mathcal{G})$. Thus, for any $k\in[K]$ and $t\in[n]$ we have,
\begin{eqnarray}
&&\sum_{j=1}^N\lfloor g_{kj}(t) E_{j}(t)\rfloor=\sum_{j=1}^N\lfloor g_{kj}(t) F_{j}(t)\rfloor\label{ret+1}\\
&&\Rightarrow|\sum_{j=1}^N g_{kj}(t) (E_{j}(t)-F_{j}(t))|\le N\label{ret+2}
\end{eqnarray}
where  (\ref{ret+2}) follows from (\ref{ret+1}) as for any real number $x$, $|x-\lfloor x\rfloor|<1$. For any $k\in[K],t\in[n],l\in\{N\}$ and any fixed values of $g_{k1}(t),\cdots,g_{k(l-1)}(t),g_{k(l+1)}(t),\cdots,g_{kM}(t)$ the random variable $g_{kl}(t)\left(E_{l}(t)-F_{l}(t)\right)$ must take values within an interval of length no more than $2{N}$. Therefore, for any $k\in[K],t\in[n],l\in[N]$ if $E_{l}(t)\neq F_{l}(t)$, then $g_{kl}(t)$ must take values in an interval of length no more than $\frac{2{N}}{|E_{l}(t)-F_{l}(t)|}$, the probability of which is no more than $\frac{2{N}f_{\max}}{|E_{l}(t)-F_{l}(t)|}$. The probability of alignment is  bounded by
\begin{eqnarray}
\mathbb{P}(\mu^{[n]}\in \mathcal{S}_{\nu^{[n]}})&\le&{\left\{\prod_{t=1,\max_{j\in N}|E_{j}(t)-F_{j}(t)|\neq0}^n\frac{2Nf_{\max}}{\max_{j\in N}|E_{j}(t)-F_{j}(t)|}\right\}}^K\\
&=&\prod_{k=1}^K~\prod_{t=1,\max_{j\in N}|E_{j}(t)-F_{j}(t)|\neq0}^n\frac{2Nf_{\max}}{\max_{j\in N}|E_{j}(t)-F_{j}(t)|}\label{pale}
\end{eqnarray}

\subsubsection{Bounding the Average Size of Aligned Image Sets}
From (\ref{vbcc+}) we have to compute the following summation,
\begin{eqnarray}
&&\mbox{E}_{\mathcal{G}}\{|S_{\nu^n}(W,\mathcal{G})|{\color{blue}\mid W=w}\}\nonumber\\
&\le&\sum_{\mu^{[n]}\in {\mathcal{Z}'}^n}\mathbb{P}(\mu^{[n]}\in \mathcal{S}_{\nu^{[n]}})\\
&\le&\sum_{\mu^{[n]}\in {\mathcal{Z}'}^n}\prod_{k=1}^K~\prod_{t=1,\max_{j\in N}|E_{j}(t)-F_{j}(t)|\neq0}^n\frac{2Nf_{\max}}{\max_{j\in N}|E_{j}(t)-F_{j}(t)|}\\
&=&\sum_{\mu_{11}^{[n]}\in {\mathcal{Z}_{11}}^n,\cdots,\mu_{Kl_K}^{[n]}\in {\mathcal{Z}}_{Kl_K}^n}\prod_{k=1}^K~\prod_{t=1,\max_{j\in N}|E_{j}(t)-F_{j}(t)|\neq0}^n\frac{2Nf_{\max}}{\max_{j\in N}|E_{j}(t)-F_{j}(t)|}\\
&\le&{(c_1+c_2\log{\bar{P}})}^{Kn}\label{boundf}
\end{eqnarray}
for some positive constants $c_1$ and $c_2$. ${\mathcal{Z}'}^n$ and $\mathcal{Z}_{kl}^{[n]}$ are defined as the support of the random variable ${\mu}^n$ and $\mu_{kl}^{[n]}$ for any $k\in[K]$ and $l\in[l_K]$.  Note that, from \eqref{jen+} and   \eqref{boundf}, the bound \eqref{xs1}  is obtained. Therefore, \eqref{dssd4} is concluded. 

\subsubsection {Proof of  \eqref{boundf} for $K=1$ and $n=1$}\label{prkn}
First let us prove  the bound \eqref{boundf} when $K=1$ and $n=1$. Without loss of generality let us drop the time index $(t)$ and assume $k=1$. Thus, our goal is to prove that,
\begin{eqnarray}
&&\sum_{|\mu_{11}-\nu_{11}|\in {\mathcal{Z}_{11}},\cdots,|\mu_{1l_1}-\nu_{1l_1}|\in {\mathcal{Z}}_{1l_1},\max_{j\in N}|E_{j}-F_{j}|\neq0}\frac{2Nf_{\max}}{\max_{j\in N}|E_{j}-F_{j}|}\nonumber\\
&&\le c_1+c_2{\log{\bar{P}}}\label{boundf++}
\end{eqnarray}
where $\mathcal{Z}_{kl}$ is defined as the set $\{0\}\cup[2MN+\lfloor MN\Delta_2{\bar{P}}^{\max_{j\in[N],k'\in I_{kl}}\min(\lambda_{kk'},\gamma_{klk'j}-\delta_{klk'j})}\rfloor]$ for any $k\in[K]$ and $l\in[l_K]$ \footnote{Note that from \eqref{+xff2f} and \eqref{+xff2f2}, we have,
\begin{eqnarray*}
|\mu_{kl}(t)-\nu_{kl}(t)|&\le&2MN+MN\Delta_2{\bar{P}}^{\max_{j\in[N],k'\in I_{kl}}\min(\lambda_{kk'},\gamma_{klk'j}-\delta_{klk'j})}.
 \end{eqnarray*}}. For any $l\in [l_1]$, the random variables $\mu_{1l}$ and $\nu_{1l}$ are derived from \eqref{+xff2f} and \eqref{+xff2f2} as, 
\begin{eqnarray}
\mu_{1l}&=&L_{1l}^{\gamma_{1lij}\delta_{1lij}}((E_{j})_{\sum_{r=1}^{i-1}\lambda_{1r}}^{\sum_{r=1}^i\lambda_{1r}}, i\in I_{1l}, j\in[N])\label{-+xff2f}\\
\nu_{1l}&=&L_{1l}^{\gamma_{1lij}\delta_{1lij}}((F_{j})_{\sum_{r=1}^{i-1}\lambda_{1r}}^{\sum_{r=1}^i\lambda_{1r}}, i\in I_{1l}, j\in[N])\label{-+xff2f2}
\end{eqnarray}
In the next step to compute the summation (\ref{boundf++}). First of all, we define $\breve{\Delta}_{lij}$ as,
\begin{eqnarray}
\breve{\Delta}_{lij}&\triangleq&((E_{j})_{\sum_{r=1}^{i-1}\lambda_{1r}}^{\sum_{r=1}^i\lambda_{1r}})^{\gamma_{1lij}}_{\delta_{1lij}}-((F_{j})_{\sum_{r=1}^{i-1}\lambda_{1r}}^{\sum_{r=1}^i\lambda_{1r}})^{\gamma_{1lij}}_{\delta_{1lij}}
\end{eqnarray}
and define the number $l^\star$ as the smallest integer where
\begin{eqnarray}
\max_{ i\in I_{1l^\star},j\in[N]}|\breve{\Delta}_{l^\star ij}|\ge2\label{conl}
\end{eqnarray}
Consider the following two cases.
\begin{enumerate}
\item{} $l^\star$ doesn't exist. \\

If there doesn't exist any $l\in[l_1]$, $ i\in I_{1l}$ and $j\in[N]$ satisfying the condition \eqref{conl}, i.e., $\forall l,i,j, l\in[l_1], i\in I_{1l}, j\in[N], \breve{\Delta}_{lij}\in\{-1,0,1\}$, then each of the variables $|\mu_{1l}-\nu_{1l}|$ is bounded by $MN\Delta_2$ as we have
\begin{eqnarray}
|\mu_{1l}-\nu_{1l}|&\le&2MN\Delta_2\label{bmn+1}
\end{eqnarray}
for any $l\in[l_k]$. Therefore, \eqref{boundf++} is true as the summation in  \eqref{boundf++} is the summation of positive numbers less than $2Nf_{\max}$ over at most ${(2MN\Delta_2)}^{l_1}$ numbers. 

\item{} $1\le l^\star\le l_1$.\\

Any number $X\in\mathcal{X}_\delta$ can be written as $(X)_{a}^{\delta}\bar{P}^a+(X)^{a}_{b}\bar{P}^b+X_b$ for any non-negative numbers $a,b$ less than $\delta$. Thus, the term $|E_j-F_j|$  can be rewritten as,
\begin{eqnarray}
&&|E_j-F_j|\nonumber\\
&=&|J_{ij}\bar{P}^{\sum_{r=1}^{i-1}\lambda_{1r}+\delta_{1l^\star ij}}+\breve{\Delta}_{l^\star ij}\bar{P}^{\sum_{r=1}^{i-1}\lambda_{1r}+\delta_{1l^\star ij}}+(E_{j})_{\sum_{r=1}^{i-1}\lambda_{1r}+\delta_{1l^\star ij}}-(F_{j})_{\sum_{r=1}^{i-1}\lambda_{1r}+\delta_{1l^\star ij}}|\nonumber\\
&=&|J_{ij}\bar{P}^{\sum_{r=1}^{i-1}\lambda_{1r}+\delta_{1l^\star ij}}+\breve{\Delta}_{l^\star ij}\bar{P}^{\sum_{r=1}^{i-1}\lambda_{1r}+\delta_{1l^\star ij}}+(E_{j})_{\sum_{r=1}^{i-1}\lambda_{1r}+\delta_{1l^\star ij}}-(F_{j})_{\sum_{r=1}^{i-1}\lambda_{1r}+\delta_{1l^\star ij}}|\nonumber\\
&\ge&(|{J}_{ij}+\breve{\Delta}_{l^\star ij}|-1)\bar{P}^{\sum_{r=1}^{i-1}\lambda_{1r}+\delta_{1l^\star ij}}\\
&\ge&(|{J}_{ij}+\breve{\Delta}_{l^\star ij}|-1)\bar{P}^{\sum_{r=1}^{i-1}\lambda_{1r}}\label{try}
\end{eqnarray}
where $J_{ij}$ is defined as,
\begin{eqnarray}
J_{ij}&=&(E_{j})_{\min(\lambda_{1i},\gamma_{1l^\star ij})}^{\sum_{r=i}^M\lambda_{1r}}\bar{P}^{\min(\lambda_{1i},\gamma_{1l^\star ij})}-(F_{j})_{\min(\lambda_{1i},\gamma_{1l^\star ij})}^{\sum_{r=i}^M\lambda_{1r}}\bar{P}^{\min(\lambda_{1i},\gamma_{1l^\star ij})}\nonumber\\
\end{eqnarray}
Moreover, define $J^{\star}_{ij}$ as the set of $\{-\bar{P}^{\min(\lambda_{1i},\gamma_{1l^\star ij})},0,\bar{P}^{\min(\lambda_{1i},\gamma_{1l^\star ij})}\}$. (\ref{try}) is true for any non-negative  real number $x$ as both of the random variables $(E_{j})_{x}$ and $(F_{j})_{x}$ are positive numbers less than $\bar{P}^{x}$. Since \eqref{try} is true for any $i\in I_{1l^\star }$, we have,
\begin{eqnarray}
&&|E_j-F_j|\nonumber\\
&\ge&\max_{i\in I_{1l^\star }}\big((|{J}_{ij}+\breve{\Delta}_{l^\star ij}|-1)\bar{P}^{\sum_{r=1}^{i-1}\lambda_{1r}}\big)\label{try2}\\
&\ge&\min_{i\in I_{1l^\star }}\bar{P}^{\sum_{r=1}^{i-1}\lambda_{1r}}\max_{i\in I_{1l^\star }}(|{J}_{ij}+\breve{\Delta}_{l^\star ij}|-1)\label{try3}\\
&\ge&\min_{i\in I_{1l^\star }}\bar{P}^{\sum_{r=1}^{i-1}\lambda_{1r}}\max_{i\in I_{1l^\star }}\min_{{J}_{ij}\in{J}^{\star}_{ij}}(|{J}_{ij}+\breve{\Delta}_{l^\star ij}|-1)\label{try3+}
\end{eqnarray}
where \eqref{try3} is true as for any two non-negative real-valued functions $f(x)$ and $g(x)$ and the set $S\subseteq\mathbb{R}$ we have,
\begin{eqnarray}
\max_{x\in S}f(x)g(x)\ge\max_{x\in S}f(x)\min_{x\in S}g(x)
\end{eqnarray}
The left side of  \eqref{boundf++} is bounded as,
\begin{eqnarray}
&&\sum_{|\mu_{11}-\nu_{11}|\in {\mathcal{Z}_{11}},\cdots,|\mu_{1l_1}-\nu_{1l_1}|\in {\mathcal{Z}}_{1l_1} ,\max_{j\in N}|E_{j}-F_{j}|\neq0}\frac{2Nf_{\max}}{\max_{j\in N}|E_{j}-F_{j}|}\nonumber\\
&\le&\sum_{|\mu_{11}-\nu_{11}|\in {\mathcal{Z}_{11}},\cdots,|\mu_{1l_1}-\nu_{1l_1}|\in {\mathcal{Z}}_{1l_1}}\frac{2Nf_{\max}}{\min_{i\in I_{1l^\star}}\bar{P}^{\sum_{r=1}^{i-1}\lambda_{1r}}(\max_{ i\in I_{1l^\star},j\in[N]}|J_{ij}-\breve{\Delta}_{l^\star ij}|-1)}\nonumber\\
\label{bmn5}\\
&=&\frac{2Nf_{\max}}{\min_{i\in I_{1l^\star}}\bar{P}^{\sum_{r=1}^{i-1}\lambda_{1r}}}\sum_{|\mu_{11}-\nu_{11}|\in {\mathcal{Z}_{11}},\cdots,|\mu_{1l_1}-\nu_{1l_1}|\in {\mathcal{Z}}_{1l_1}}\frac{1}{\max_{ i\in I_{1l^\star},j\in[N]}|J_{ij}-\breve{\Delta}_{l^\star ij}|-1}\nonumber\\
\label{bmn6}\\
&\le&\frac{2Nf_{\max}}{\min_{i\in I_{1l^\star}}\bar{P}^{\sum_{r=1}^{i-1}\lambda_{1r}}}\sum_{|\mu_{11}-\nu_{11}|\in {\mathcal{Z}_{11}},\cdots,|\mu_{1l_1}-\nu_{1l_1}|\in {\mathcal{Z}}_{1l_1}}\nonumber\\
&&\frac{1}{\max_{ i\in I_{1l^\star},j\in[N]}\min_{J_{ij}\in J^{\star}_{ij}}|J_{ij}-\breve{\Delta}_{l^\star ij}|-1}\nonumber\\
\label{bmn61}\\
&\le&\frac{2Nf_{\max}}{\min_{i\in I_{1l^\star}}\bar{P}^{\sum_{r=1}^{i-1}\lambda_{1r}}}\big(\sum_{|J+\mu_{1l^\star }-\nu_{1l^\star }|<3MN,|\mu_{11}-\nu_{11}|\in {\mathcal{Z}_{11}},\cdots,|\mu_{1l_1}-\nu_{1l_1}|\in {\mathcal{Z}}_{1l_1}} 1\nonumber\\
&&+\sum_{3MN\le|J+\mu_{1l^\star }-\nu_{1l^\star }|,|\mu_{11}-\nu_{11}|\in {\mathcal{Z}_{11}},\cdots,|\mu_{1l_1}-\nu_{1l_1}|\in {\mathcal{Z}}_{1l_1} }\frac{MN\Delta_2}{|{J}+\mu_{1l^\star }-\nu_{1l^\star }|-3MN+1}\big)\nonumber\\
\label{bmn7}\\
&\le&\frac{2Nf_{\max}}{\min_{i\in I_{1l^\star}}\bar{P}^{\sum_{r=1}^{i-1}\lambda_{1r}}}\big(3MN\times\big(\prod_{l=1}^{l^\star-1}2MN\Delta_2\big)\times\prod_{l=l^\star+1}^{l_1}|\mathcal{Z}_{1l}|\nonumber\\
&&+\sum_{3MN\le|J+\mu_{1l^\star }-\nu_{1l^\star }|,|\mu_{1l^\star}-\nu_{1l^\star}|\in {\mathcal{Z}_{1l^\star}}}\frac{MN\Delta_2}{|{J}+\mu_{1l^\star }-\nu_{1l^\star }|-3MN+1}\nonumber\\
&&\times\big(\prod_{l=1}^{l^\star-1}2MN\Delta_2\big)\times\prod_{l=l^\star+1}^{l_1}|\mathcal{Z}_{1l}|\big)
\label{bmn73}\\
&=&\frac{2Nf_{\max}}{\min_{i\in I_{1l^\star}}\bar{P}^{\sum_{r=1}^{i-1}\lambda_{1r}}}\times\big(\prod_{l=1}^{l^\star-1}2MN\Delta_2\big)\times\big(\prod_{l=l^\star+1}^{l_1}|\mathcal{Z}_{1l}|\big)\big(3MN+\sum_{\hat{z}\in {\bar{\mathcal{Z}}_{1l^\star}} }\frac{MN\Delta_2}{\hat{z}}\big)\label{bmn77}\\
&\le&\frac{2Nf_{\max}{(2MN\Delta_2)}^{l^\star-1}\prod_{l=l^\star+1}^{l_1}
3MN\Delta_2{\bar{P}}^{\max_{k'\in I_{1l}}\lambda_{1k'}{\min(\lambda_{1k'},\gamma_{1lk'j}-\delta_{1lk'j})}}
}{\min_{i\in I_{1l^\star}}\bar{P}^{\sum_{r=1}^{i-1}\lambda_{1r}}}\nonumber\\
&&\times\big(3MN+MN\Delta_2\ln(1+3MN\Delta_2{\bar{P}})\big)\label{bmn8}\\
&\le&2Nf_{\max}{(3MN\Delta_2)}^{l_1-1}(3MN+MN\Delta_2\ln(1+3MN\Delta_2{\bar{P}}))\label{bmn9}
\end{eqnarray}
where $J$ and $\hat{J}_{ij}$ are defined as $L_{1l^\star}(\hat{J}_{ij})=\sum_{i\in I_{1l^\star}, j\in[N]}\lfloor h_{l^\star ij}\hat{J}_{ij}\rfloor$ and $\hat{J}_{ij}=\underset{J_{ij}\in J^{\star}_{ij}}{\operatorname{argmin}}|J_{ij}-\breve{\Delta}_{l^\star ij}|$ (Note~ that ~$J$ ~is a constant number). ~$\bar{\mathcal{Z}}_{1l^\star}$ ~is~ defined as the set of  $[MN+\lfloor 2MN\Delta_2{\bar{P}}^{\max_{j\in[N],k'\in I_{1l^\star }}\min(\lambda_{1k'},\gamma_{1l^\star k'j}-\delta_{1l^\star k'j})}\rfloor]$. $h_{l^\star ij}$ are also defined as the coefficients of the linear combination $L_{1l^\star}$.  Note that for $l_1=1$, the product $\prod_{l=2}^{l_1}\mid{\mathcal{Z}_{1l}}\mid$ is the empty product, with the value 1. Note that the denominator in \eqref{bmn5} cannot be zero as from  \eqref{conl} we have, $|\breve{\Delta}_{l^\star ij}|\ge2$. \eqref{bmn5}  yields from (\ref{try}) and \eqref{bmn7} is true as from \eqref{-+xff2f} and \eqref{-+xff2f2}, we have,
\begin{eqnarray}
&&|\mu_{1l^\star }-\nu_{1l^\star }-\sum_{i\in I_{1l^\star }, j\in[N]}h_{l^\star ij}\breve{\Delta}_{l^\star ij}|\le MN\\
&&\rightarrow |J+\mu_{1l^\star }-\nu_{1l^\star }-\sum_{i\in I_{1l^\star }, j\in[N]}h_{l^\star ij}(\hat{J}_{ij}+\breve{\Delta}_{l^\star ij})|\le 2MN\label{rarro}
\end{eqnarray}
\eqref{rarro} is true as $J$ is equal to $\sum_{i\in I_{1l^\star }, j\in[N]}\lfloor h_{l^\star ij}\hat{J}_{ij}\rfloor$. Thus from \eqref{rarro}, the inequality \eqref{bmn7} is concluded as,
\begin{eqnarray}
 |J+\mu_{1l^\star }-\nu_{1l^\star }|\le MN\Delta_2\max_{i\in I_{1l^\star }, j\in[N]}\min_{J_{ij}\in J^{\star}_{ij}}|J_{ij}+\breve{\Delta}_{l^\star ij}-1|+3MN\label{rarro2}
\end{eqnarray}
\eqref{bmn73} is obtained from \eqref{bmn+1}. \eqref{bmn77} follows similar to \eqref{klb2} and \eqref{bmn8} is concluded as the partial sum of harmonic series can be bounded above by logarithmic function i.e., $\sum_{i=1}^n\frac{1}{n}\le 1+\ln{n}$. Finally, \eqref{bmn9} is obtained from \eqref{conth4+} setting $k=1$,
\begin {eqnarray}
\sum_{l^\star< l\le l_1}{\max_{j\in[N],k'\in I_{1l}}\min(\lambda_{1k'},\gamma_{1lk'j}-\delta_{1lk'j})}\le {\sum_{1\le k'<\min_{m\in I_{1l^\star}}m} \lambda_{1k'}}\label{conth4++}
\end{eqnarray}

\end{enumerate}

\subsubsection {Proof of  \eqref{boundf}}
Now we prove  the bound \eqref{boundf} for the general $K$ and $n$. Our goal is to prove that,
\begin{eqnarray}
&&\sum_{\mu_{11}^{[n]}\in {\mathcal{Z}_{11}}^n,\cdots,\mu_{Kl_K}^{[n]}\in {\mathcal{Z}}_{Kl_K}^n}\prod_{k=1}^K~\prod_{t=1,\max_{j\in N}|E_{j}(t)-F_{j}(t)|\neq0}^n\frac{2Nf_{\max}}{\max_{j\in N}|E_{j}(t)-F_{j}(t)|}\nonumber\\
&\le&{(c_1+c_2\log{\bar{P}})}^{Kn}\label{boundf++}
\end{eqnarray}
Similar to \ref{prkn} let us define $\breve{\Delta}_{klij}(t)$, $J_{kij}(t)$, $J^{\star}_{kij}(t)$, $\hat{J}_{kij}(t)$ and $J_{k}(t)$ as
\begin{eqnarray}
\breve{\Delta}_{klij}(t)&\triangleq&((E_{j}(t))_{\sum_{r=1}^{i-1}\lambda_{kr}}^{\sum_{r=1}^i\lambda_{kr}})^{\gamma_{klij}}_{\delta_{klij}}-((F_{j}(t))_{\sum_{r=1}^{i-1}\lambda_{kr}}^{\sum_{r=1}^i\lambda_{kr}})^{\gamma_{klij}}_{\delta_{klij}}\\
J_{kij}(t)&=&(E_{j}(t))^{-\min(\lambda_{ki},\gamma_{kl^\star_k(t) ij})+\sum_{r=i}^M\lambda_{kr}}\bar{P}^{\min(\lambda_{ki},\gamma_{kl^\star_k(t) ij})}\nonumber\\
&&-(F_{j}(t))^{-\min(\lambda_{ki},\gamma_{kl^\star_k(t) ij})+\sum_{r=i}^M\lambda_{kr}}\bar{P}^{\min(\lambda_{ki},\gamma_{kl^\star_k(t) ij})}\nonumber\\
J^{\star}_{kij}(t)&=&\{-\bar{P}^{\min(\lambda_{ki},\gamma_{kl^\star_k(t) ij})},0,\bar{P}^{\min(\lambda_{ki},\gamma_{kl^\star_k(t) ij})}\}\\
\hat{J}_{kij}(t)&=&\underset{J_{kij}(t)\in J^{\star}_{kij}(t)}{\operatorname{argmin}}|J_{kij}(t)-\breve{\Delta}_{kl^\star_k(t) ij}(t)|\\
J_{k}(t)&=&L_{kl^\star_k(t)}(t)(\hat{J}_{kij}(t))
\end{eqnarray}
and define the number $l^\star_k(t)$ as the smallest integer where
\begin{eqnarray}
\max_{ i\in I_{kl^\star_k(t)},j\in[N]}|\breve{\Delta}_{kl^\star_k(t) ij}(t)|\ge2\label{conl+}
\end{eqnarray}
Therefore, similar to \ref{prkn} the left side of  \eqref{boundf++} is bounded as,
\begin{eqnarray}
&&\sum_{\mu_{11}^{[n]}\in {\mathcal{Z}_{11}}^n,\cdots,\mu_{Kl_K}^{[n]}\in {\mathcal{Z}}_{Kl_K}^n}\prod_{k=1}^K~\prod_{t=1,\max_{j\in N}|E_{j}(t)-F_{j}(t)|\neq0}^n\frac{2Nf_{\max}}{\max_{j\in N}|E_{j}(t)-F_{j}(t)|}\nonumber\\
&\le&\sum_{\mu_{11}^{[n]}\in {\mathcal{Z}_{11}}^n,\cdots,\mu_{Kl_K}^{[n]}\in {\mathcal{Z}}_{Kl_K}^n}\prod_{k=1}^K~\prod_{t=1,\max_{j\in N}|E_{j}(t)-F_{j}(t)|\neq0}^n\nonumber\\
&&\frac{2Nf_{\max}}{\min_{i\in I_{kl^\star_k(t)}}\bar{P}^{\sum_{r=1}^{i-1}\lambda_{kr}}(\max_{ i\in I_{kl^\star_k(t)},j\in[N]}|J_{kij}(t)-\breve{\Delta}_{kl^\star_k(t) ij}(t)|-1)}\label{bmn1q}\\
&\le&\sum_{\mu_{11}^{[n]}\in {\mathcal{Z}_{11}}^n,\cdots,\mu_{Kl_K}^{[n]}\in {\mathcal{Z}}_{Kl_K}^n}\prod_{k=1}^K~\prod_{t=1,3MN\le|{J_k(t)}+\mu_{kl^\star_k(t) }-\nu_{kl^\star_k(t) }|}^n\frac{2Nf_{\max}}{\min_{i\in I_{kl^\star_k(t)}}\bar{P}^{\sum_{r=1}^{i-1}\lambda_{kr}}}\nonumber\\
&&\times\frac{MN\Delta_2}{|{J_k(t)}+\mu_{kl^\star_k(t) }-\nu_{kl^\star_k(t) }|-3MN+1}\label{bmn2q}\\
&=&\prod_{k=1}^K~\prod_{t=1}^n\frac{2Nf_{\max}}{\min_{i\in I_{kl^\star_k(t)}}\bar{P}^{\sum_{r=1}^{i-1}\lambda_{kr}}}\nonumber\\
&&\times\big(\sum_{|\mu_{k1}(t)-\nu_{k1}(t)|\in {\mathcal{Z}_{k1}},\cdots,|\mu_{kl_k}(t)-\nu_{kl_k}(t)|\in {\mathcal{Z}}_{kl_k},|{J_k(t)}+\mu_{kl^\star_k(t) }-\nu_{kl^\star_k(t) }|<3MN} 1\nonumber\\
&&+\sum_{|\mu_{k1}(t)-\nu_{k1}(t)|\in {\mathcal{Z}_{k1}},\cdots,|\mu_{kl_k}(t)-\nu_{kl_k}(t)|\in {\mathcal{Z}}_{kl_k},3MN\le|{J_k(t)}+\mu_{kl^\star_k(t) }-\nu_{kl^\star_k(t) }|}\nonumber\\
&&\frac{MN\Delta_2}{|{J_k(t)}+\mu_{kl^\star_k(t) }-\nu_{kl^\star_k(t) }|-3MN+1}\big)\label{bmn3q}\\
&\le&\prod_{k=1}^K~\prod_{t=1}^n\frac{2Nf_{\max}}{\min_{i\in I_{kl^\star_k(t)}}\bar{P}^{\sum_{r=1}^{i-1}\lambda_{kr}}}\nonumber\\
&&\times\big(\prod_{l=1}^{l^\star_k(t)-1}2MN\Delta_2\big)\times\prod_{l=l^\star_k(t)+1}^{l_k}|\mathcal{Z}_{kl}|\big(3MN+\sum_{\hat{z}\in {\bar{\mathcal{Z}}_{kl^\star_k(t)}} }\frac{MN\Delta_2}{\hat{z}}\big)\label{bmn4q}\\
&\le&\prod_{k=1}^K~\prod_{t=1}^n2Nf_{\max}{(3MN\Delta_2)}^{l_k-1}(3MN+MN\Delta_2\ln(1+3MN\Delta_2{\bar{P}}))\label{bmn5q}\\
&=&(2Nf_{\max})^{Kn}{(3MN\Delta_2)}^{n\sum_{k=1}^Kl_k-Kn}(3MN+MN\Delta_2\ln(1+3MN\Delta_2{\bar{P}}))^{Kn}\label{bmn6q}
\end{eqnarray}
where (\ref{bmn3q}) follows from interchange of the summation and the product \footnote{ Note that for the arbitrary functions $f_1(x),f_2(x),\cdots,f_n(x)$ and the arbitrary sets of numbers $S_1,S_2,\cdots,S_n$ we have,
\begin{align}
&\sum_{a_1\in S_1,a_2\in S_2,\cdots,a_n\in S_n}\prod_{t=1}^nf_t(a_t)\nonumber\\
=&\sum_{a_1\in S_1}\sum_{a_2\in S_2}\cdots\sum_{a_n\in S_n}\prod_{t=1}^nf_t(a_t)\\
=&\sum_{a_1\in S_1}f_1(a_1)\times\sum_{a_2\in S_2}f_2(a_2)\times\cdots\times\sum_{a_n\in S_n}f_n(a_n)\label{ret}\\
=&\prod_{t=1}^n\sum_{a_t\in S_t}f_t(a_t)
\end{align}}.

\subsection{Proof of Theorem \ref{theoremIC}}\label{PIC} Consider the two user $(M_1,M_2,N_1,N_2)=(5,5,2,3)$ MIMO IC  with $(\alpha_{11},\alpha_{12},\alpha_{21},\alpha_{22})=(1,\frac{3}{4},\frac{2}{3},1)$ and $\beta_{12}=\frac{1}{4},\beta_{21}=\frac{1}{3}$ levels of partial CSIT. The bounds $d_1\le2$ and $d_2\le3$ follow from the single user bounds. So, let us prove the bound $d_1+d_2\le3+\frac{7}{9}$ with the aid of sum-set inequalities. 
\begin{enumerate}
\item{} Writing Fano's Inequality for the first receiver we have,
\begin{eqnarray}
nR_1&\le& I(\bar{\bf Y}^{[n]}_{1};\bar{\mathbf{X}}^{[n]}_{1}\mid\mathcal{G})\nonumber\\
&=&H(\bar{\bf Y}^{[n]}_{1}\mid \mathcal{G})-H(\bar{\bf Y}^{[n]}_{1}\mid \bar{\mathbf{X}}^{[n]}_{1},\mathcal{G})\label{BCet}
\end{eqnarray}
Multiplying the number $3$ to \eqref{BCet} we have,
\begin{eqnarray}
&&3nR_1\nonumber\\
&\le& 2H(\bar{\bf Y}^{[n]}_{1}\mid \mathcal{G})-2H(\bar{\bf Y}^{[n]}_{1}\mid \bar{\mathbf{X}}^{[n]}_{1},\mathcal{G})+ H(\bar{\bf Y}^{[n]}_{1}\mid \mathcal{G})-H(\bar{\bf Y}^{[n]}_{1}\mid \bar{\mathbf{X}}^{[n]}_{1},\mathcal{G})\\
&=& 2H(\bar{\bf Y}^{[n]}_{1}\mid \mathcal{G})-2H(\bar{\bf Y}^{[n]}_{1}\mid \bar{\mathbf{X}}^{[n]}_{1},\mathcal{G})\nonumber\\
&&+ H((\bar{\bf Y}^{[n]}_{1})_{\frac{2}{3}},(\bar{\bf Y}^{[n]}_{1})_{\frac{2}{3}}^{1}\mid \mathcal{G})-H((\bar{\bf Y}^{[n]}_{1})_{\frac{2}{3}},(\bar{\bf Y}^{[n]}_{1})_{\frac{2}{3}}^{1}\mid \bar{\mathbf{X}}^{[n]}_{1},\mathcal{G})\label{BCet1}\\
&=& 2H(\bar{\bf Y}^{[n]}_{1}\mid \mathcal{G})-2H(\bar{\bf Y}^{[n]}_{1}\mid \bar{\mathbf{X}}^{[n]}_{1},\mathcal{G})\nonumber\\
&&+ H((\bar{\bf Y}^{[n]}_{1})_{\frac{2}{3}}^{1}\mid \mathcal{G})-H((\bar{\bf Y}^{[n]}_{1})_{\frac{2}{3}}^{1}\mid \bar{\mathbf{X}}^{[n]}_{1},\mathcal{G})\nonumber\\
&&+ H((\bar{\bf Y}^{[n]}_{1})_{\frac{2}{3}}\mid (\bar{\bf Y}^{[n]}_{1})_{\frac{2}{3}}^{1},\mathcal{G})-H((\bar{\bf Y}^{[n]}_{1})_{\frac{2}{3}}\mid (\bar{\bf Y}^{[n]}_{1})_{\frac{2}{3}}^{1},\bar{\mathbf{X}}^{[n]}_{1},\mathcal{G})\label{BCet2}\\
&\le& \frac{16}{3}n\log{\bar{P}}-2H(\bar{\bf Y}^{[n]}_{1}\mid \bar{\mathbf{X}}^{[n]}_{1},\mathcal{G})-H((\bar{\bf Y}^{[n]}_{1})_{\frac{2}{3}}\mid (\bar{\bf Y}^{[n]}_{1})_{\frac{2}{3}}^{1},\bar{\mathbf{X}}^{[n]}_{1},\mathcal{G})\nonumber\\
&&+ H((\bar{\bf Y}^{[n]}_{1})_{\frac{2}{3}}^{1}\mid \mathcal{G})-H((\bar{\bf Y}^{[n]}_{1})_{\frac{2}{3}}^{1}\mid \bar{\mathbf{X}}^{[n]}_{1},\mathcal{G})\label{BCet3}\\
&\le& \frac{16}{3}n\log{\bar{P}}-2H(\bar{\bf Y}^{[n]}_{1}\mid \bar{\mathbf{X}}^{[n]}_{1},\mathcal{G})-H((\bar{\bf Y}^{[n]}_{1})_{\frac{2}{3}}\mid (\bar{\bf Y}^{[n]}_{1})_{\frac{2}{3}}^{1},\bar{\mathbf{X}}^{[n]}_{1},\mathcal{G})\nonumber\\
&&+H((\bar{\bf Y}^{[n]}_{1})_{\frac{2}{3}}^{1}\mid \bar{\mathbf{X}}^{[n]}_{2},\mathcal{G})\label{BCet4}\\
&\le& \frac{16}{3}n\log{\bar{P}}-2H((\bar{\mathbf{X}}^{[n]}_{2c})^1_{\frac{1}{2}})+H((\bar{\bf Y}^{[n]}_{1})_{\frac{2}{3}}^{1}\mid \bar{\mathbf{X}}^{[n]}_{2},\mathcal{G})+n~o(\log{\bar{P}})\label{BCet5}
\end{eqnarray}
where (\ref{BCet1}) follows from Definition \ref{powerlevel} and   (\ref{BCet2})  is true from  the chain rule. (\ref{BCet3}) is concluded as the entropy of a random variable is bounded by logarithm of the cardinality of it, i.e., $H((\bar{\bf Y}^{[n]}_{1})_{\frac{2}{3}}\mid (\bar{\bf Y}^{[n]}_{1})_{\frac{2}{3}}^{1},\bar{\mathbf{X}}^{[n]}_{1},\mathcal{G})\le \frac{4}{3}n\log{\bar{P}}$, $H(\bar{\bf Y}^{[n]}_{1}\mid \mathcal{G})\le 2n\log{\bar{P}}$.  \eqref{BCet4} is true as for any random variable $t$ and independent random variables $w_1$ and $w_2$ we have, $I(t;w_1)\le I(t;w_1\mid w_2)$. As a result, we have $I((\bar{\bf Y}^{[n]}_{1})_{\frac{2}{3}}^{1};\bar{\mathbf{X}}^{[n]}_{1}\mid\mathcal{G})\le I( (\bar{\bf Y}^{[n]}_{1})_{\frac{2}{3}}^{1};\bar{\mathbf{X}}^{[n]}_{1}\mid \bar{\mathbf{X}}^{[n]}_{2},\mathcal{G})$. \eqref{BCet5} yields from summation of  (\ref{BCet4}) and (\ref{firstlem}) from Lemma \ref{lemma1}. 
\item{}  Writing Fano's Inequality for the second receiver we have,
\begin{eqnarray}
&&nR_2\nonumber\\
&\le& I(\bar{\bf Y}^{[n]}_{2};\bar{\mathbf{X}}^{[n]}_{2}\mid\bar{\mathbf{X}}^{[n]}_{1},\mathcal{G})\nonumber\\
&=&H(\bar{\bf Y}^{[n]}_{2}\mid\bar{\mathbf{X}}^{[n]}_{1},\mathcal{G})\label{r15e}\\
&&nR_2\nonumber\\
&\le& I(\bar{\bf Y}^{[n]}_{2};\bar{\mathbf{X}}^{[n]}_{2}\mid\mathcal{G})\nonumber\\
&=&H(\bar{\bf Y}^{[n]}_{2}\mid\mathcal{G})-H(\bar{\bf Y}^{[n]}_{2}\mid\bar{\mathbf{X}}^{[n]}_{2},\mathcal{G})\label{r14e}
\end{eqnarray}
Let us remind from \eqref{rre3} that the received signal at the second receiver, i.e., $\bar{\mathbf{Y}}_{2}(t)$ is expressed as, 
\begin{eqnarray}
\bar{\mathbf{Y}}_{2}(t)&=&L_{2}^b(t)\left((\bar{\mathbf{X}}_{1a}(t))^1_{\frac{1}{3}}\bigtriangledown \bar{\mathbf{X}}_{2c}(t)\bigtriangledown (\bar{\bf X}_{1c}(t))^1_{\frac{2}{3}}\right)\label{rre3+t}
\end{eqnarray}
summing   \eqref{r15e} twice and \eqref{r14e}  together, we have,
\begin{eqnarray}
&&3nR_2\nonumber\\
&\le& H(\bar{\bf Y}^{[n]}_{2}\mid\mathcal{G})-H(\bar{\bf Y}^{[n]}_{2}\mid\bar{\mathbf{X}}^{[n]}_{2},\mathcal{G})+2H(\bar{\bf Y}^{[n]}_{2}\mid\bar{\mathbf{X}}^{[n]}_{1},\mathcal{G})\label{BCet6}\\
&\le& {3}n\log{\bar{P}}-H(\bar{\bf Y}^{[n]}_{2}\mid\bar{\mathbf{X}}^{[n]}_{2},\mathcal{G})+2H(\bar{\bf Y}^{[n]}_{2}\mid\bar{\mathbf{X}}^{[n]}_{1},\mathcal{G})\label{BCet7}\\
&=& {3}n\log{\bar{P}}-H(\bar{\bf Y}^{[n]}_{2}\mid\bar{\mathbf{X}}^{[n]}_{2},\mathcal{G})+2H(\bar{\bf X}^{[n]}_{2c})\label{BCet8}\\
&\le& {3}n\log{\bar{P}}-H((\bar{\bf Y}^{[n]}_{1})_{\frac{2}{3}}^{1}\mid \bar{\mathbf{X}}^{[n]}_{2},\mathcal{G})+2H(\bar{\bf X}^{[n]}_{2c})+n~o(\log{\bar{P}})\label{BCet9}\\
&=& {3}n\log{\bar{P}}-H((\bar{\bf Y}^{[n]}_{1})_{\frac{2}{3}}^{1}\mid \bar{\mathbf{X}}^{[n]}_{2},\mathcal{G})+2H((\bar{\bf X}^{[n]}_{2c})^1_{\frac{1}{2}})+2H((\bar{\bf X}^{[n]}_{2c})_{\frac{1}{2}}\mid(\bar{\bf X}^{[n]}_{2c})^1_{\frac{1}{2}})\nonumber\\
&&+n~o(\log{\bar{P}})\label{BCet10}\\
&\le& 6n\log{\bar{P}}-H((\bar{\bf Y}^{[n]}_{1})_{\frac{2}{3}}^{1}\mid \bar{\mathbf{X}}^{[n]}_{2},\mathcal{G})+2H((\bar{\bf X}^{[n]}_{2c})^1_{\frac{1}{2}})+n~o(\log{\bar{P}})\label{BCet11}
\end{eqnarray}
\eqref{BCet7} is concluded as the entropy of a random variable is bounded by logarithm of the cardinality of it, i.e., $H(\bar{\bf Y}^{[n]}_{2}\mid\mathcal{G})\le {3}n\log{\bar{P}}$. \eqref{BCet8} follows from \eqref{rre3+t} and \eqref{BCet10} yields from the chain rule. \eqref{BCet11} is concluded similar to \eqref{BCet7}  as $H((\bar{\bf X}^{[n]}_{2c})_{\frac{1}{2}}\mid(\bar{\bf X}^{[n]}_{2c})^1_{\frac{1}{2}})\le \frac{3}{2}n\log{\bar{P}}$. Finally, \eqref{BCet9} is obtained as,
\begin{eqnarray}
&&H((\bar{\bf Y}^{[n]}_{1})_{\frac{2}{3}}^{1}\mid \bar{\mathbf{X}}^{[n]}_{2},\mathcal{G})\nonumber\\
&=&H( (\bar{{\bf X}}^{[n]}_{1c})^1_{\frac{2}{3}})\label{mnx1}\\
&=&H( (\bar{{\bf X}}^{[n]}_{1c})^1_{\frac{2}{3}}\mid\bar{\mathbf{X}}^{[n]}_{2},\mathcal{G})\label{kkj1}\\
&\le&H(\bar{\bf Y}^{[n]}_{2}\mid\bar{\mathbf{X}}^{[n]}_{2},\mathcal{G})+n~o(\log{\bar{P}})\label{kkj2}
\end{eqnarray}
\eqref{mnx1} yields from \eqref{rrre2}. (\ref{kkj1}) follows from independence of $ (\bar{{\bf X}}^{[n]}_{1c})^1_{\frac{2}{3}}$ and $\bar{\mathbf{X}}^{[n]}_{2}$, and (\ref{kkj2}) is true as any $3$ components of $\bar{\bf Y}^{[n]}_{2}$ is a bounded density linear combination of random variables including components of and $(\bar{{\bf X}}^{[n]}_{1c})^1_{\frac{2}{3}}$ from (\ref{rre3}). To further clarify it, let us present the following illustration of Theorem \ref{Theorem AIS04}. For random variable $W$ independent of $\mathcal{G}$, any $n$-letter real-valued random variables $r_1^{[n]},r_2^{[n]},r_3^{[n]}$ independent of $\mathcal{G}$ and $n$-letter real-valued random variable $s_1^{[n]},s_2^{[n]},s_3^{[n]},s_4^{[n]}$  where  for any $t\in[n]$,
\begin{eqnarray}
s_1(t)&=&L_1^{\vec{\gamma}\vec{\delta}}(r_1(t),r_2(t),r_3(t))\label{grg1}\\
s_2(t)&=&L_2^{\vec{\gamma}\vec{\delta}}(r_1(t),r_2(t),r_3(t))\label{grg2}\\
s_3(t)&=&L_3^b(r_1(t),r_2(t),r_3(t))\label{grg4}\\
s_4(t)&=&L_4^b(r_1(t),r_2(t),r_3(t))\label{grg5}
\end{eqnarray}
we have
\begin{eqnarray}
H(s_1^{[n]},s_2^{[n]}\mid W,\mathcal{G})\le H(s_3^{[n]},s_4^{[n]}\mid W,\mathcal{G})+n~o(\log{\bar{P}})\label{grg3}.
 \end{eqnarray}
 (\ref{kkj2}) is concluded from \eqref{grg3} \footnote{ This also could be concluded from Theorem 1 in \cite{Arash_Jafar_PN}.}.
\item{} Summing  (\ref{BCet5}) and (\ref{BCet11}) together we have,
\begin{eqnarray}
3nR_1+3nR_2&\le&\frac{34}{3}n\log{\bar{P}}+n~o(\log{\bar{P}})\label{kkj4}
\end{eqnarray}
Dividing (\ref{kkj4}) by $3\log{\bar{P}}$, $d_1+d_2\le3+\frac{7}{9}$ is concluded.
\end{enumerate}

In order to prove the region \eqref{L_new1}, the bound $\frac{d_1}{2}+\frac{d_2}{3}\le{\frac{3}{2}}$ is proved as follows.

\begin{enumerate}
\item{} Writing Fano's Inequality for the first and second receivers we have,
\begin{eqnarray}
nR_1&\le& I(\bar{\bf Y}^{[n]}_{1};\bar{\mathbf{X}}^{[n]}_{1}\mid\mathcal{G})\nonumber\\
&=&H(\bar{\bf Y}^{[n]}_{1}\mid \mathcal{G})-H(\bar{\bf Y}^{[n]}_{1}\mid \bar{\mathbf{X}}^{[n]}_{1},\mathcal{G})\label{fano-1}\\
nR_2&\le& I(\bar{\bf Y}^{[n]}_{2};\bar{\mathbf{X}}^{[n]}_{2}\mid\bar{\mathbf{X}}^{[n]}_{1},\mathcal{G})\nonumber\\
&=&H(\bar{\bf Y}^{[n]}_{2}\mid \bar{\mathbf{X}}^{[n]}_{1},\mathcal{G})\label{fano-2}
\end{eqnarray}
summing  \eqref{fano-1}  three times and \eqref{fano-2} twice we have,
\begin{eqnarray}
&&3nR_1+2nR_2\nonumber\\
&\le&3H(\bar{\bf Y}^{[n]}_{1}\mid \mathcal{G})-3H(\bar{\bf Y}^{[n]}_{1}\mid \bar{\mathbf{X}}^{[n]}_{1},\mathcal{G})+2H(\bar{\bf Y}^{[n]}_{2}\mid \bar{\mathbf{X}}^{[n]}_{1},\mathcal{G})\label{fano-3}\\
&\le& 6n\log{\bar{P}}-3H(\bar{\bf Y}^{[n]}_{1}\mid \bar{\mathbf{X}}^{[n]}_{1},\mathcal{G})+2H(\bar{\bf Y}^{[n]}_{2}\mid \bar{\mathbf{X}}^{[n]}_{1},\mathcal{G})\label{fano-4}\\
&=& 6n\log{\bar{P}}-3H(\bar{\bf Y}^{[n]}_{1}\mid \bar{\mathbf{X}}^{[n]}_{1},\mathcal{G})+2H(\bar{  Y}^{[n]}_{21},\bar{  Y}^{[n]}_{22},\bar{ Y}^{[n]}_{23}\mid \bar{\mathbf{X}}^{[n]}_{1},\mathcal{G})\label{fano-5}\\
&\le& 6n\log{\bar{P}}-3H(\bar{\bf Y}^{[n]}_{1}\mid \bar{\mathbf{X}}^{[n]}_{1},\mathcal{G})+H(\bar{  Y}^{[n]}_{21},\bar{  Y}^{[n]}_{22}\mid \bar{\mathbf{X}}^{[n]}_{1},\mathcal{G})\nonumber\\
&&+H(\bar{  Y}^{[n]}_{21},\bar{ Y}^{[n]}_{23}\mid \bar{\mathbf{X}}^{[n]}_{1},\mathcal{G})+H(\bar{  Y}^{[n]}_{22},\bar{ Y}^{[n]}_{23}\mid \bar{\mathbf{X}}^{[n]}_{1},\mathcal{G})\label{fano-6}\\
&\le& 9n\log{\bar{P}}+n~o(\log{\bar{P}})\label{fano-7}
\end{eqnarray}
where (\ref{fano-4}) is true as the entropy of a random variable is bounded by logarithm of the cardinality of it, i.e., $H(\bar{\bf Y}^{[n]}_{1}\mid \mathcal{G})\le 2n\log{\bar{P}}$. (\ref{fano-6}) is concluded from sub-modularity properties of entropy function, i.e., for any $m$ random variables $\{X_1,X_2,\cdots,X_m\}$ where we define $X_{k+m}$ as $X_k$ for positive numbers of $k$ we have, 
\begin{eqnarray}
nH(X_1,\cdots,X_m)\le \sum_{i=1}^mH(X_i,X_{i+1},\cdots,X_{i+n})
\end{eqnarray}
if $n\le m$. To prove \eqref{fano-7}, consider any of the three entropies $H(\bar{  Y}^{[n]}_{21},\bar{  Y}^{[n]}_{22}\mid \bar{\mathbf{X}}^{[n]}_{1},\mathcal{G})$, $H(\bar{  Y}^{[n]}_{21},\bar{  Y}^{[n]}_{23}\mid \bar{\mathbf{X}}^{[n]}_{1},\mathcal{G})$ and $H(\bar{  Y}^{[n]}_{22},\bar{  Y}^{[n]}_{23}\mid \bar{\mathbf{X}}^{[n]}_{1},\mathcal{G})$. For instance, consider $H(\bar{  Y}^{[n]}_{21},\bar{  Y}^{[n]}_{23}\mid \bar{\mathbf{X}}^{[n]}_{1},\mathcal{G})$ and bound it as,
\begin{eqnarray}
&&H(\bar{  Y}^{[n]}_{21},\bar{  Y}^{[n]}_{23}\mid \bar{\mathbf{X}}^{[n]}_{1},\mathcal{G})\nonumber\\
&=&H((\bar{  Y}^{[n]}_{21})_{\frac{1}{2}}^1,(\bar{  Y}^{[n]}_{23})_{\frac{1}{2}}^1\mid \bar{\mathbf{X}}^{[n]}_{1},\mathcal{G})+H((\bar{  Y}^{[n]}_{21})_{\frac{1}{2}},(\bar{  Y}^{[n]}_{23})_{\frac{1}{2}}\mid (\bar{  Y}^{[n]}_{21})_{\frac{1}{2}}^1,(\bar{  Y}^{[n]}_{23})_{\frac{1}{2}}^1,\bar{\mathbf{X}}^{[n]}_{1},\mathcal{G})\label{fano-9}\\
&\le&H((\bar{  Y}^{[n]}_{21})_{\frac{1}{2}}^1,(\bar{  Y}^{[n]}_{23})_{\frac{1}{2}}^1\mid \bar{\mathbf{X}}^{[n]}_{1},\mathcal{G})+n~\log{\bar{P}}\\
&\le&H(\bar{\bf Y}^{[n]}_{1}\mid \bar{\mathbf{X}}^{[n]}_{1},\mathcal{G})+n\log{\bar{P}}+n~o(\log{\bar{P}})\label{fano-10}
\end{eqnarray}
\eqref{fano-9} yields from the chain rule and \eqref{fano-10} is true as conditioned on $\bar{\mathbf{X}}^{[n]}_{1}$, the random variables $(\bar{  Y}_{21}(t))_{\frac{1}{2}}^1$ and $(\bar{  Y}_{23}(t))_{\frac{1}{2}}^1$ are bounded density linear combinations of random variables  $(\bar{{ X}}^{[n]}_{23})^1_{\frac{1}{2}}$,  $(\bar{{ X}}^{[n]}_{24})^1_{\frac{1}{2}}$ and  $(\bar{{ X}}^{[n]}_{25})^1_{\frac{1}{2}}$ while $\bar{  Y}_{11}(t)$ and $\bar{  Y}_{12}(t)$ are bounded density linear combinations of random variables $(\bar{{ X}}^{[n]}_{21})^1_{\frac{1}{4}}$,  $(\bar{{ X}}^{[n]}_{22})^1_{\frac{1}{4}}$, $(\bar{{ X}}^{[n]}_{23})^1_{\frac{1}{2}}$,  $(\bar{{ X}}^{[n]}_{24})^1_{\frac{1}{2}}$ and  $(\bar{{ X}}^{[n]}_{25})^1_{\frac{1}{2}}$, see (\ref{rre3}). Thus, \eqref{fano-9} is concluded similar to  (\ref{kkj2}). Dividing (\ref{fano-7}) by $6\log{\bar{P}}$, $\frac{d_1}{2}+\frac{d_2}{3}\le{\frac{3}{2}}$ is concluded.
\end{enumerate}

\section{Proof of Lemma \ref{lemma1}} \label{appendix1}
As $(\bar{\mathbf{X}}^{[n]}_{2c})^1_{\frac{1}{2}}$ is independent from  $\bar{\mathbf{X}}^{[n]}_{1}$, (\ref{firstlem}) can be written as,
\begin{eqnarray}
2H((\bar{\mathbf{X}}^{[n]}_{2c})^1_{\frac{1}{2}}\mid\bar{\mathbf{X}}^{[n]}_{1}){+}H( (\bar{\bf Y}^{[n]}_{1})_{\frac{2}{3}}^{1}\mid\bar{\mathbf{X}}^{[n]}_{1},\mathcal{G})&{\le}&3H(\bar{\bf Y}^{n}_{1}\mid \bar{\mathbf{X}}^{[n]}_{1},\mathcal{G})+n~o~(\log{\bar{P}})\label{sse}
\end{eqnarray}
(\ref{sse}) follows from the chain rule, i.e.,
\begin{eqnarray}
&&H(\bar{\bf Y}^{[n]}_{1}\mid \bar{\mathbf{X}}^{[n]}_{1},\mathcal{G})\nonumber\\
&=&H((\bar{\bf Y}^{[n]}_{1})_{\frac{2}{3}}^{1},(\bar{\bf Y}^{n}_{1})_{\frac{2}{3}}\mid\bar{\mathbf{X}}^{[n]}_{1},\mathcal{G})\nonumber\\
&=&H((\bar{\bf Y}^{[n]}_{1})_{\frac{2}{3}}^{1}\mid\bar{\mathbf{X}}^{[n]}_{1},\mathcal{G})+H((\bar{\bf Y}^{n}_{1})_{\frac{2}{3}}\mid(\bar{\bf Y}^{[n]}_{1})_{\frac{2}{3}}^{1},\mathcal{G})
\end{eqnarray}
Starting from the left side of (\ref{sse}) we have,
\begin{eqnarray}
&&2H((\bar{\mathbf{X}}^{[n]}_{2c})^1_{\frac{1}{2}}\mid\bar{\mathbf{X}}^{[n]}_{1}){+}H( (\bar{\bf Y}^{[n]}_{1})_{\frac{2}{3}}^{1}\mid\bar{\mathbf{X}}^{[n]}_{1},\mathcal{G})\nonumber\\
&=&2H((\bar{{X}}^{[n]}_{23})^1_{\frac{1}{2}},(\bar{{X}}^{[n]}_{24})^1_{\frac{1}{2}},(\bar{{X}}^{[n]}_{25})^1_{\frac{1}{2}}\mid\bar{\mathbf{X}}^{[n]}_{1}){+}H( (\bar{\bf Y}^{[n]}_{1})_{\frac{2}{3}}^{1}\mid\bar{\mathbf{X}}^{[n]}_{1},\mathcal{G})\label{sse2}\\
&\le&2H((\bar{{X}}^{[n]}_{23})^1_{\frac{1}{2}},(\bar{{X}}^{[n]}_{24})^1_{\frac{1}{2}},(\bar{{X}}^{[n]}_{25})^1_{\frac{1}{2}}, (\bar{\bf Y}^{[n]}_{1})_{\frac{2}{3}}^{1}\mid\bar{\mathbf{X}}^{[n]}_{1},\mathcal{G}){+}H( (\bar{\bf Y}^{[n]}_{1})_{\frac{2}{3}}^{1}\mid\bar{\mathbf{X}}^{[n]}_{1},\mathcal{G})\label{sse3}\\
&=&2H((\bar{{X}}^{[n]}_{23})^1_{\frac{1}{2}},(\bar{{X}}^{[n]}_{24})^1_{\frac{1}{2}},(\bar{{X}}^{[n]}_{25})^1_{\frac{1}{2}}\mid(\bar{\bf Y}^{[n]}_{1})_{\frac{2}{3}}^{1},\bar{\mathbf{X}}^{[n]}_{1},\mathcal{G}){+}3H( (\bar{\bf Y}^{[n]}_{1})_{\frac{2}{3}}^{1}\mid\bar{\mathbf{X}}^{[n]}_{1},\mathcal{G})\label{sse4}\\
&\le&H((\bar{{X}}^{[n]}_{23})^1_{\frac{1}{2}},(\bar{{X}}^{[n]}_{24})^1_{\frac{1}{2}}\mid(\bar{\bf Y}^{[n]}_{1})_{\frac{2}{3}}^{1},\bar{\mathbf{X}}^{[n]}_{1},\mathcal{G})+H((\bar{{X}}^{[n]}_{23})^1_{\frac{1}{2}},(\bar{{X}}^{[n]}_{25})^1_{\frac{1}{2}}\mid(\bar{\bf Y}^{[n]}_{1})_{\frac{2}{3}}^{1},\bar{\mathbf{X}}^{[n]}_{1},\mathcal{G})\nonumber\\
&&+H((\bar{{X}}^{[n]}_{24})^1_{\frac{1}{2}},(\bar{{X}}^{[n]}_{25})^1_{\frac{1}{2}}\mid(\bar{\bf Y}^{[n]}_{1})_{\frac{2}{3}}^{1},\bar{\mathbf{X}}^{[n]}_{1},\mathcal{G})+3H((\bar{\bf Y}^{[n]}_{1})_{\frac{2}{3}}^{1}\mid\bar{\mathbf{X}}^{[n]}_{1},\mathcal{G})\label{sse5}\\
&=&H((\bar{{X}}^{[n]}_{23})^1_{\frac{1}{2}},(\bar{{X}}^{[n]}_{24})^1_{\frac{1}{2}},(\bar{\bf Y}^{[n]}_{1})_{\frac{2}{3}}^{1}\mid\bar{\mathbf{X}}^{[n]}_{1},\mathcal{G})+H((\bar{{X}}^{[n]}_{23})^1_{\frac{1}{2}},(\bar{{X}}^{[n]}_{25})^1_{\frac{1}{2}},(\bar{\bf Y}^{[n]}_{1})_{\frac{2}{3}}^{1}\mid\bar{\mathbf{X}}^{[n]}_{1},\mathcal{G})\nonumber\\
&&+H((\bar{{X}}^{[n]}_{24})^1_{\frac{1}{2}},(\bar{{X}}^{[n]}_{25})^1_{\frac{1}{2}},(\bar{\bf Y}^{[n]}_{1})_{\frac{2}{3}}^{1}\mid\bar{\mathbf{X}}^{[n]}_{1},\mathcal{G})\label{sse6}\\
&\le&3H(\bar{\bf Y}^{[n]}_{1}\mid \bar{\mathbf{X}}^{[n]}_{1},\mathcal{G})+n~o~(\log{\bar{P}})\label{sse7}
\end{eqnarray}
(\ref{sse2}) follows from the definition of $\bar{\mathbf{X}}^{[n]}_{2c}$. (\ref{sse4}) and (\ref{sse6}) is true from the chain rule. (\ref{sse5}) follows similar to \eqref{fano-6} from sub-modularity properties of entropy function. Finally, (\ref{sse7}) is true from Theorem  \ref{Theorem AIS03} as any of the three entropies in \eqref{sse6} are less than $H(\bar{\bf Y}_{1}^{[n]}\mid \bar{\mathbf{X}}^{[n]}_{1},\mathcal{G})+n~o~(\log{\bar{P}})$, i.e., 
\begin{eqnarray}
 H((\bar{{X}}^{[n]}_{24})^1_{\frac{1}{2}},(\bar{{X}}^{[n]}_{25})^1_{\frac{1}{2}},(\bar{\bf Y}^{[n]}_{1})_{\frac{2}{3}}^{1}\mid\bar{\mathbf{X}}^{[n]}_{1},\mathcal{G})&\le& H(\bar{\bf Y}_{1}^{[n]}\mid \bar{\mathbf{X}}^{[n]}_{1},\mathcal{G})+n~o~(\log{\bar{P}})\label{thg1}
\end{eqnarray}
To further illuminate how (\ref{thg1}) is concluded from Theorem \ref{Theorem AIS04}, define $Z_1(t),Z_2(t),Z_{11}(t),Z_{12}(t),Z_{21}(t),Z_{22}(t)$ and $W$ from \eqref{rrre2} for all $t\in[n]$ as,
\begin{eqnarray}
Z_1(t)&=&\bar{ Y}_{11}(t)=L_{1}^b(t)(\bar{\bf X}_{1c}(t))\nonumber\\
&&+L_{2}^b(t)\left((\bar{{X}}_{21}(t))^1_{\frac{1}{4}},(\bar{{X}}_{22}(t))^1_{\frac{1}{4}},(\bar{{X}}_{23}(t))^1_{\frac{1}{2}},(\bar{{X}}_{24}(t))^1_{\frac{1}{2}},(\bar{{X}}_{25}(t))^1_{\frac{1}{2}}\right)\\
Z_2(t)&=&\bar{ Y}_{12}(t)=L_{3}^b(t)(\bar{\bf X}_{1c}(t))\nonumber\\
&&+L_{4}^b(t)\left((\bar{{X}}_{21}(t))^1_{\frac{1}{4}},(\bar{{X}}_{22}(t))^1_{\frac{1}{4}},(\bar{{X}}_{23}(t))^1_{\frac{1}{2}},(\bar{{X}}_{24}(t))^1_{\frac{1}{2}},(\bar{{X}}_{25}(t))^1_{\frac{1}{2}}\right)\\
Z_{11}(t)&=&(\bar{ Y}_{11}(t))_{\frac{2}{3}}^1=L_{1}^b(t)((\bar{\bf X}_{1c}(t))^1_{\frac{2}{3}})+L_{2}^b(t)\left((\bar{{X}}_{21}(t))^1_{\frac{11}{12}},(\bar{{X}}_{22}(t))^1_{\frac{11}{12}}\right)\\
Z_{21}(t)&=&(\bar{ Y}_{12}(t))_{\frac{2}{3}}^1=L_{3}^b(t)((\bar{\bf X}_{1c}(t))^1_{\frac{2}{3}})+L_{4}^b(t)\left((\bar{{X}}_{21}(t))^1_{\frac{11}{12}},(\bar{{X}}_{22}(t))^1_{\frac{11}{12}}\right)\\
Z_{12}(t)&=&(\bar{{X}}_{24}(t))^1_{\frac{1}{2}}\\
Z_{22}(t)&=&(\bar{{X}}_{25}(t))^1_{\frac{1}{2}}\\
W&=&\bar{\mathbf{X}}^{[n]}_{1}
\end{eqnarray}
where  $\lambda_1,\lambda_2$ and the singleton sets of $I_1,I_2$ are derived as
\begin {eqnarray}
\lambda_{ij}=\frac{1}{2}&,&\forall i,j\in\{1,2\}\\
I_{11}=I_{21}=\{2\}&,&I_{12}=I_{22}=\{1\}
\end{eqnarray}
So, the condition \eqref{t3con3} is satisfied and \eqref{thg1} is concluded from Theorem \ref{Theorem AIS04}, i.e.,
\begin {eqnarray}
H(Z_{11}^{[n]},Z_{12}^{[n]},Z_{21}^{[n]},Z_{22}^{[n]}\mid\bar{\mathbf{X}}^{[n]}_{1},\mathcal{G})&\le&H(Z_{1}^{[n]},Z_{2}^{[n]}\mid\bar{\mathbf{X}}^{[n]}_{1},\mathcal{G})+n~o~(\log{\bar{P}})
\end{eqnarray}

\bibliographystyle{IEEEtran}
\bibliography{Thesis}

\begin{thebibliography}{10}
\providecommand{\url}[1]{#1}
\csname url@samestyle\endcsname
\providecommand{\newblock}{\relax}
\providecommand{\bibinfo}[2]{#2}
\providecommand{\BIBentrySTDinterwordspacing}{\spaceskip=0pt\relax}
\providecommand{\BIBentryALTinterwordstretchfactor}{4}
\providecommand{\BIBentryALTinterwordspacing}{\spaceskip=\fontdimen2\font plus
\BIBentryALTinterwordstretchfactor\fontdimen3\font minus
  \fontdimen4\font\relax}
\providecommand{\BIBforeignlanguage}[2]{{%
\expandafter\ifx\csname l@#1\endcsname\relax
\typeout{** WARNING: IEEEtran.bst: No hyphenation pattern has been}%
\typeout{** loaded for the language `#1'. Using the pattern for}%
\typeout{** the default language instead.}%
\else
\language=\csname l@#1\endcsname
\fi
#2}}
\providecommand{\BIBdecl}{\relax}
\BIBdecl

\bibitem{Ruzsa}
I.~Z. Ruzsa, ``Sumsets and entropy,'' \emph{Random Structures and Algorithms},
  vol.~34, no.~1, pp. 1--10, 2009.

\bibitem{Cover_Thomas}
T.~M. Cover and J.~A. Thomas, \emph{Elements of Information Theory}.\hskip 1em
  plus 0.5em minus 0.4em\relax Wiley, 1991.

\bibitem{Etkin_Tse_Wang}
R.~Etkin, D.~Tse, and H.~Wang, ``{Gaussian interference channel capacity to
  within one bit},'' \emph{IEEE Transactions on Information Theory}, vol.~54,
  no.~12, pp. 5534--5562, 2008.

\bibitem{Avestimehr_Diggavi_Tse}
A.~Avestimehr, S.~Diggavi, and D.~Tse, ``Wireless network information flow: A
  deterministic approach,'' \emph{IEEE Trans. on Inf. Theory}, vol.~57, pp.
  1872--1905, 2011.

\bibitem{Cadambe_Jafar_int}
V.~Cadambe and S.~Jafar, ``{Interference Alignment and the Degrees of Freedom
  of the $K$ user Interference Channel},'' \emph{IEEE Transactions on
  Information Theory}, vol.~54, no.~8, pp. 3425--3441, Aug. 2008.

\bibitem{Madiman}
I.~Kontoyiannis and M.~Madiman, ``Sumset and inverse sumset inequalities for
  differential entropy and mutual information,'' \emph{IEEE Transactions on
  Information Theory}, vol.~60, no.~8, pp. 4503--4514, 2014.

\bibitem{Etkin_Ordentlich}
R.~Etkin and E.~Ordentlich, ``The degrees-of-freedom of the {K-User Gaussian}
  interference channel is discontinuous at rational channel coefficients,''
  \emph{IEEE Trans. on Information Theory}, vol.~55, pp. 4932--4946, Nov. 2009.

\bibitem{Jafar_FnT}
S.~Jafar, ``{Interference Alignment: A New Look at Signal Dimensions in a
  Communication Network},'' in \emph{Foundations and Trends in Communication
  and Information Theory}, 2011, pp. 1--136.

\bibitem{Lapidoth_Shamai_Wigger_BC}
A.~Lapidoth, S.~Shamai, and M.~Wigger, ``On the capacity of fading {MIMO}
  broadcast channels with imperfect transmitter side-information,'' in
  \emph{Proceedings of 43rd Annual Allerton Conference on Communications,
  Control and Computing}, Sep. 28-30, 2005.

\bibitem{Arash_Jafar_PN}
A.~G. Davoodi and S.~A. Jafar, ``Aligned image sets under channel uncertainty:
  Settling conjectures on the collapse of degrees of freedom under finite
  precision {CSIT},'' \emph{IEEE Transactions on Information Theory}, vol.~62,
  no.~10, pp. 5603--5618, 2016.

\bibitem{Arash_Jafar_TC}
------, ``{Transmitter Cooperation under Finite Precision CSIT:A GDoF
  Perspective},'' \emph{IEEE Transactions on Information Theory}, 2016.

\bibitem{Arash_Jafar_IC}
------, ``Generalized {D}egrees of {F}reedom of the {S}ymmetric {$K$}-{U}ser
  {I}nterference {C}hannel under {F}inite {P}recision {CSIT},'' \emph{arXiv
  preprint arXiv:1601.06463}, 2016.

\bibitem{Arash_Bofeng_Jafar_BC}
A.~G. Davoodi, B.~Yuan, and S.~A. Jafar, ``{GDoF} of the {MISO BC}: Bridging
  the gap between finite precision and perfect {CSIT},'' \emph{arXiv preprint
  arXiv:1602.02203}, 2016.

\bibitem{Arash_Jafar_Coherence}
A.~G. Davoodi and S.~A. Jafar, ``Network coherence time matters -- aligned
  image sets and the degrees of freedom of interference networks with finite
  precision {CSIT} and perfect {CSIR},'' \emph{arXiv preprint
  arXiv:1705.02775}, May 2017.

\bibitem{Arash_Jafar_MIMOsym_ArXiv}
------, ``Aligned image sets and the {GDoF} of symmetric {MIMO} interference
  channel with partial {CSIT},'' \emph{arXiv preprint arXiv:1705.00769}, April.
  2017.

\bibitem{Gou_Jafar}
T.~Gou and S.~A. Jafar, ``Optimal use of current and outdated channel state
  information: Degrees of freedom of the \textsc{MISO BC} with mixed
  \textsc{CSIT},'' \emph{IEEE Communication Letters}, vol.~16, no.~7, pp. 1084
  -- 1087, July 2012.

\bibitem{Bofeng_Arash_Jafar_ArXiv}
B.~Yuan, A.~G. Davoodi, and S.~A. Jafar, ``{DoF} region of the {MIMO}
  interference channel with partial {CSIT},'' \emph{Available on ArXiv}, April.
  2017.

\bibitem{Arash_Jafar_MIMOBC_Region}
A.~Gholami~Davoodi and S.~A. Jafar, ``Degrees of freedom region of general
  {$(M,N_1,N_2)$} {MIMO} broadcast channel with arbitrary levels of partial
  {CSIT}: Sum-set inequalities application,'' \emph{Available on ArXiv}, 2017.

\bibitem{subm}
A.~Schrijver, ``Combinatorial optimization: polyhedra and efficiency,''
  \emph{Springer Science and Business Media}, vol.~24, Dec 2002.

\bibitem{Hao_Rassouli_Clerckx}
C.~Hao, B.~Rassouli, and B.~Clerckx, ``Achievable {DoF} regions of {MIMO}
  networks with imperfect {CSIT},'' \emph{http://arxiv.org/abs/1603.07513},
  vol. abs/1603.07513, 2016.

\bibitem{stein}
E.~M. Stein and R.~Shakarchi, ``Real analysis: Measure theory, integration, and
  hilbert spaces,'' \emph{Princeton University Press}, 2009.

\end{thebibliography}

\end{document}